\documentclass[3p,twocolumn]{elsarticle}
\usepackage{amsmath, amssymb}
\usepackage{graphicx}
\usepackage{listings}
\usepackage{xcolor}
\usepackage{hyperref}
\usepackage{float}
\usepackage{booktabs}
\usepackage{tikz}
\usepackage{natbib}
\usepackage{multirow}
\usepackage{appendix}
\usepackage{subcaption}
\usepackage[absolute,overlay]{textpos}
\setcitestyle{open={(},authoryear,close={)}}
\journal{Astroparticle Physics}
\begin{document}
    \def\figureautorefname{Fig.}
    \def\tableautorefname{Tab.}
    \def\sectionautorefname{Sec.}
    \def\subsectionautorefname{Sec.}
    \def\subsubsectionautorefname{Sec.}
    \def\equationautorefname{Eq.}
    \def\appendixautorefname{} 
    \def\appendixname{App.}
    \begin{textblock*}{16cm}(2.75cm,26.65cm) 
        \begin{footnotesize}
            \noindent \textit{Astroparticle Pysics's internal reference: ??????, volume: ???} \newline
            \textit{International Standard Serial Number (ISSN): ????-????} \newline
            \textit{Digital Object Identifier (DOI): \url{https://doi.org/??.????/j.astropartphys.????.??????}}
        \end{footnotesize}
    \end{textblock*}
    \begin{frontmatter}
        \title{
            Simulating an imaging atmospheric radio telescope to observe cosmic gamma rays and cosmic neutrinos 
        }
        \author[mpik]{Sebastian Achim Mueller}
        \author[mpik,imprs]{Anne Timmermans\corref{corresponding_author_email}}
        \author[imapp,nikhef]{Juan Ammerman-Yebra}
        \author[imapp,nikhef]{Harm Schoorlemmer}
        \cortext[corresponding_author_email]{anne.timmermans@mpi-hd.mpg.de}
        \affiliation[mpik]{
            organization={Max-Planck-Institute for Nuclear Physics},
            addressline={Saupfercheckweg\,1},
            city={Heidelberg},
            postcode={69117},
            state={Germany}
        }
        \affiliation[imprs]{
            organization={Fellow of the International Max Planck Research School for Astronomy and Cosmic Physics at the University of Heidelberg (IMPRS-HD)}
        }
        \affiliation[imapp]{
            organization={IMAPP Radboud University Nijmegen},
            city={Nijmegen},
            state={The Netherlands}
        }
        \affiliation[nikhef]{
            organization={Nationaal Instituut voor Kernfysica en Hoge Energie Fysica (NIKHEF)},
            addressline={Science Park},
            city={Amsterdam},
            state={The Netherlands}
        }
        \begin{abstract}
            Observations of cosmic gamma~rays with energies above the  10$^{12}$ electronvolts (TeV) regime are often a direct probe of events where our theoretical predictions are challenged by extreme conditions.
Because high energetic cosmic gamma~rays are rare, one needs collection areas with the size of soccer fields (10$^{5}$\,m$^2$) or more to gather significant counting statistics in short time.
The imaging atmospheric Cherenkov telescope detects cosmic gamma~rays in such large areas and currently offers the best reconstruction power for the cosmic particle's type, energy, and direction in particular.

However, with only about 1,000\,hours of dark and clear nights every year, the imaging atmospheric Cherenkov~telescope only has a duty cycle of about $12\%$.
To overcome this limitation, we propose a novel imaging atmospheric radio telescope.
By observing radio emission in the $10\,$GHz regime from air showers, the imaging atmospheric radio telescope could observe 24/7, independent of the day night cycle, and almost independent of the weather.
The novelty here is the high resolution imaging of the air shower's radio emission which might result in high resolution images as one finds them in imaging atmospheric Cherenkov~telescopes.
We present a technique to simulate the image formation in the radio telescope using wave mechanics instead of conventional ray tracing.
The imaging atmospheric radio telescope could increase our access more than eight-fold to an accurately reconstructed high energy gamma~ray~sky.
Further, the imaging atmospheric radio telescope might also eight fold the observation power for cosmic neutrinos in the so called Earth-skimming technique.

        \end{abstract}
    \end{frontmatter}
    \newcommand{\MirrorFocalLength}{f}
    \newcommand{\MirrorDiameter}{D}
    \newcommand{\MirrorInnerDiameter}{D_\text{inner}}
    \newcommand{\ThinLensImageDistance}{b}
    \newcommand{\ThinLensObjectDistance}{g}
    \newcommand{\CameraScreenDistance}{d}
    \newcommand{\MirrorPointCloud}{\mathcal{C}_\text{mirror}}
    \newcommand{\MirrorScatterCenterComponent}{p}
    \newcommand{\MirrorScatterCenterVector}{\vec{\MirrorScatterCenterComponent{}}}
    \newcommand{\NumMirrorScatterCenters}{M}
    \newcommand{\MirrorIndex}{m}
    \newcommand{\MirrorTotalArea}{A_\text{mirror}}
    \newcommand{\CameraPointCloud}{\mathcal{C}_\text{camera}}
    \newcommand{\CameraScatterCenterComponent}{q}
    \newcommand{\FeedHornScatterCenterVector}{\vec{\CameraScatterCenterComponent{}}}
    \newcommand{\NumFeedHorns}{K}
    \newcommand{\CameraIndex}{n}
    \newcommand{\CameraFeedHornKey}{\text{feed-horn}}
    \newcommand{\CameraFeedHornArea}{A_\CameraFeedHornKey{}}
    \newcommand{\NumScatterCentersPerFeedHorn}{N}
    \newcommand{\ElectricField}{\vec{\epsilon}}
    \newcommand{\CorsikaCoreas}{\texttt{CORSIKA-CoREAS}}
    \newcommand{\Aires}{\texttt{ZHAireS}}
    \section{Introduction}
    \label{SecIntroduction}
    The observation of cosmic gamma rays offers unique insights into the most violent processes in the universe.
    Often, gamma rays are emitted in transient catastrophes on cosmic scales \citep{aleksic2014blackholelightning, tavani2011discovery, ackermann2014fermi} which are best observed with detectors that can gather statistics in a timely manner.
    Further, most gamma ray emissions show a power law energy spectra where higher energies are much rarer than lower energies.
    In combination, timing the high energy gamma ray sky in and beyond the Tera electron Volt regime demands detectors with large collection areas.
    \subsection{Current Methods}
        \label{SubSecCurrentMethods}
        Today, one can detect cosmic gamma rays either directly in space or indirectly with the help of Earth's atmosphere.
        Energetic cosmic particles initiate cascades of secondary particles when they interact with matter, such as a particle detector in space.
        %
        %
        %
        %
        Detectors in space \citep{acero2015fermi3fgl, tavani2011discovery} offer good reconstruction power and an excellent purity for gamma rays in the pool of abundant cosmic rays.
        But despite their high cost, detectors in space only have collection areas with the size of a desk (1\,m$^2$) and thus limit our timely access to the high energy sky.
        %

        Alternatively, one can wait for the cosmic particle to interact with Earth's atmosphere and then observe the cascade of secondary particles in an air shower.
        Even when the energy of the cosmic particle is too low for secondary particles to reach the ground, the air shower still emits Cherenkov, radio, and fluorescence photons, and in some cases muons, which reach the ground where they illuminate a pool with the size of soccer fields ($>$10$^{5}$\,m$^2$).
        The fact that a detector on the ground can observe an air shower while the detector's own physical size can be smaller than the pool illuminated by the air shower makes such indirect atmospheric detectors potentially cost-effective.
        There exist different indirect atmospheric detection methods.
        Often they have complementary advantages and sometimes they are combined \citep{cao2019large} to observe the same air showers simultaneously.
        Two methods here are of particular interest for this study.

        First, the imaging atmospheric Cherenkov telescope \citep{weekes1989observation, aharonian2006observations, tridon2010magic} makes high-resolution ($\approx$ 0.1$^{\circ}$) videos of the air showers which it faces head-on in its narrow field-of-view ($\approx$ 5$^{\circ}$ - 8$^{\circ} $).
        Here, the energy threshold can be as low as $\approx\,10^{-2}$\,TeV \citep{abdalla2018first} and with the proposed Cherenkov plenoscope potentially as low as $\approx\,10^{-3}$\,TeV \citep{mueller2024exploring}.
        The high-resolution imaging is powerful to separate air showers induced by gamma rays from air showers induced by cosmic rays \citep{hillas1985cerenkov} which allows the imaging atmospheric Cherenkov technique to collect the next purest sample of cosmic gamma rays after space based detectors.
        Its cost effectiveness paired with its low energy threshold allowed the imaging atmospheric Cherenkov telescope to lead the next generation's observations in the Tera electron Volt regime \citep{cta2013introducing}.
        However, with only $\approx 1,000$\,hours of dark and clear nights every year, the duty cycle is only $\approx 12\%$.

        Second, there is the atmospheric radio technique which in most cases can observe with a duty cycle close to $\approx 100\%$.
        Also in the radio technique, there are imaging telescope prototypes which have operated in the 1--10\,GHz radio regime.
        The experiments AMBER \citep{gorham2007observations}, CROME \citep{smida2014first}, and MIDAS \citep{alvarez2013midas} all imaged the radio emission of air showers in directional bins (pixels). In the case of CROME, there exist coincident reconstructions of events from the ground based particle detector array KASCADE-Grande \citep{werner2013phd}.
        The experiments used imaging mirrors with diameters in the $3\,$m regime with estimated     energy thresholds for cosmic rays in the $10^{4}$ to $10^{6}$\,TeV regime.
        Although a high energy threshold of $10^{4}$\,TeV \citep{werner2013phd} is unfavorable for observing cosmic gamma rays, the potential to multiply the duty cycle eightfold is tempting.
        Furthermore, such an imaging atmospheric radio telescope could be effective for detection of Earth-skimming neutrinos \citep{otte2019studies} with high energies in the $10^{5}$\,TeV regime.

        Seeing the power of imaging in the atmospheric Cherenkov technique, one might ask whether there could be a dedicated imaging atmospheric radio telescope with sufficient angular resolution to benefit from powerful reconstruction techniques in the image space \citep{hillas1985cerenkov}.
        In this study, we want to explore the prospects of a high resolution imaging atmospheric radio telescope by implementing a realistic system, while details of the implications of the observed emission are discussed in another article \cite{ammermanyebra2026imagingradiowaveemissionextensive}.
    \subsection{Proposing an Imaging Radio~Telescope}
        \label{SecProposing}
        To collect and image the radio emission of an air shower, two physical limits restrict the telescope's minimum and maximum mirror diameter.

        First, Airy estimates the minimum mirror diameter which is required to potentially reach an angular resolution of
        \begin{eqnarray}
            \Theta{}_\text{Airy-full} &=& 2 \arcsin{ \left( 1.22 \frac{c}{\nu \MirrorDiameter{}} \right) }
            \label{EqAiry}
        \end{eqnarray}
        at a given radio frequency $\nu$. Here $c$ is the speed of light.
        Airy's estimate assumes that the mirror's focal length is much
        larger than the mirror's diameter (far field approximation). It uses the interference pattern created by the mirror's disk-like aperture (Bessel function) and defines the angle contained
        between the first order zero points as $\Theta_\text{Airy-full}$.
        Experience in the atmospheric Cherenkov method shows that a telescope's angular resolution is preferably $\Theta{}_\text{full} \lessapprox{} 0.3^\circ$ to begin reconstructions of the cosmic particle's properties, see \autoref{TabCherenkovTelescopeThetaFull}.
        \begin{table}
            \begin{tabular}{lrr}
                $\Theta{}_\text{full} / (1)^{\circ}$ & Instrument & \\
                \midrule
                0.50 & Whipple & \citep{weekes1989observation} \\
                0.16 & H.E.S.S. I & \citep{aharonian2006observations} \\
                0.12 & CAT France & \citep{barrau1998cat} \\
                0.10 & MAGIC & \citep{tridon2010magic} \\
                0.07 & H.E.S.S. II & \citep{abdalla2018first} \\
            \end{tabular}
            \caption{
                Angular resolutions of some Cherenkov telescopes.
            }
            \label{TabCherenkovTelescopeThetaFull}
        \end{table}

        Second, the narrowing depth-of-field limits the maximum useful mirror diameter before the images start to show significant blurring.
        In \citep{bernlohr2013monte}, one finds an estimate which suggests that mirror diameters beyond $D = 23.0$\,m will only yield diminishing returns in reconstruction power for a Cherenkov telescope 2,000\,m above sea level.

\begin{figure}{}
            \centering
            \includegraphics[width=1.0\columnwidth]{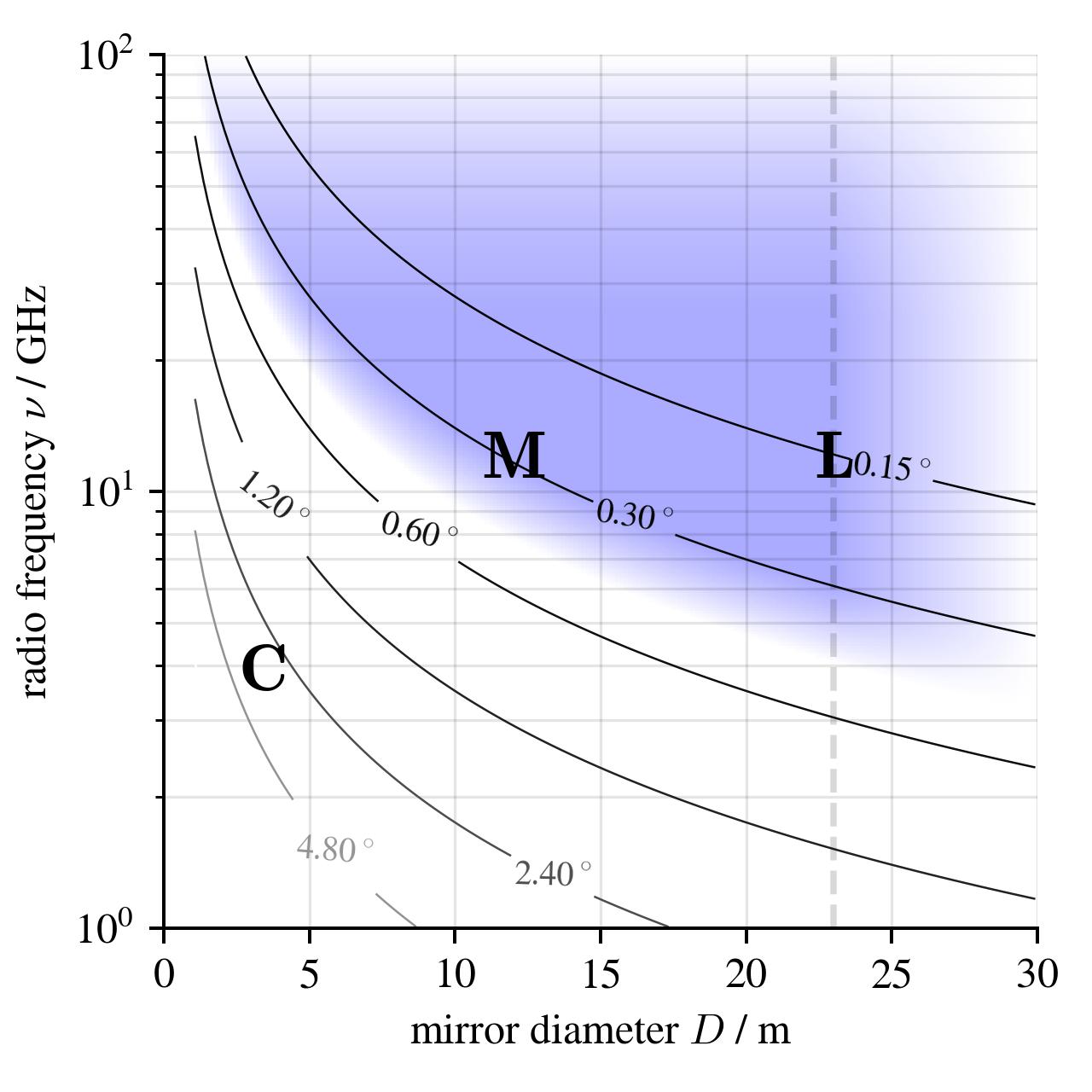}
            \caption{
                Airy and the narrowing depth-of-field limit the minimum and maximum diameter of an imaging radio telescope for air showers.
                Contour lines show angular resolution $\Theta{}_\text{Airy-full}$ according to \autoref{EqAiry}.
                The blue region for $\Theta_\text{full} < 0.3^\circ$, $D < 23\,$m, and $\nu < 25\,$GHz indicates the preferable regime to image air showers.
                Vertical, grey, dashed line at $D=23$\,m marks the depth-of-field limit suggested by \citep{bernlohr2013monte}.
                $\mathbf{C}$, $\mathbf{M}$, and $\mathbf{L}$ mark the three telescopes
                Crome, Medium, and Large.
            }
            \label{FigMirrorLimits}
        \end{figure}

        \autoref{FigMirrorLimits} shows how Airy and the narrowing depth-of-field limit the options for an imaging atmospheric radio telescope.
        
        A third limitation comes from the attenuation of radio light in earth's atmosphere.
        In \autoref{FigMirrorLimits}, the region of interest fades out for frequencies above $\approx$ 25\,GHz as here the specific attenuation of radio light caused by O$_2$ and H$_2$O in Earth's atmosphere increases to and exceeds $\approx 0.1$(km)$^{-1}$ \citep{liebe1993propagation}.

\subsection{Comparing Signal and Background}
    \label{SecSignalAndNoise}
        To estimate the energy threshold of a cosmic gamma-ray event for an atmospheric imaging radio telescope, one can analyze the output of the telescope's low-noise-blocks (LNBs), which detect the incoming radio signal.  
        An LNB produces an electrical output not only in response to the incident radio signal but also due to intrinsic noise from its amplifiers and associated electronics.
        
        \subsubsection*{Background}
            In the 5--15\,GHz frequency range, the dominant contribution to the system noise arises from the LNB itself.  
            The LNB considered here is typical of inexpensive wideband satellite converters, with a noise figure $\lesssim 1$\,dB over the 10.7-12.75\,GHz band \citep{inverto:2021:lnb}.  
            This corresponds to an effective receiver noise temperature of $\sim 50$-$75$\,K, which dominates all other background contributions.  
            For comparison, CROME's 5\,GHz C-band system used higher-cost commercial LNBs with a noise temperature of only 50\,K \citep{werner2013phd}.
            Additional contributions come from Galactic synchrotron emission, the cosmic microwave background, and atmospheric emission, which together add a few kelvin under dry conditions and up to $\sim 15$\,K in humid weather \citep{CondonRansom+2016}.  
            Including all contributions, the total system noise temperature is at most $T_\text{sys} \sim 100$\,K.
            
            Wideband commercial LNBs for satellite television (ASTRA, EUTELSAT, TUERKSAT; IEEE-K$_u$-band) span a bandwidth of $\Delta\nu = 2.05$\,GHz, from $\nu_\text{start} \approx 10.70$\,GHz to $\nu_\text{stop} \approx 12.75$\,GHz.  
            The corresponding output power due to system noise is therefore
            \begin{eqnarray}
            P_\text{background} &=& k_\text{B} \, T_\text{sys} \, \Delta\nu \\
            &\approx& 2.8 \times 10^{-12}\,\mathrm{W},
            \end{eqnarray}
            where $k_\text{B}$ is the Boltzmann constant. This establishes the baseline against which any incoming cosmic gamma-ray-induced radio signal must be detected.

        \subsubsection*{Signal}
        The antenna in the example low-noise-block for the IEEE-K$_u$-band ($\sim$ 12 - 18 GHz) has an effective area of
        \begin{eqnarray}
            A_\text{antenna} &=& \frac{c^2}{4 \pi \nu^2}\\
            &\approx& 7.2\times10^{-5}\text{m}^2.
        \end{eqnarray}
        Thus, in order for the signal to be of comparable power to the background, the signal's electric field must have an areal power density of
        \begin{eqnarray}
            S_\text{antenna} &=& \frac{P_\text{background}}{A_\text{antenna}}\\
            \label{EqIdealAntennaEffectiveArea}
            &\approx& 4.0\times10^{-8}\text{W}\text{m}^{-2}
        \end{eqnarray}
        at the low-noise-block's antenna.
        Now an imaging mirror with a diameter of $D=12\,$m can have a geometric gain of
        \begin{eqnarray}
            G_\text{geometry} &=& \frac{\pi (D/2)^2}{A_\text{antenna}}\\
            &\approx& 1.6\times10^{6}
        \end{eqnarray}
        With shadowing of the camera, transmission losses at the feed horns,
        the overall gain of a telescope should still be half of the geometric gain
        \begin{eqnarray}
            G &=& G_\text{geometry}/2\\
            &\approx& 8.0\times10^{5}.
        \end{eqnarray}
        This implies that such a telescope might detect a signal from an electric field with an areal power density of
        \begin{eqnarray}
            S_\text{mirror} &=& \frac{S_\text{antenna}}{G}\\
            &\approx& 5.0\times10^{-14}\text{W}\text{m}^{-2}
        \end{eqnarray}
        on its mirror.
        The areal power density of an electric field is
        \begin{eqnarray}
            S &=& \frac{\epsilon^2}{Z}
        \end{eqnarray}
        where for $Z$ one can assume the vacuum impedance $Z_0 = 120\pi\,\Omega$.
        This implies that the signal electric field must have a strength of
        \begin{eqnarray}
            \epsilon_\text{mirror} &=& \sqrt{Z_0 S_\text{mirror}}\\
                &\approx& 4\mu \text{V}\text{m}^{-1}.
        \end{eqnarray}
        Using the simulations described in section \ref{SecAirshowerImages}, an energy threshold in the order of 1--5\,PeV is expected.

        This estimate is roughly in line with observations of CROME and KASCADE-Grande \cite{werner2013phd}.
        CROME is $D=3.4\,$m in diameter and operates in the IEEE-C-band ($\nu_\text{start} \approx 3.4$\,GHz to $\nu_\text{stop} \approx 4.2$\,GHz).
        CROME was able to detect signals with a factor 6.4 ($8$\,dB) over its background from air showers with energies in the 50\,PeV regime.

        This estimate supports the expectation that cosmic particles with energies in the peta electronvolt regime can also be detected and reconstructed
        with a high resolution atmospheric imaging radio telescope.
    \section{Imaging Atmospheric Radio Telescope}
    \label{SecAskaryanTelescope}
    The imaging atmospheric radio telescope detects radio light with frequencies in the GHz regime.
    Just like the Cherenkov~telescope, the imaging atmospheric radio telescope is made out of an imaging mirror and a camera.
    Unlike telescopes in radio astronomy, the imaging atmospheric radio telescope requires focal-ratios of $f/D \geq 1$ to have a sharp image across a field-of-view that is wide enough (multiple 1$^\circ$) to contain the radio light of an air shower.
    A parabolic mirror collects the incoming radio light from the sky and from the atmosphere and images it onto the camera's screen.
    The camera screen is a dense array of sensors for radio light.
    A sensor for radio light in the GHz regime can be made out of a feed horn, and a successive antenna.
    The feed~horns help to increase the sensor's geometric fill factor similar to the light guides on a Cherenkov~telescope \citep{winston2018nonimaging}.
    Due to the wave mechanics of radio light, the feed~horns are often characterized by their geometric gain and their Fresnel reflection losses from insufficient impedance matching \citep{balanis2015antenna}.

    To simulate the imaging of the atmospheric radio~telescope, one has to predict all the individual electric~fields present at the feed~horn's successive antennas.
    To make sure that one's simulation is stable with regard to the physical telescope size and frequency regime, one can discuss three specific imaging atmospheric radio telescopes.
\subsection{Telescopes: $\mathbf{C}$rome, $\mathbf{M}$edium, and $\mathbf{L}$arge}
    \label{SecTelescopeName}
    One can introduce three different imaging atmospheric radio telescopes named
    $\mathbf{C}$rome, $\mathbf{M}$edium, and $\mathbf{L}$arge to discuss specific designs with different sizes and different astrophysical performances.
    See \autoref{TabTelescopesOptics} for optical and geometric specifications.

    First, the 3.4\,m $\mathbf{C}$rome is the smallest telescope and adopted from the `Cosmic-Ray Observation via Microwave Emission' experiment located in Karlsruhe, Germany \citep{smida2014first}.
    This $\mathbf{C}$rome here has the same mirror geometry\footnote{Prdelin 1344 series} as the Karlsruhe CROME, and uses the same feed~horn and low-noise-block geometry\footnote{Norsat 8215F} \citep{werner2013phd}.
    However, the $\mathbf{C}$rome here has a different, denser and a more numerous arrangement of feed horns in its camera screen.
    See \autoref{FigCromeGeometry}.

    Next a 12\,m $\mathbf{M}$edium telescope is the first access to
    decent resolution (0.27$^\circ$) imaging of air showers in the GHz regime.
    The feed horn of $\mathbf{M}$edium, as well as its low-noise-block fit into a diameter of 5.6\,cm which is a compact commercial design\footnote{Inverto 40mm wideband LNB IDLP-WDB02-OOPRO-OPP with both horizontal and vertical channels}.
    Unlike $\mathbf{C}$rome, $\mathbf{M}$edium has a larger focal-ratio of $f/D=1$ to improve its off axis image quality.
    See \autoref{FigMediumGeometry}.

    Finally a 23\,m $\mathbf{L}$arge telescope allows high resolution imaging (0.15$^\circ$) of air showers in the GHz regime.
    Its mirror size is close to the upper limit of the narrowing depth-of-field and its focal-ratio is a comfortable $f/D=1.4$ what allows a rather wide (6.5$^\circ$) and flat field-of-view.
    See \autoref{FigLargeGeometry}.

    \begin{table}
        \begin{tabular}{lrrr}
            Telescopes & $\mathbf{C}$ & $\mathbf{M}$ & $\mathbf{L}$\\
            \toprule
            Mirror &  &  &  \\
            \midrule
outer diameter &  &  &  \\
$D$\,/\,m & 3.35 & 12.00 & 23.00 \\
focal-length &  &  &  \\
$f$\,/\,m & 1.19 & 12.00 & 32.20 \\
focal-ratio &  &  &  \\
$f\,/\,D$ & 0.36 & 1.00 & 1.40 \\
inner diameter &  &  &  \\
$D_\text{inner}$\,/\,m & 0.77 & 2.40 & 5.66 \\
area &  &  &  \\
$A_\text{mirror}$\,/\,m$^{2}$ & 8.3 & 108.6 & 390.3 \\
num. scatters &  &  &  \\
$M$ & 723 & 637 & 1922 \\
scatter area &  &  &  \\
$A_\text{SM}$\,/\,(cm)$^{2}$ & 115 & 1704 & 2031 \\
                \midrule
                Camera &  &  &  \\
                \midrule
diameter &  &  &  \\
$D_\text{camera}$\,/\,m & 0.39 & 1.05 & 3.66 \\
field-of-view &  &  &  \\
$\Theta_\text{fov}$\,/\,(1$^\circ$) & 18.4 & 5.0 & 6.5 \\
feed horn diameter &  &  &  \\
$D_\text{feed-horn}$\,/\,cm & 6.66 & 5.60 & 8.40 \\
feed horn field-of-view &  &  &  \\
$\Theta_\text{feed-horn-full}$\,/\,(1$^\circ$) & 3.20 & 0.27 & 0.15 \\
num. feed horns &  &  &  \\
& 31 & 313 & 1723 \\
feed horn area &  &  &  \\
$A_\text{feed-horn}$\,/\,(cm)$^{2}$ & 38.41 & 27.16 & 61.11 \\
focus distance &  &  &  \\
$g$\,/\,km & 10.0 & 10.0 & 10.0 \\
distance to mirror &  &  &  \\
$d$\,/\,m & 1.19 & 12.01 & 32.30 \\
num. scatters per feed horn &  &  &  \\
$N$ & 7 & 7 & 7 \\
scatter center area &  &  &  \\
$A_\text{SF}$\,/\,(cm)$^{2}$ & 5.49 & 3.88 & 8.73 \\
                \midrule
                Frequency regime &  &  &  \\
                \midrule
$\nu_\text{start}$\,/\,GHz & 3.40 & 10.70 & 12.75 \\
$\nu_\text{stop}$\,/\,GHz & 4.20 & 10.70 & 12.75 \\
IEEE band & C & K$_u$ & K$_u$ \\
            \bottomrule
        \end{tabular}
        \caption{
            Optical specifications of the three telescopes $\mathbf{C}$rome, $\mathbf{M}$edium, and $\mathbf{L}$arge.
        }
        \label{TabTelescopesOptics}
    \end{table}
    \begin{figure}{}
        \subfloat[mirror]{
            \centering
            \includegraphics[width=0.8\columnwidth]{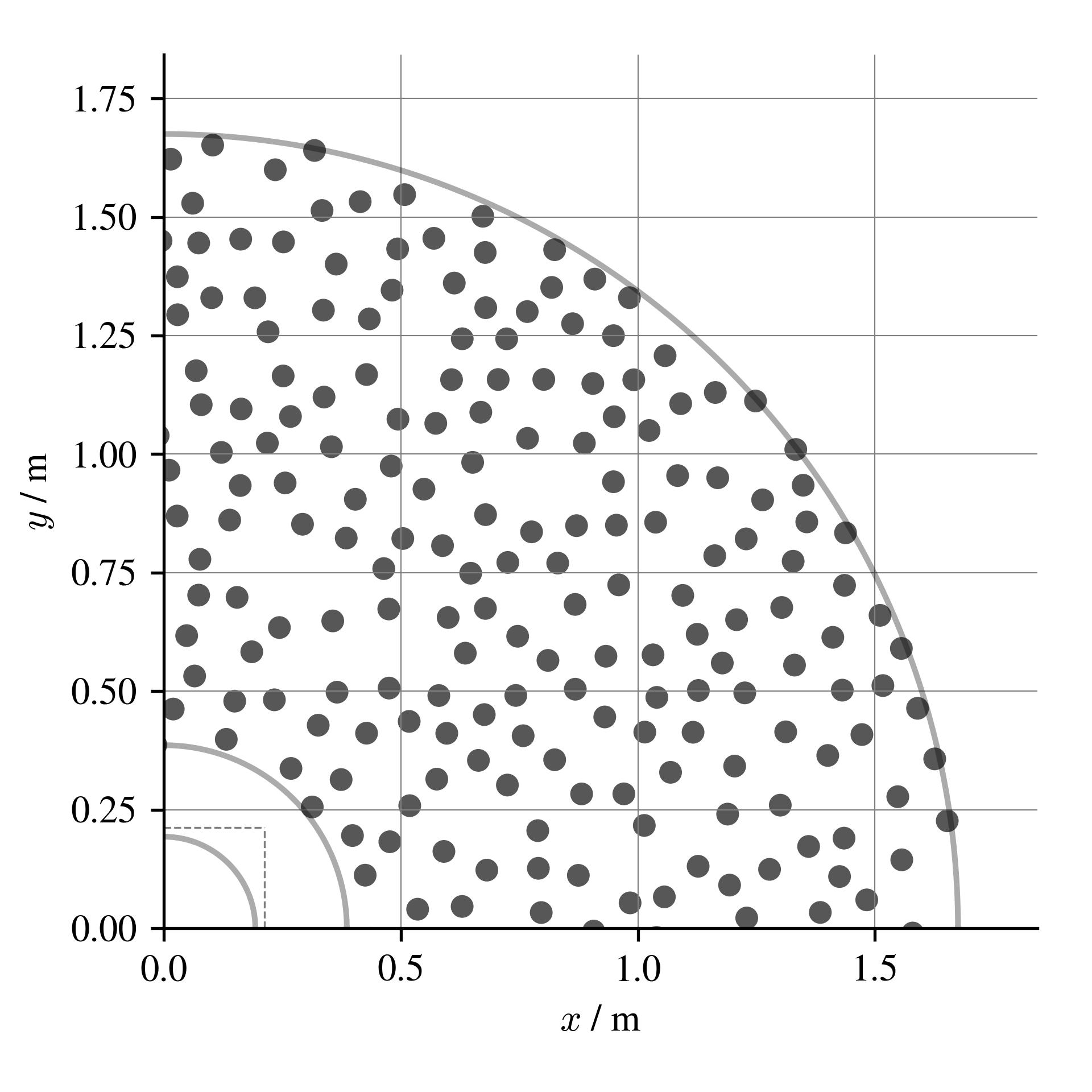}
            \label{FigCromeGeometryCloseUpMirror}
        }
        \\
        \subfloat[camera]{
            \centering
            \includegraphics[width=0.8\columnwidth]{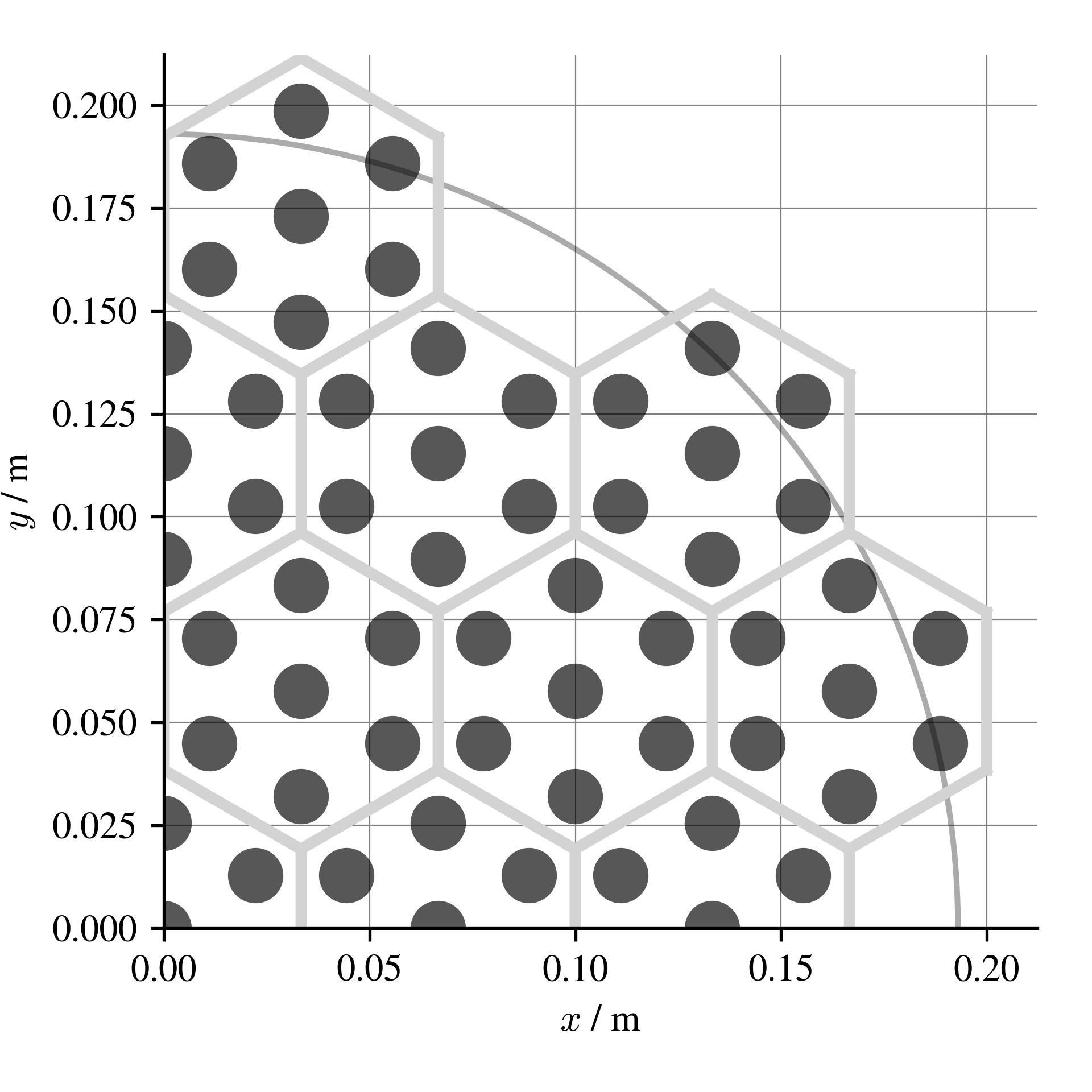}
            \label{FigCromeGeometryCloseUpCamera}
        }
        \caption{
            $\mathbf{C}$rome close-up on mirror and camera.
            Projection in the $x$-$y$-plane, where the $z$-axis is the optical axis.
            See \autoref{FigCromeGeometry} for $x$-$z$-projection.
        }
        \label{FigCromeGeometryCloseUp}
    \end{figure}
    \begin{figure}{}
        \centering
        \includegraphics[width=1\columnwidth]{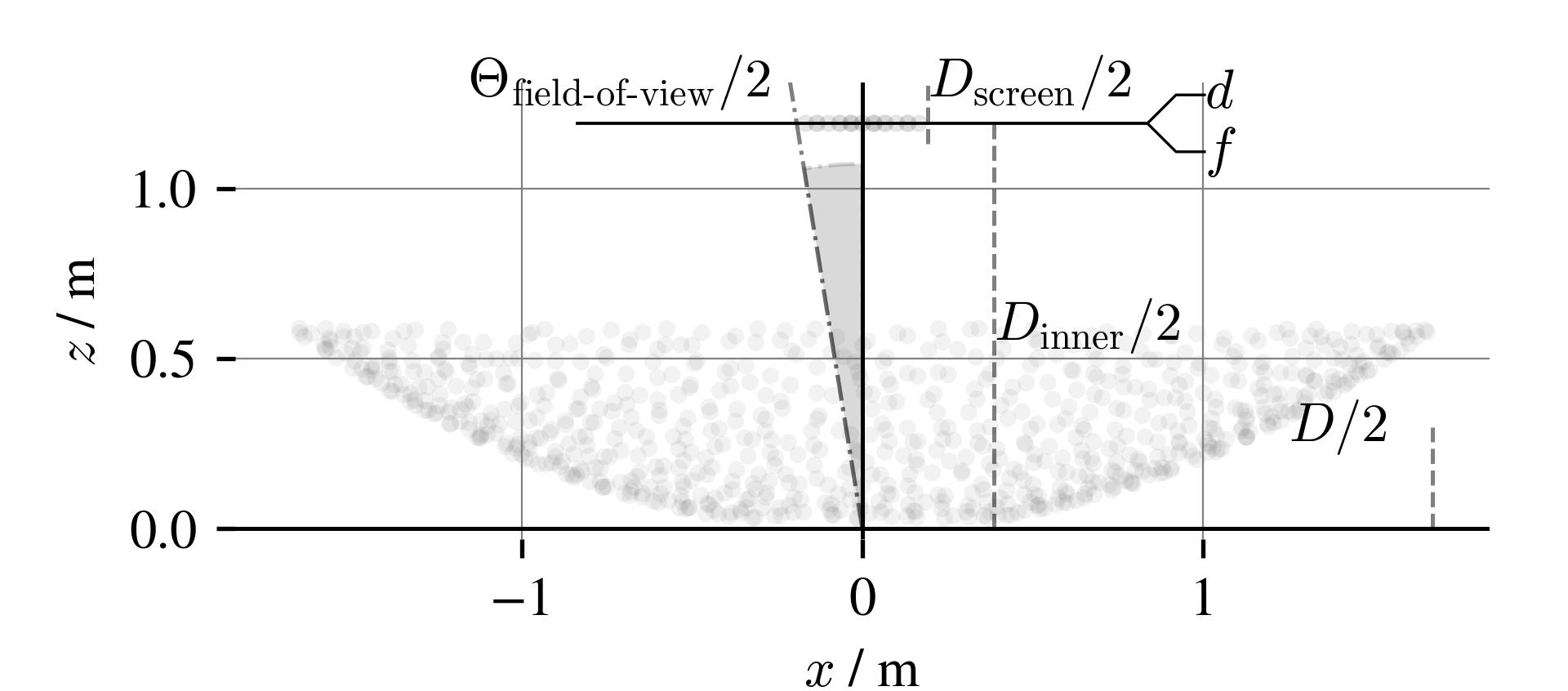}
        \caption{
            $\mathbf{C}$rome telescope geometry.
            Projection in the $x$-$z$-plane, where the $z$-axis is the optical axis.
            The mirror and the camera are represented using the Huygens scatter centers used in the simulations.
            Compare \autoref{TabTelescopesOptics}.
            See \autoref{FigCromeGeometryCloseUp} for $x$-$y$-projection.
        }
        \label{FigCromeGeometry}
    \end{figure}
    \begin{figure}{}
        \centering
        \includegraphics[width=1\columnwidth]{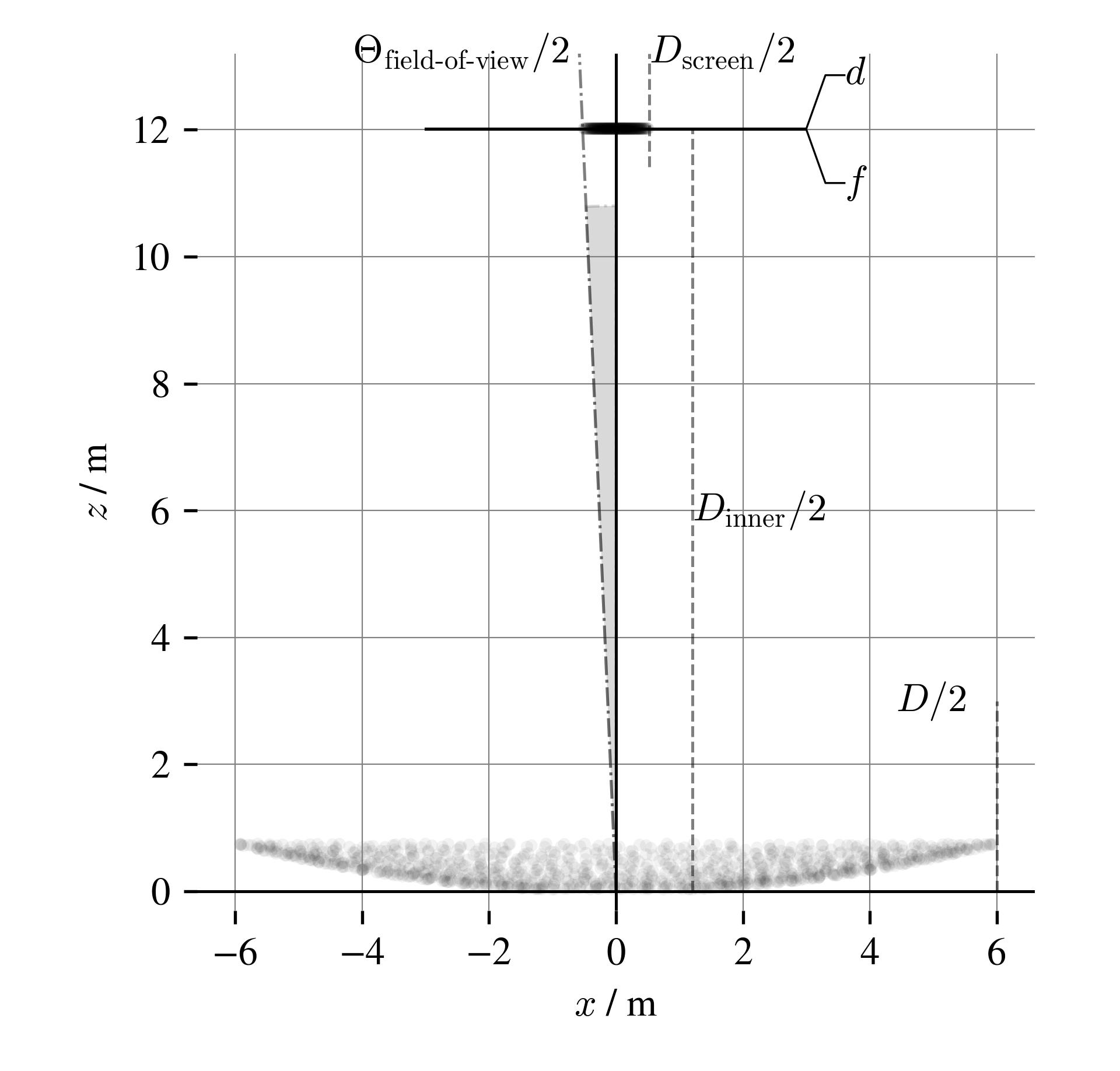}
        \caption{
            $\mathbf{M}$edium telescope geometry.
            See \autoref{FigCromeGeometry} for axis description.
            Compare \autoref{TabTelescopesOptics}.
        }
        \label{FigMediumGeometry}
    \end{figure}
    \begin{figure}{}
        \centering
        \includegraphics[width=1\columnwidth]{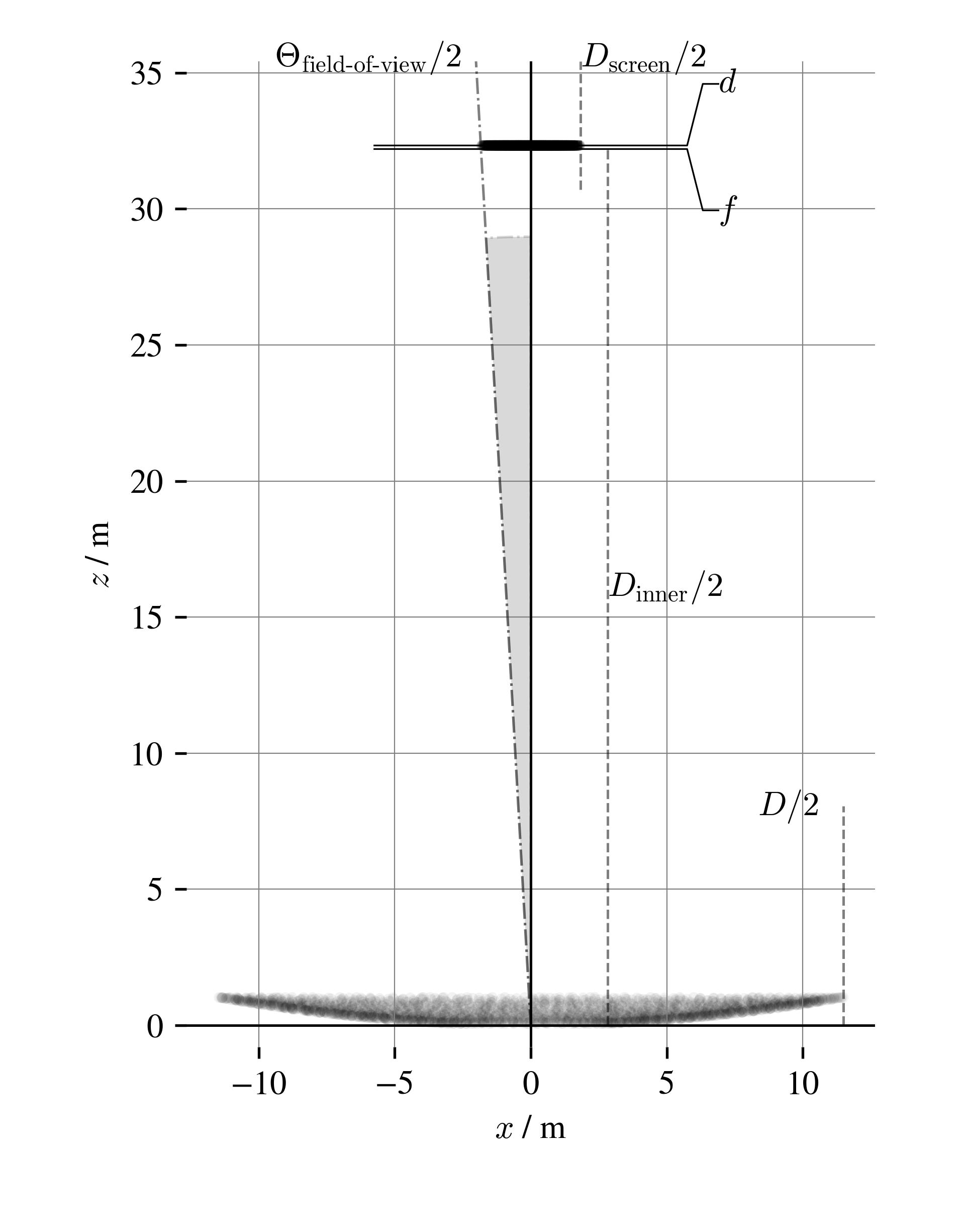}
        \caption{
            $\mathbf{L}$arge telescope geometry.
            See \autoref{FigCromeGeometry} for axis description.
            Compare \autoref{TabTelescopesOptics}.
        }
        \label{FigLargeGeometry}
    \end{figure}
    \section{Huygens image formation}
    \label{SecHuygensImageFormation}
    We propose to simulate image formation in the imaging atmospheric radio telescope using Huygens' principle.
    Notably, this method does not assume any particular shape of the incoming wavefront — plane, spherical, or otherwise. Instead, the scatter centers propagate the full complex electromagnetic field from existing air-shower radio simulations to the focal plane, naturally accounting for interference, diffraction, phase delays, and partial coherence.
    The telescope mirror and camera are represented by point clouds of Huygens scatter centers.
    Each scatter center acts as a secondary source that re-emits the incident complex field as a spherical wavelet according to the Huygens–Fresnel principle.
    Air shower simulations such as \CorsikaCoreas{} \citep{huege2013simulating} and \Aires{} \citep{alvarez2012monte,sciutto2023aires} output the electric field $\ElectricField{}(\vec{x}, t)$ versus the time $t$ at a given position $\vec{x}$.
    To feed the electric field output of existing air shower simulations into the telescope simulation one simply asks the air shower simulations for the electric fields at the positions of the mirror's scatter centers.
    \subsection{Telescope}
        \label{SecHuygensPointCloudTelescope}
        \subsubsection*{Mirror}
            The imaging atmospheric radio telescope's mirror can be made out of a metal surface.
            The point cloud
            \begin{eqnarray}
                \MirrorPointCloud{} &=& \begin{Bmatrix}
                     \MirrorScatterCenterVector{}_1,
                     \MirrorScatterCenterVector{}_2,
                     \dots,
                     \MirrorScatterCenterVector{}_\NumMirrorScatterCenters{}
                \end{Bmatrix}
                \label{EqMirrorScatterCenterDefinition}
            \end{eqnarray}
            of $\NumMirrorScatterCenters{}$ Huygens scatter center positions $\MirrorScatterCenterVector{}$ describes the surface of this mirror.
            To make an imaging mirror, one places the scatter centers on a paraboloid with focal-length $\MirrorFocalLength{}$ and its optical axis being the $z$-axis so that the components $\MirrorScatterCenterVector{} = (\MirrorScatterCenterComponent{}_x, \MirrorScatterCenterComponent{}_y, \MirrorScatterCenterComponent{}_z)^T$ fulfill
            \begin{eqnarray}
                \MirrorScatterCenterComponent{}_z &=& \frac{1}{4 \MirrorFocalLength{}} (\MirrorScatterCenterComponent{}_x^2 + \MirrorScatterCenterComponent{}_y^2).
                \label{EqMirrorScatterCenterPositionsParabola}
            \end{eqnarray}
            Perpendicular to the mirror's optical axis is the $x$-$y$-plane where the positions of the mirror's scatter centers are restricted
            \begin{eqnarray}
                 \MirrorInnerDiameter{}/2 < \sqrt{ \MirrorScatterCenterComponent{}_x^2 + \MirrorScatterCenterComponent{}_y^2 } \leq \MirrorDiameter{}/2
                \label{EqMirrorScatterCenterPositionsAnnulus}
            \end{eqnarray}
            such that only an annulus between the mirror's outer diameter $\MirrorDiameter{}$ and an inner diameter $\MirrorInnerDiameter{}$ is populated.
            Fig.\,\autoref{FigCromeGeometryCloseUpMirror} shows the mirror's scatter centers in $\mathbf{C}$rome being located in an annulus.
            Restricting the placement of scatter centers in the inner part of the mirror approximates the shadowing caused by the telescope's camera.
            Thus $\MirrorInnerDiameter{}$ should be at least the diameter of the camera.
            With the inner shadowing, the effective mirror area is
            \begin{eqnarray}
                 \MirrorTotalArea{} &=& \pi \left(\frac{\MirrorDiameter{}}{2}\right)^2 - \pi \left(\frac{\MirrorInnerDiameter{}}{2}\right)^2
                \label{EqMirrorTotalArea}
            \end{eqnarray}
            and the average mirror area associated with one of the mirror's $\NumMirrorScatterCenters{}$ scatter centers is by design
            \begin{eqnarray}
                 A_\text{SM} &=& \frac{\MirrorTotalArea{}}{\NumMirrorScatterCenters{}}.
                \label{EqMirrorScatterArea}
            \end{eqnarray}
            For the number of scatter centers in the mirror one has to find a balance between the compute effort and the risk of introducing artifacts.
            Idealy one would fill the mirror with scatter centers each having only the ideal antenna area of $A_\text{antenna}$ from \autoref{EqIdealAntennaEffectiveArea}.
            Unfortunately, the resulting compute effort of such a high areal density is daunting.
            One compromise is to have about as many scatter centers in the mirror as one has scatter centers in the screen.
            To reduce the potential of artificially introducing interference patterns in the images, one can adopt strategies \citep{mort2016analysing} to populate the mirror's scatter centers in apparently random patterns, see \,\autoref{FigCromeGeometryCloseUpMirror}.
            See \autoref{SecArrangingScatterCenters} for a discussion of possible strategies we have tried and later \autoref{SecArtifacts} to see example artifacts.
        \subsubsection*{Camera}
\begin{figure}{}
                \centering
                \includegraphics[width=0.75\columnwidth]{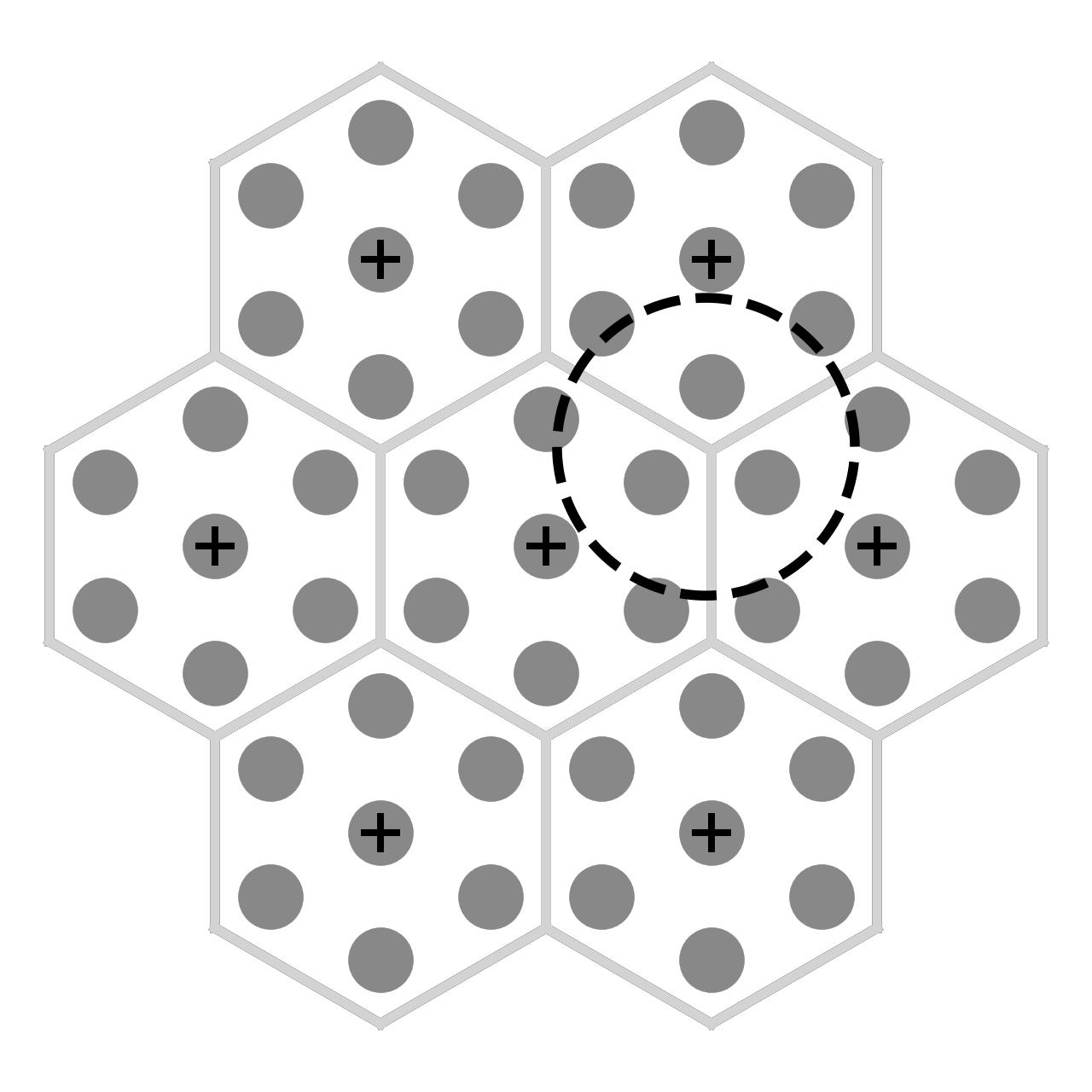}
                \caption{
                    Gray hexagons are openings of feed horns.
                    Black `$+$' mark the centers of the feed horns.
                    Grey dots, of which there are $\NumScatterCentersPerFeedHorn = 7$ in each feed horn, mark mark the Huygens scatter centers $\FeedHornScatterCenterVector{}$ listed in \autoref{EqCameraScatterCenterDefinition}.
                    Black, dashed circle marks the mirror's point spread function which is a good match for the feed horn size but might suffer from poor sampling if Huygens scatter centers were only put in the center `$+$' of each feed horn.
                }
                \label{FigFeedHornMesh}
            \end{figure}

            Just like the mirror, one represents the camera with a point cloud
            \begin{eqnarray}
                \CameraPointCloud{} &=& \begin{Bmatrix}
                     \FeedHornScatterCenterVector{}_1,
                     \FeedHornScatterCenterVector{}_2,
                     \dots,
                     \FeedHornScatterCenterVector{}_{(\NumFeedHorns{} \times \NumScatterCentersPerFeedHorn{})}.
                \end{Bmatrix}
                \label{EqCameraScatterCenterDefinition}
            \end{eqnarray}
            using scatter positions $\FeedHornScatterCenterVector{} = (\CameraScatterCenterComponent{}_x, \CameraScatterCenterComponent{}_y, \CameraScatterCenterComponent{}_z)^T$.
            However, these scatter positions do not represent the $\NumFeedHorns{}$ feed horns directly.
            One finds that it can be beneficial to have more Huygens scatter centers than feed horns in the camera's screen.
            The reason for this is shown in \autoref{FigFeedHornMesh} where one over samples the feed horn positions in the camera screen by a factor of $\NumScatterCentersPerFeedHorn = 7$.
            If the areal sampling in the camera screen is too sparse, the bulk of the reflected radio light coming from a far distant source can end up in between scatter centers where it will go unnoticed.
            To avoid this, the areal sampling density in the camera screen should be sufficient to have multiple scatter centers in the mirror's point spread function.

            One arranges the components $\CameraScatterCenterComponent{}_x$ and $\CameraScatterCenterComponent{}_y$ in a roughly uniform grid to represent each of the $\NumFeedHorns{}$ feed horns with $\NumScatterCentersPerFeedHorn$ scatter centers.
            Thus the effective area of a scatter center in the camera screen is
            \begin{eqnarray}
                 A_\text{SF} &=& \frac{\CameraFeedHornArea{}}{\NumScatterCentersPerFeedHorn{}}
                \label{EqFeedHornScatterArea}
            \end{eqnarray}
            where $\CameraFeedHornArea{}$ is the area of the feed horn's entrance opening.

            To focus the telescope to a specific object-distance $\ThinLensObjectDistance{}$, one sets all the feed horn positions in $z$ to the screen distance $\CameraScreenDistance{}=\ThinLensImageDistance{}$
            \begin{eqnarray}
                {\CameraScatterCenterComponent{}_z}_i &=& \CameraScreenDistance{}, \,\, \forall i
                \label{EqCameraFocus}
            \end{eqnarray}
            with $b$ being the image distance  according to the thin lens equation
            \begin{eqnarray}
                1/\MirrorFocalLength{} &=& 1/\ThinLensImageDistance{} + 1/\ThinLensObjectDistance{}.
                \label{EqThinLens}
            \end{eqnarray}
            \autoref{FigLargeGeometry} shows the small but important difference between the focal-length $f$ and the screen distance $\CameraScreenDistance{}=\ThinLensImageDistance{}$ for a focus set to $g=10\,$km.
    \section{Image formation}
    Imaging is done by propagating the scattered electric fields from the mirror towards the camera screen where the electric fields are super positioned.

    However, as simple as Huygens' principle seems to be, it can be challenging to conserve the energy propagated from the mirror to the camera's screen when one has to compromise the resolution of one's simulations due to limited compute power.
    Fortunately it turns out, one can ensure the conservation of energy by scaling the electric fields arriving in the screen using a global and scalar calibration factor.
    Once the energy calibration factor is known, one can simulate the images in the screen made by arbitrary electric fields.

    From the screen on, a final step simulates the feed horns to obtain the single electric field present at the successive antenna behind the feed horn.
    \subsection{Huygens principle}
    \label{SecHuygensPrinciple}
        The electric field present at the $\CameraIndex{}$-th scatter center $\FeedHornScatterCenterVector{}_\CameraIndex{}$ in the camera's screen can be computed by adding up the $\NumMirrorScatterCenters{}$ electric fields present at the mirror's scatter centers
        \begin{eqnarray}
            \ElectricField{}(\FeedHornScatterCenterVector{}_\CameraIndex{}, t) &=& \sqrt{w} \sum_{\MirrorIndex{}}^{\NumMirrorScatterCenters{}} \sqrt{v_{\MirrorIndex{}, \CameraIndex{}}} \ElectricField{}(\MirrorScatterCenterVector{}_\MirrorIndex{}, t + \tau_{\MirrorIndex{},\CameraIndex{}})
            \label{EqHuygensSuperposition}
        \end{eqnarray}
        while shifting the electric fields in time with a time-of-flight delay $\tau_{\MirrorIndex{},\CameraIndex{}}$ and applying the weights $w$ and $v$ to the mirror's electric fields.
        The time-of-flight delay
        \begin{eqnarray}
            \tau_{\MirrorIndex{},\CameraIndex{}} &=& \frac{\left| \MirrorScatterCenterVector{}_\MirrorIndex{} - \FeedHornScatterCenterVector{}_\CameraIndex{} \right|}{c} \eta_\text{atmosphere}
            \label{EqTimeOfFlightDelay}
        \end{eqnarray}
        is the time it takes for the radio light to travel from the scatter center $\MirrorScatterCenterVector{}_\MirrorIndex{}$ in the mirror to the scatter center $\FeedHornScatterCenterVector{}_\CameraIndex{}$ in the camera.
        Here $\eta_\text{atmosphere}$ is the refractive index of the atmosphere between the telescope's mirror and screen.

        The weights for the electric field $w$ and $v$ are proportional to energy and thus are in square roots in \autoref{EqHuygensSuperposition} as energy or power are proportional to $\left|\,\ElectricField{}\,\right|^2$.

        One can separate the weights into a component $v_{\MirrorIndex{},\CameraIndex{}}$ what depends on the actual combination of the mirror's scatter center $\MirrorIndex{}$ and the screen's scatter center $\CameraIndex{}$, and a global component $w$ which is independent of the scatter combination.
        \subsubsection*{Spherical waves and the weight $v_{m,n}$}
\begin{figure}{}
            \centering
            \includegraphics[width=1.0\columnwidth]{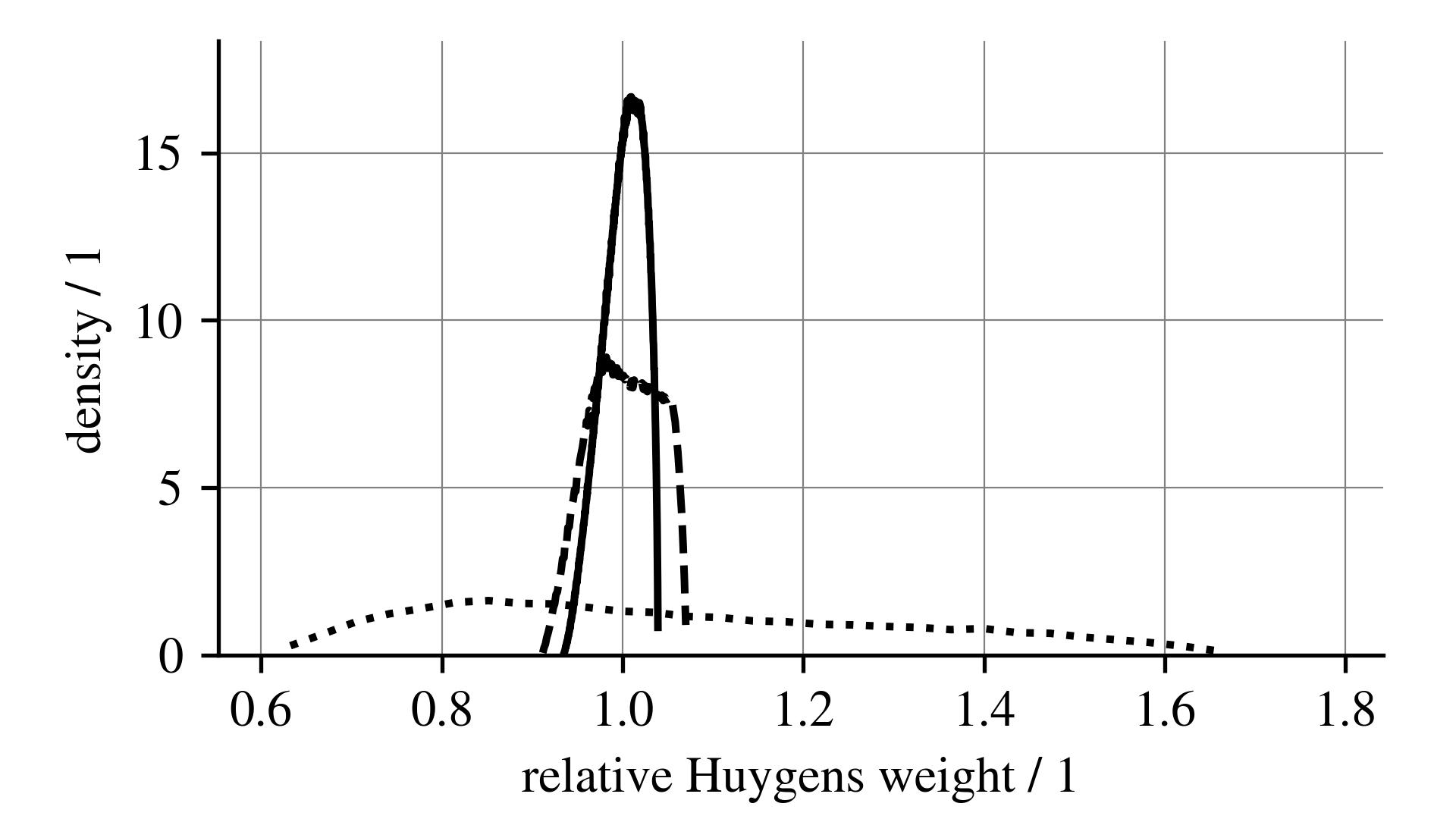}
            \caption{
                Distribution of the electric field weights $v_{m,n}$ for the telescopes $\mathbf{C}$rome (dotted), $\mathbf{M}$edium (dashed), and $\mathbf{L}$arge (line).
                With its short focal-ratio and wide field-of-view, $\mathbf{C}$rome introduces a significant spread among the areas of its scatter spheres.
            }
            \label{FigWeightScatterSphere}
        \end{figure}

        As one has constructed the point clouds for the mirror and the screen to be rather uniform, one can assume that all the $M$ scatter centers in the mirror represent an equal size of mirror area $A_\text{SM}$.
        Same one can assume that all the $N$ scatter centers of a feed horn represent an equal size of the feed horns entrance area $A_\text{SF}$.
        With the represented areas being independent of the scatter combination, the weight $v_{m,n}$ only describes how the areal power density $s$ coming from a mirror scatter center $\vec{p}_m$ vanishes with the distance to the receiving scatter center $\vec{q}_n$.
        The areal power density vanishes with the distance because Huygens principle assumes that its scatter centers emit spherical waves.
        The area of the sphere centered around the mirror scatter center $\vec{p}_m$ and having the receiving scatter center $\vec{q}_n$ on its surface is
        \begin{eqnarray}
            {A_\text{scatter-sphere}}_{m,n} &=& 4 \pi \left| \vec{p}_m - \vec{q}_n \right|^2.
            \label{EqScatterSphere}
        \end{eqnarray}
        To discuss the weights $v$ as relative weights, one calculates the mean area E$[{A_\text{scatter-sphere}}]$ of all the telescope's scatter spheres and defines
        \begin{eqnarray}
            v_{m,n} &=& \frac{
                \text{E}[A_\text{scatter-sphere}]
            }{
                {A_\text{scatter-sphere}}_{m,n}
            }.
            \label{EqWeightScatterSphere}
        \end{eqnarray}
        \autoref{FigWeightScatterSphere} shows the distribution of the weights $v_{m,n}$ for the telescopes $\mathbf{C}$rome, $\mathbf{M}$edium, and $\mathbf{L}$arge.
        One finds that on the telescopes $\mathbf{M}$edium and $\mathbf{L}$arge, the spread among the sizes of their scatter spheres is rather small.
        Only the geometry of $\mathbf{C}$rome introduces a noticable spread among the weights $v_{m,n}$.
        \subsubsection*{Geometry and the global weight $w$}
        The global weight $w$, which is applied to all the mirror's electric fields in \autoref{EqHuygensSuperposition}, is crucial to conserve the energy transported from the mirror to the screen in a way that one would expect from an imaging mirror.
        Geometry suggests that $w$ is proportional to three different components.
        First,
        \begin{eqnarray}
            w &\propto& \frac{
                A_\text{SM}
            }{
                A_\text{SF}
            }
            \label{EqWeightEnergyArea}
        \end{eqnarray}
        is proportional to the ratio of the scatter center areas, in the mirror and the screen.
        Second,
        \begin{eqnarray}
            w &\propto& \frac{
                1
            }{
                M
            }
            \label{EqWeightEnergyCounting}
        \end{eqnarray}
        is proportional to the inverse number of scatter centers in the mirror.
        This is because \autoref{EqHuygensSuperposition} does not sum a quantity proportional to energy or power but electric fields.
        Thus to conserve energy, the electric fields in \autoref{EqHuygensSuperposition} must be weighted with $\sqrt{1/M}$.
        And third,
        \begin{eqnarray}
            w &\propto& \frac{
                Q
            }{
                N
            }
            \label{EqWeightEnergyPointContainment}
        \end{eqnarray}
        depends on the mirror's point spread function and on how much of this point spread function can be contained in a feed horn.
        Here we define $Q$ to be the fraction of radio light coming from a far distant point source which can be contained in a single feed horn.
        The factor $1/N$ is because we define the containment $Q$ to be relative to one feed horn but \autoref{EqHuygensSuperposition} is about a feed horn scatter center of which there are $N$ in each feed horn.
        As $Q$ is only about the ratio of energy, it can be estimated with the initial Huygens imaging step in \autoref{EqHuygensSuperposition} even when the weight $w$ is not yet known and set to e.g. $w=1$.
        \autoref{TabTelescopesPsfFeedHornContainment} shows the containment factor $Q$ for the three telescopes $\mathbf{C}$rome, $\mathbf{M}$edium, and $\mathbf{L}$arge.
        \begin{table}
            \begin{tabular}{lrrr}
                 & $\mathbf{C}$ & $\mathbf{M}$ & $\mathbf{L}$\\
                \toprule
                $Q/1$ & 0.52 & 0.81 & 0.84 \\
            \end{tabular}
            \caption{
                The fraction of energy $Q$ contained in the central feed horn of the three example telescopes for radio light coming in a plane wave traveling parallel to the mirror's optical axis.
                This is estimated using high resolution images similar to the most right panels in \autoref{FigGuideStars}.
            }
            \label{TabTelescopesPsfFeedHornContainment}
        \end{table}

        With geometry's suggestions applied, the weight
        \begin{eqnarray}
            w &=& \frac{ A_\text{SM} }{ A_\text{SF} } \frac{1}{M} \frac{Q}{N} u
            \label{EqWeightEnergy}
        \end{eqnarray}
        conserves the energy transported from the mirror to the camera screen up to a missing  factor of $u \approx 6$ which seems to be independent of the area or time related resolutions in our simulations.
        \autoref{TabEnergyMissingFactor} shows the missing factor $u$ from \autoref{EqWeightEnergy} for the telescopes $\mathbf{C}$rome, $\mathbf{M}$edium, and $\mathbf{L}$arge.

        Currently, the origin of the missing factor $u$ is not fully understood.
        It is suspected that the missing factor $u$ is largely due to a physical mirror emits only into the forward hemisphere and with a characteristic angular dependence whereas the implementation presented here assumes an isotropic emission.
        However, in order to do gamma ray astronomy it is not relevant to understand the factors in the weight $w$ as long as one can calibrate $w$ to make the Huygens based image formation conserve the energy in a way that one would expect from an imaging mirror.
        \begin{table}
            \begin{tabular}{lrrr}
                 & $\mathbf{C}$ & $\mathbf{M}$ & $\mathbf{L}$\\
                \toprule
                $u/1$ & 4.7 & 5.8 & 6.7 \\
            \end{tabular}
            \caption{
                The missing energy factor $u$ which is required to roughly conserve the energy transported by the mirror to the camera screen.
            }
            \label{TabEnergyMissingFactor}
        \end{table}
    \subsection{Energy calibration}
        \label{SecEnergyCalibration}
\begin{figure*}{}
            \centering
            \includegraphics[width=1\textwidth]{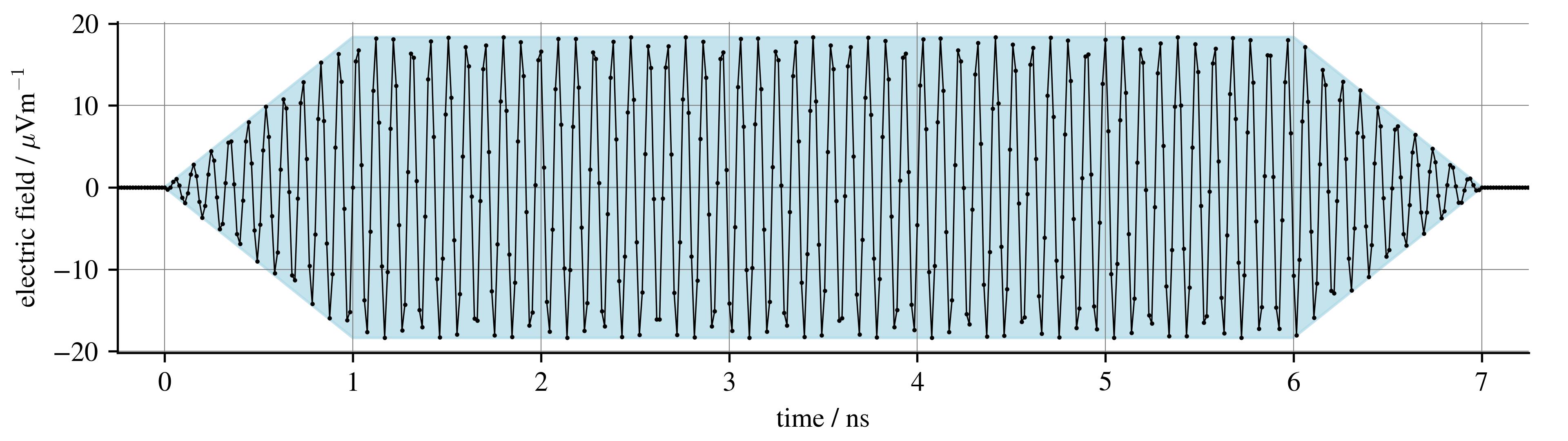}
            \caption{
                The electric field amplitude of an example plane wave package used by our calibration source.
                Blue area is the hull of the amplitude showing the 1\,ns ramp up, the 5\,ns plateau, and the 1\,ns ramp down.
                Small black dots mark the sampling moments in time.
                Black line guides the eye to see the underlying sine wave what here has a frequency of $\nu = 10.32\,$GHz.
                The wave's plateau reaches an areal power density of $4.5\times10^{-13}$\,Wm$^{-2}$.
                By chance, this corresponds to a source emitting 1\,W (cell phone) in a distance of 422\,km (altitude of the International Space Station).
                Over time, this plane wave package will deposit an areal energy density of $2.7\times10^{-21}$\,Jm$^{-2}$.
                According to simulations in section \ref{SecAirshowerImages} this plateau electric field amplitude of $\approx 18\,\mu$Vm$^{-1}$ corresponds roughly to the radio emission found in a 2\,GHz bandwidth coming from an air shower induced by an order 100 \,TeV cosmic ray.
            }
            \label{FigPlaneWavePackage}
        \end{figure*}

        To calibrate the energy transport of the Huygens based image formation, one simulates many observations of plane wave packages.
        A plane wave package corresponds to a radio signal coming from a far distant source.
        The plane waves have frequencies uniformly distributed in the telescope's sensitive frequency range, and directions randomly drawn inside and outside the telescopes field-of-views.
        Also the energy and polarization angles in the wave packages vary.
        \autoref{FigPlaneWavePackage} shows the electric field amplitude of an example plane wave package used in our simulations.
        From the statistics of many observations of such plane wave packages, one gathers the statistics to estimate the global energy conservation factor $w$ from \autoref{EqHuygensSuperposition}.
        For the energy calibration based on plane waves which resemble far distant stars, the camera screens of the telescopes here are focused to an object distance $g=\infty$ such that their screens are in a distance to the mirror of $d=f$.

        To be precise: The input, or expected energy here is defined as the energy carried by the plane wave package what theoretically could be collected in the mirror's effective area $A_\text{mirror}$ from \autoref{EqMirrorTotalArea}.
        The output, or actual energy is the energy carried by the average electric fields present at the entrance opening of specific feed horns, see \autoref{SecSimulatingTheFeedHorn}.
        For this, one only picks specific feed horns in the proximity of the brightest spot found in the image.
        The proximity radius here is chosen to be the mirror's Airy disk diameter $\Theta_\text{Airy-full}$ from \autoref{EqAiry} what usually includes in the order of $\approx 10$ feed horns.
        As this is about electric fields at the entrance opening of the feed horns, possible transmission or reflection losses caused by the feed horns are not taken into account here.
        This is only about the Huygens image formation from the mirror to the screen.

        The figure panels \autoref{FigCromePerformanceEnergy}, \autoref{FigMediumPerformanceEnergy}, and \autoref{FigLargePerformanceEnergy} show the resulting energy conservation in the three telescopes $\mathbf{C}$rome, $\mathbf{M}$edium, and $\mathbf{L}$arge after the global weight $w$ has been estimated.
        In case, one uses \autoref{EqWeightEnergy} to express the global weight $w$ using our suggested components plus the missing scale factor $u$, one finds the value for $u$ in \autoref{TabEnergyMissingFactor}.
        One can estimate $u$ by running the observations of the plane wave packages with $u=1$ and then applying a linear fit to the then resulting curves corresponding to the panels \autoref{FigCromePerformanceEnergy}, \autoref{FigMediumPerformanceEnergy}, and \autoref{FigLargePerformanceEnergy}.
        With the fit one can estimate the missing energy factor in the center of the field-of-view using observations which are scattered across the entire reasonable useful field-of-view.
        As can be seen in the resulting panels \autoref{FigCromePerformanceEnergy}, \autoref{FigMediumPerformanceEnergy}, and \autoref{FigLargePerformanceEnergy}, the energy conservation is mostly below 1.0 and peaks in the center of the field-of-view.

        Clearly, there is room for improvement here.
        But currently we just want to explore if a Huygens based image formation with the imaging atmospheric radio  telescope is possible at all.
    \subsection{Combining the scatter centers of a feed horn}
        \label{SecSimulatingTheFeedHorn}
        Finally, the $\NumScatterCentersPerFeedHorn{}$ electric fields at the entrance opening of the feed horn must be combined into a single electric field which is present at the successive antenna behind the feed horn.
        Ideally, one would simulate the propagation of the electric fields through the feed horn based on the geometry of the feed horn.
        However, we failed to simulate the feed horn using Huygen's principle.
        As a fall back, one can combine the $\NumScatterCentersPerFeedHorn{}$ input electric fields of the feed horn into one output electric field such that the energy carried by the single output electric field is the sum of the energies carried by the $\NumScatterCentersPerFeedHorn{}$ input electric fields.
        To do so, one can define the function
        \begin{eqnarray}
            \vec{\text{P}}[\vec{x}, p] &=& \frac{\vec{x}}{\left| \vec{x} \right|} \left| \vec{x} \right|^p
            \label{EqPowerElement}
        \end{eqnarray}
        to raise the magnitude of a vector $\vec{x}$ to a given power $p$.
        With this one adds the squared electric field amplitudes
        \begin{eqnarray}
            {\vec{\epsilon}_\text{feed-horn}} =
            \sqrt{\frac{A_\text{SF}}{A_\text{feed-horn}}}
            \vec{\text{P}}\left[
                \sum_{n}^{N} \vec{\text{P}}
                \left[
                    {{\vec{\epsilon}}_{\text{SF}_n}},
                    2
                \right],
                1/2
            \right]
            \label{EqFeedHornCombine}
        \end{eqnarray}
        and applies the square root to the sum.
        The result is a single electric field ${\vec{\epsilon}_\text{feed-horn}}$ present at a virtual scatter center that is associated with the feed horn's entrance area $A_\text{feed-horn} = N A_\text{SF}$.
        One might think of this approximation as the average electric field present at the entrance of the feed horn.
        At least in the frequency domain, the combination procedure presented in \autoref{EqFeedHornCombine} does not seem to induce unexpected artifacts as \autoref{FigMediumMulitExampleFrequency} and \autoref{FigLargeMulitExampleFrequency} suggest.
        To get the final electric field present at the successive antenna behind the feed horn one still has to multiply the electric field amplitude with the feed horn's geometric gain factor, and potentially a transmission loss factor due to Fresnel reflection.
    \section{Results of Huygens image formation}
\begin{figure}{}
        \subfloat[energy conservation]{
            \centering
            \includegraphics[width=1\columnwidth]{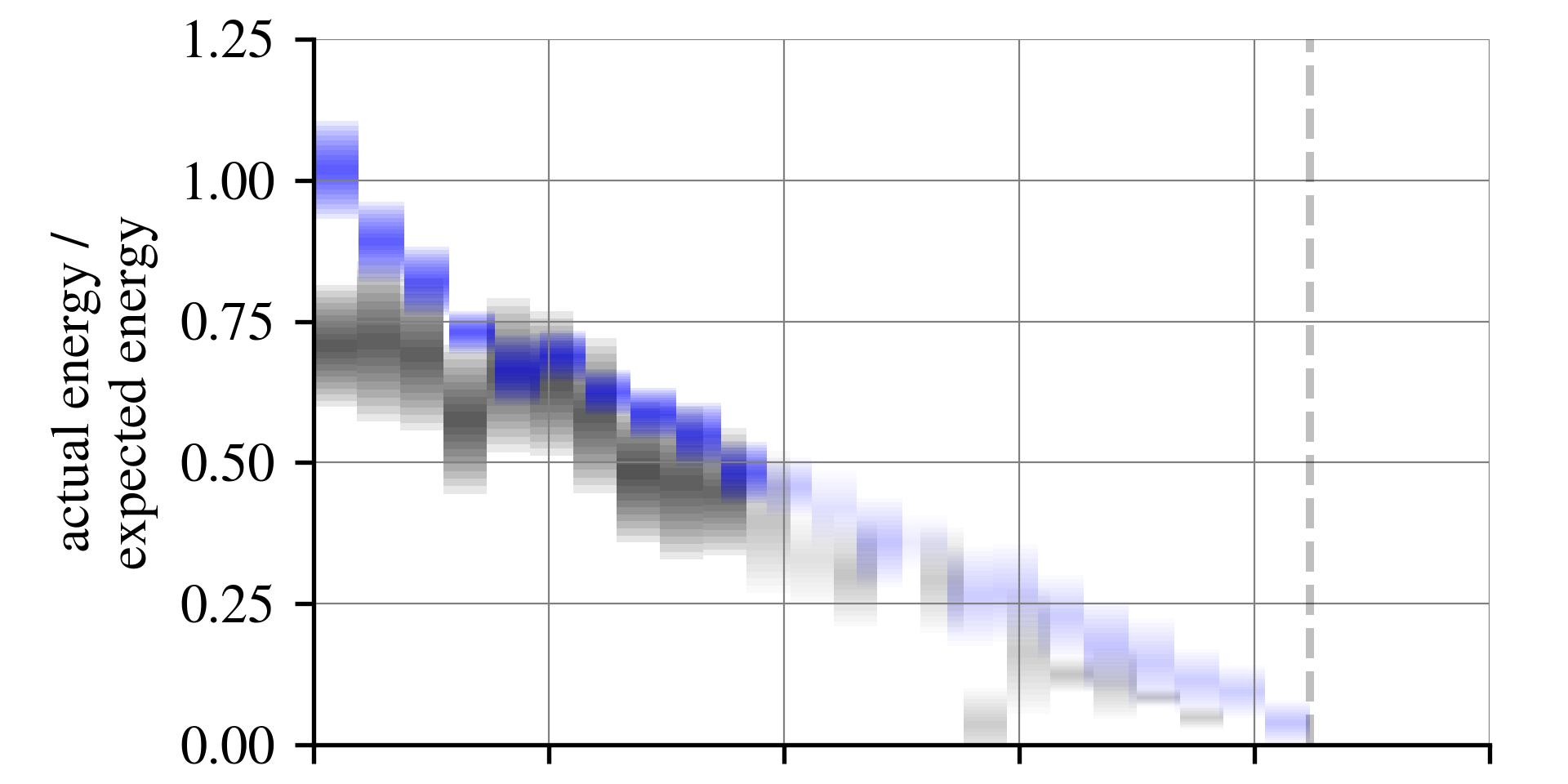}
            \label{FigCromePerformanceEnergy}
        }\\
        \subfloat[spread]{
            \centering
            \includegraphics[width=1\columnwidth]{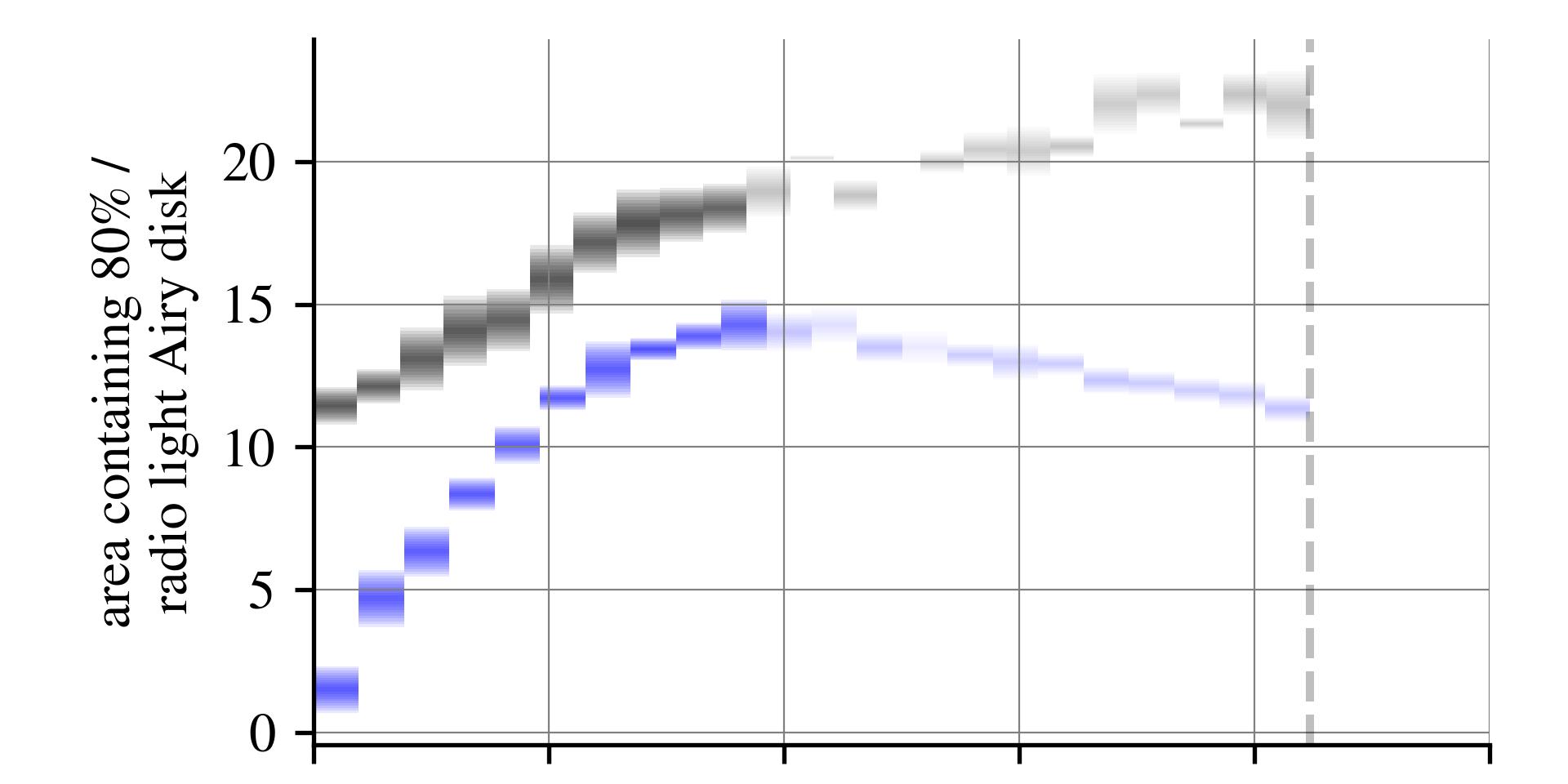}
            \label{FigCromePerformanceSpread}
        }\\
        \subfloat[distortion]{
            \centering
            \includegraphics[width=1\columnwidth]{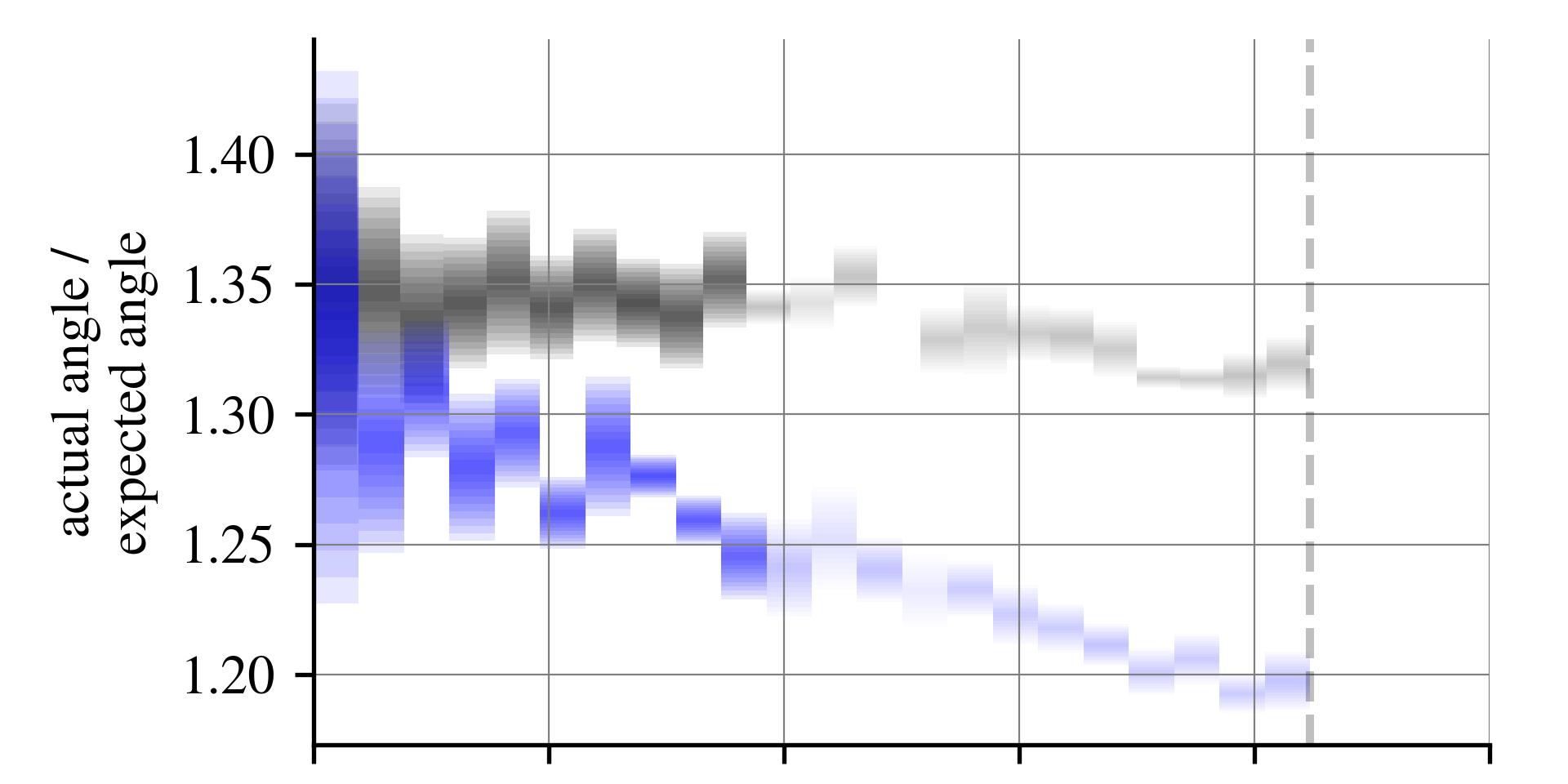}
            \label{FigCromePerformanceDistortion}
        }\\
        \subfloat[statistics]{
            \centering
            \includegraphics[width=1\columnwidth]{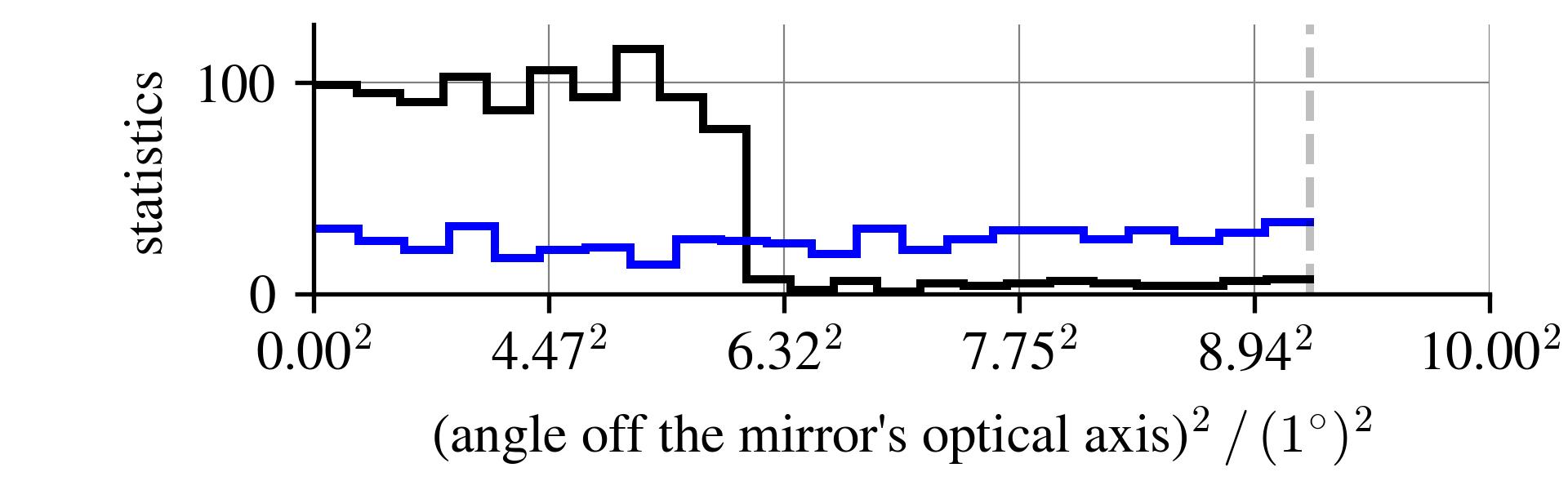}
        }
        \caption{
            $\mathbf{C}$rome telescope performance in black.
            Blue is the expectation based on geometric optics.
            The angle off the optical axis is squared here so that the bins have equal solid angles in the sky.
        }
        \label{FigCromePerformance}
    \end{figure}

\begin{figure}{}
        \subfloat[energy conservation]{
            \centering
            \includegraphics[width=1\columnwidth]{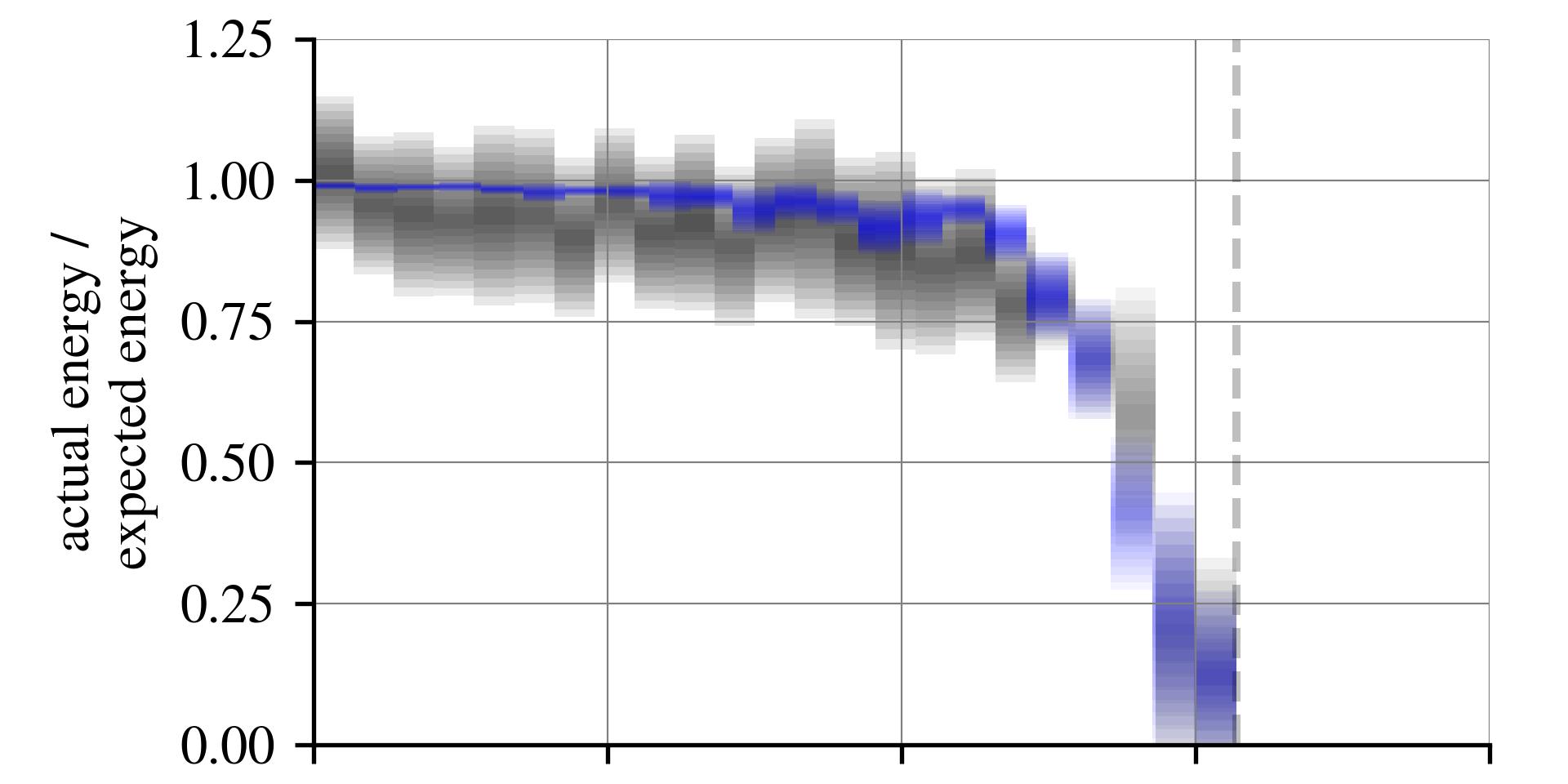}
            \label{FigMediumPerformanceEnergy}
        }\\
        \subfloat[spread]{
            \centering
            \includegraphics[width=1\columnwidth]{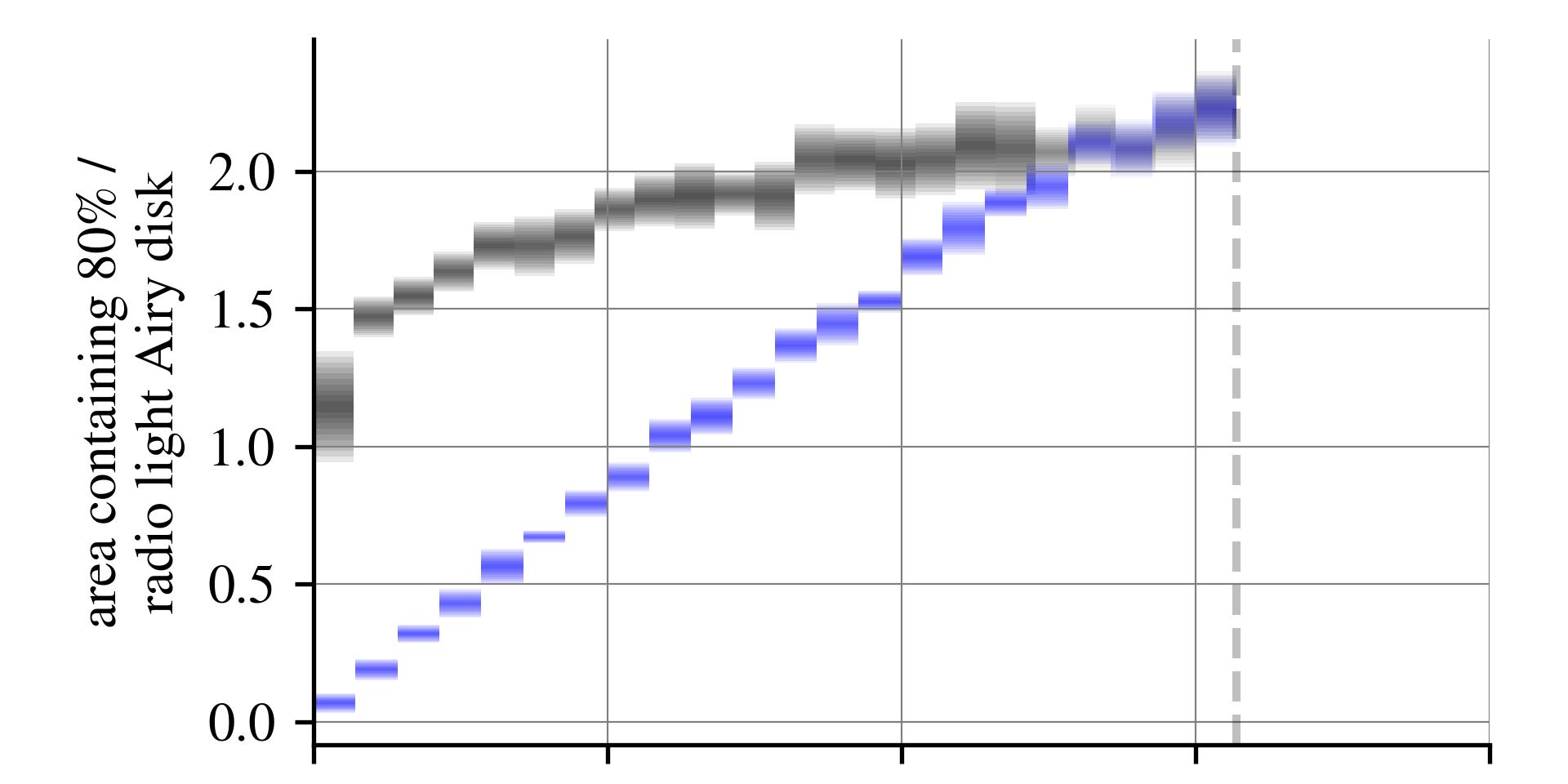}
            \label{FigMediumPerformanceSpread}
        }\\
        \subfloat[distortion]{
            \centering
            \includegraphics[width=1\columnwidth]{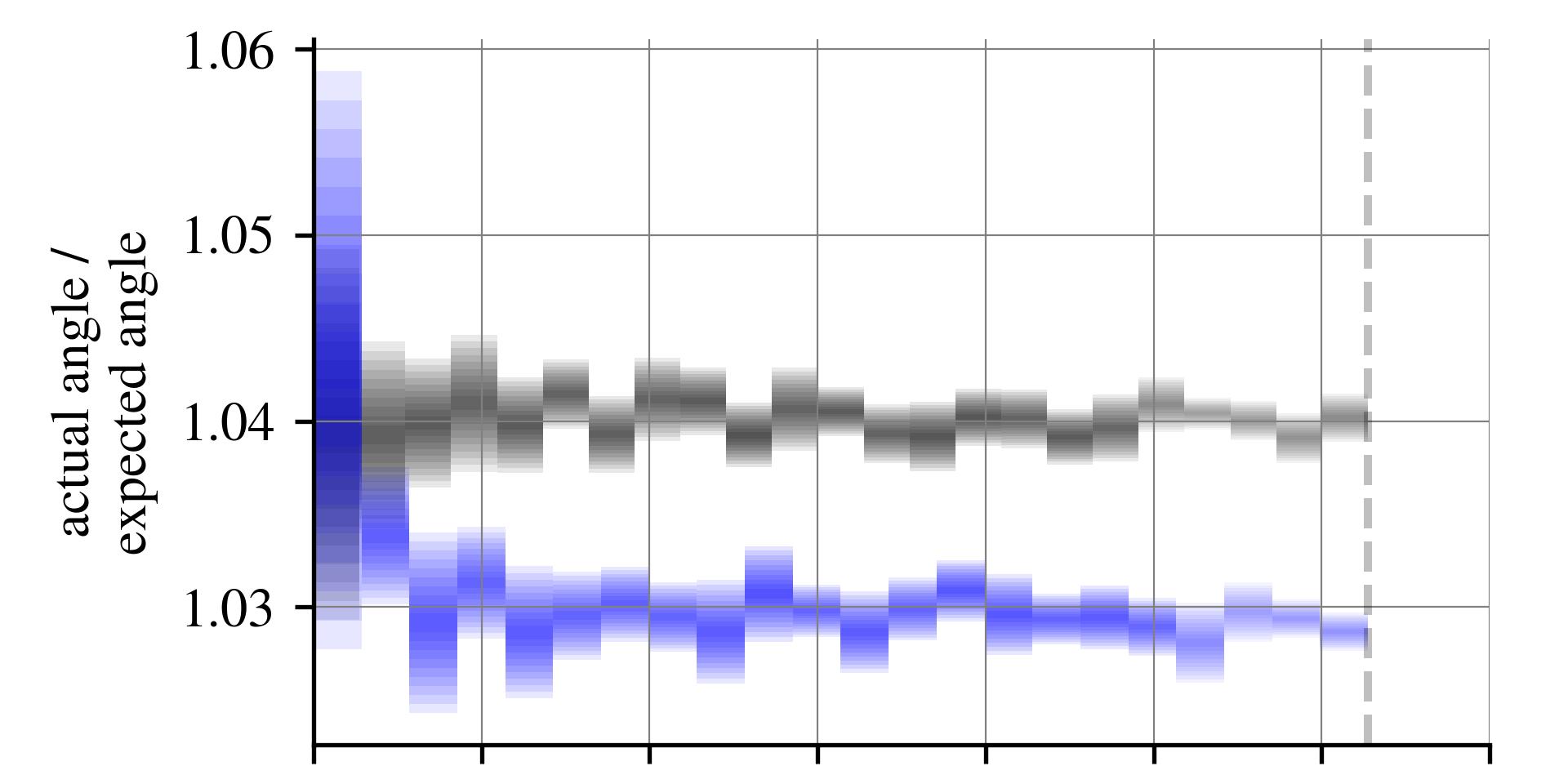}
            \label{FigMediumPerformanceDistortion}
        }\\
        \subfloat[statistics]{
            \centering
            \includegraphics[width=1\columnwidth]{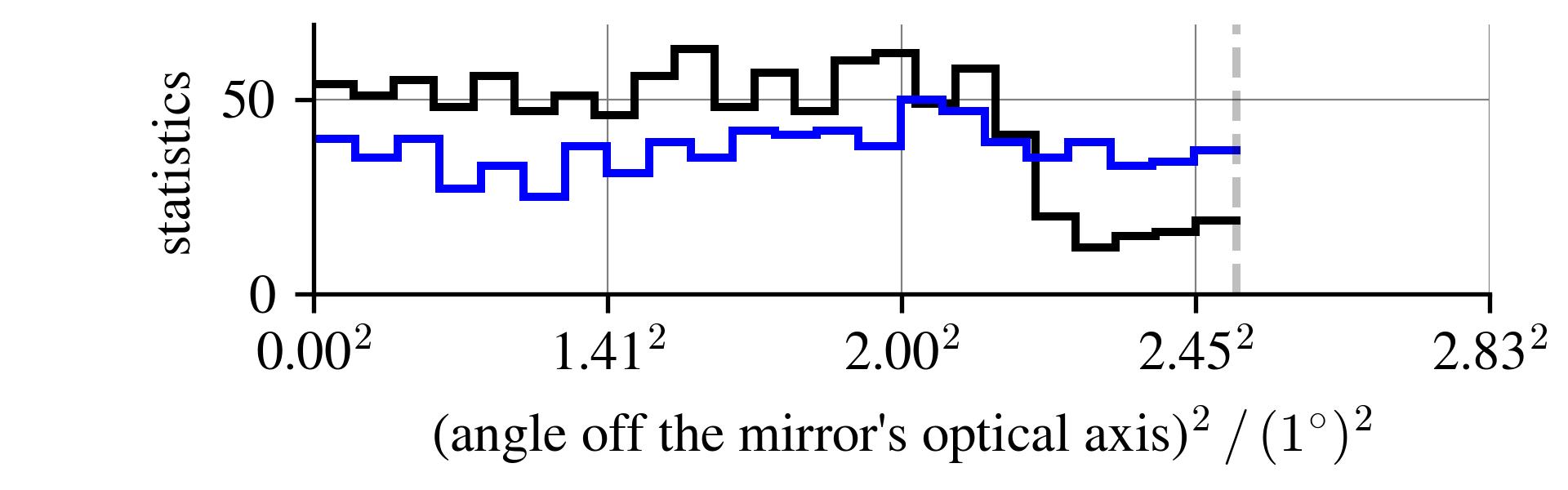}
        }
        \caption{
            $\mathbf{M}$edium telescope performance.
            Blue is the expectation based on geometric optics.
        }
        \label{FigMediumPerformance}
    \end{figure}

\begin{figure}{}
        \subfloat[energy conservation]{
            \centering
            \includegraphics[width=1\columnwidth]{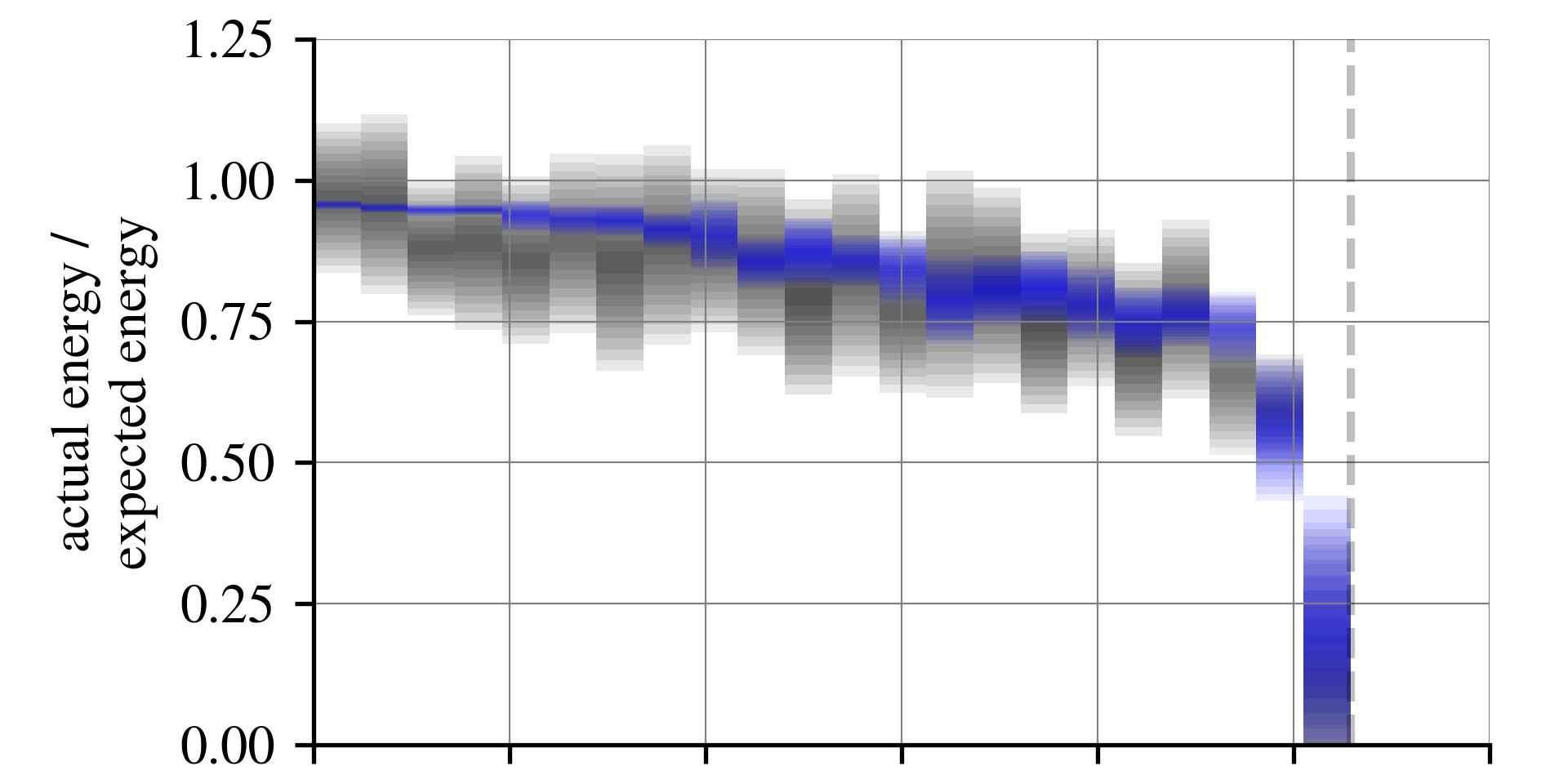}
            \label{FigLargePerformanceEnergy}
        }\\
        \subfloat[spread]{
            \centering
            \includegraphics[width=1\columnwidth]{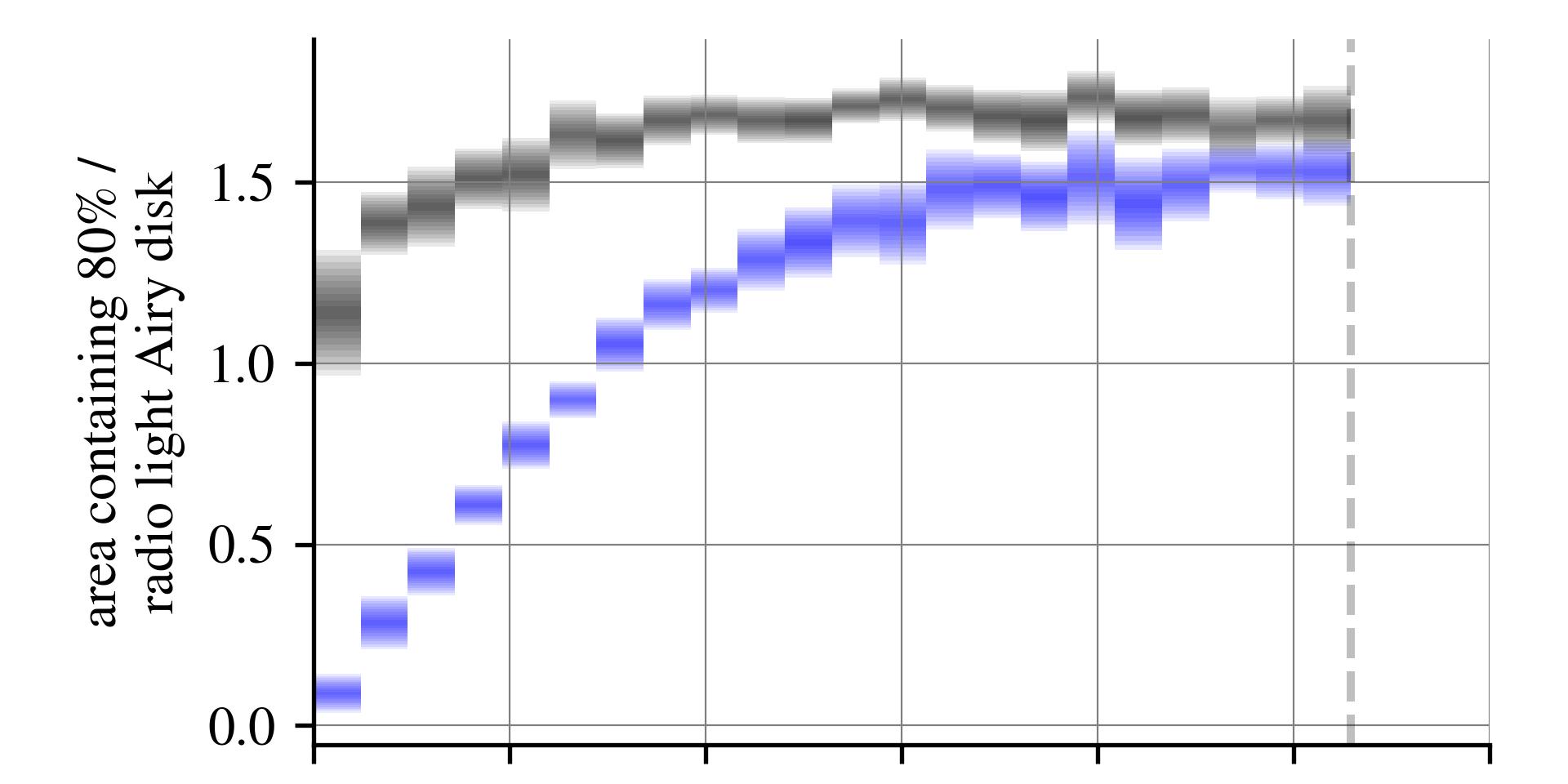}
            \label{FigLargePerformanceSpread}
        }\\
        \subfloat[distortion]{
            \centering
            \includegraphics[width=1\columnwidth]{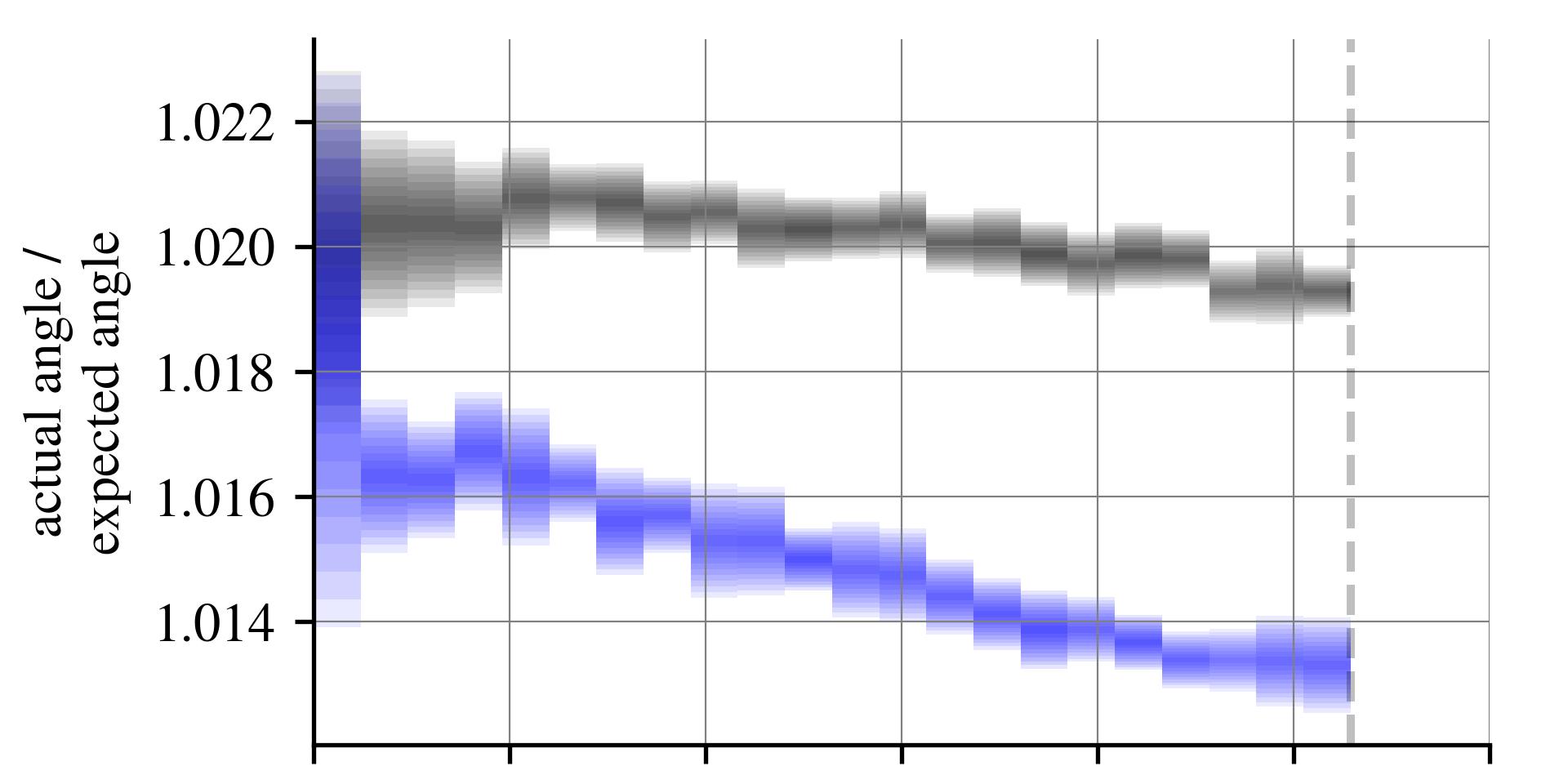}
            \label{FigLargePerformanceDistortion}
        }\\
        \subfloat[statistics]{
            \centering
            \includegraphics[width=1\columnwidth]{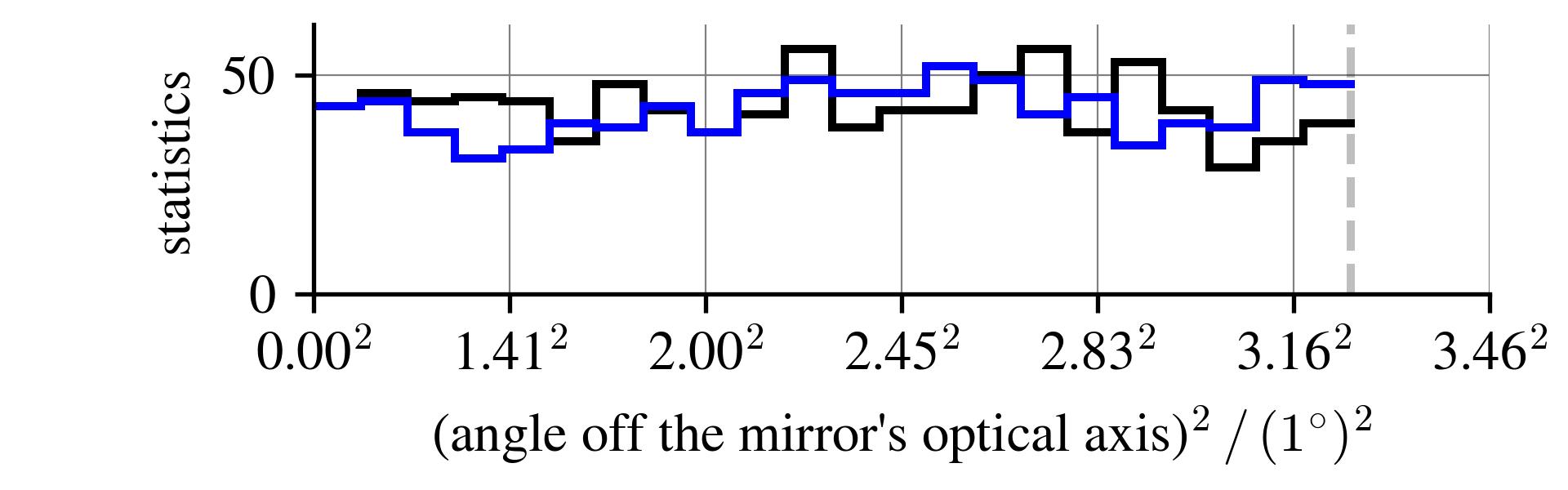}
        }
        \caption{
            $\mathbf{L}$arge telescope performance.
            Blue is the expectation based on geometric optics.
        }
        \label{FigLargePerformance}
    \end{figure}

    The figures \autoref{FigCromePerformance}, \autoref{FigMediumPerformance}, and \autoref{FigLargePerformance} show the simulated properties and performance of the three telescopes $\mathbf{C}$rome, $\mathbf{M}$edium, and $\mathbf{L}$arge in black.
    To interpret the performance of the Huygens based image formation one can compare it to two other methods.

    First, one can compare the Huygens based image formation to the far field approximation for wave mechanics, such as Airy's limit.
    In case of the telescope, a `far field' approximation assumes that the focal-length was $f = \infty$ infinite.
    From this one expects that the mirror's smallest possible point spread function is Airy's disk and that it shows similar rings of constructive and destructive interference as Airy's Bessel function predicts.

    Second, one can compare the Huygens based image formation to geometric optics based on ray tracing where phase and interference are neglected.
    From this one expects that the point spread function suffers from the parabolic mirror's aberration and distortion which become stronger with the angle off the optical axis.
    One also expects these effects to scale with the telescope's focal-ratio $f/D$.

    Thus in total, one expects the Huygens based image formation to show features from both these extreme cases.
    The blue curves in \autoref{FigCromePerformance}, \autoref{FigMediumPerformance}, and \autoref{FigLargePerformance} are based on ray tracing where one uses the same definitions and extraction algorithms to define energy containment, spread, and distortion.
    \subsection{Conserving Energy}
        \label{SecResultsConservingEnergy}
        With the definition of expected and actual energy discussed in \autoref{SecEnergyCalibration}, the panels \autoref{FigCromePerformanceEnergy}, \autoref{FigMediumPerformanceEnergy}, and \autoref{FigLargePerformanceEnergy} show how the Huygens based image formation (black) compares to ray tracing (blue).
        For all three telescopes, one finds that the Huygens based image formation is able to reproduce the energy conservation expectations based on ray tracing, but it also shows that the Huygens image formation shown here fluctuates more noticeably.

        Unlike the Large telescope, Crome and Medium show a steady decline in energy conservation towards larger angles off the optical axis before the steep drop off at the edge of the instrumented field-of-view.
        This decline is caused by Crome's and Medium's lower focal-ratios $f/D \leq 1$ which makes their point spread functions grow quickly beyond the group of feed horns which we defined to carry the expected energy, see \autoref{SecEnergyCalibration}.
        Especially in the case of Crome it is apparent that its mirror was really only meant for a single feed horn in its focal-point and not for imaging in a wider field-of-view.
    \subsection{Point spread function}
        \label{SecPointSpreadFunction}
        One can describe the confusion of an imaging optics using the size of the spot it projects onto a screen for an incoming plane wave.
        The light distribution of this spot is called the point spread function.
        The figure panels \autoref{FigCromePerformanceSpread}, \autoref{FigMediumPerformanceSpread}, and \autoref{FigLargePerformanceSpread} show the area of the spot which contains 80\% of the light coming from a plane wave.
        The areas obtained with the Huygens image formation are shown in black, while areas obtained with ray tracing are shown in blue. 
        For the Medium and for the Large telescope one finds a remarkable confirmation of the expectations based on the far field approximation and geometric optics.

\begin{figure*}{}
        \subfloat[$\mathbf{C}$rome telescope, compare (\citet{werner2013phd}, Fig.\,5.16)]{
            \centering
            \includegraphics[width=0.3\textwidth]{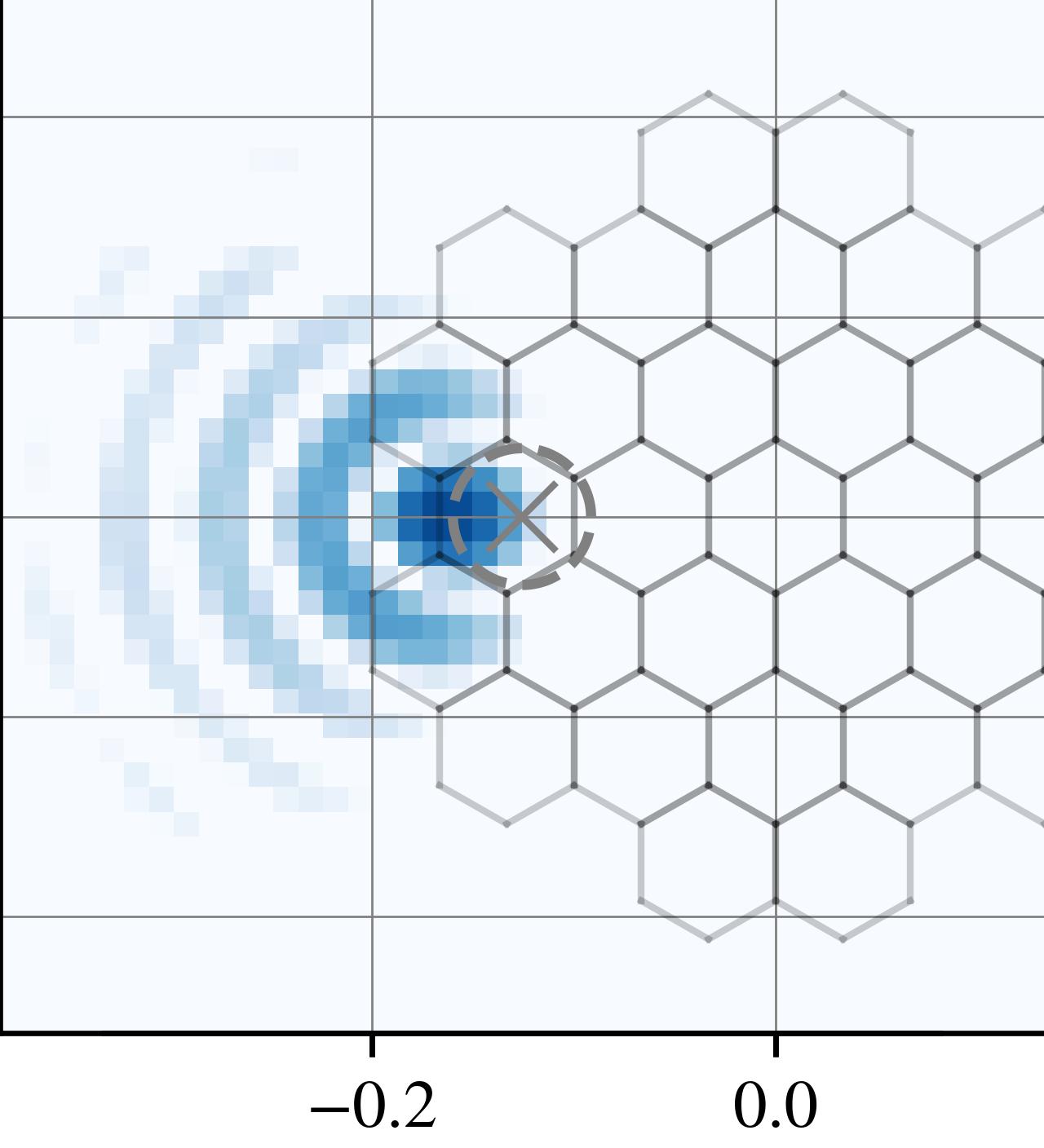}
            \includegraphics[width=0.3\textwidth]{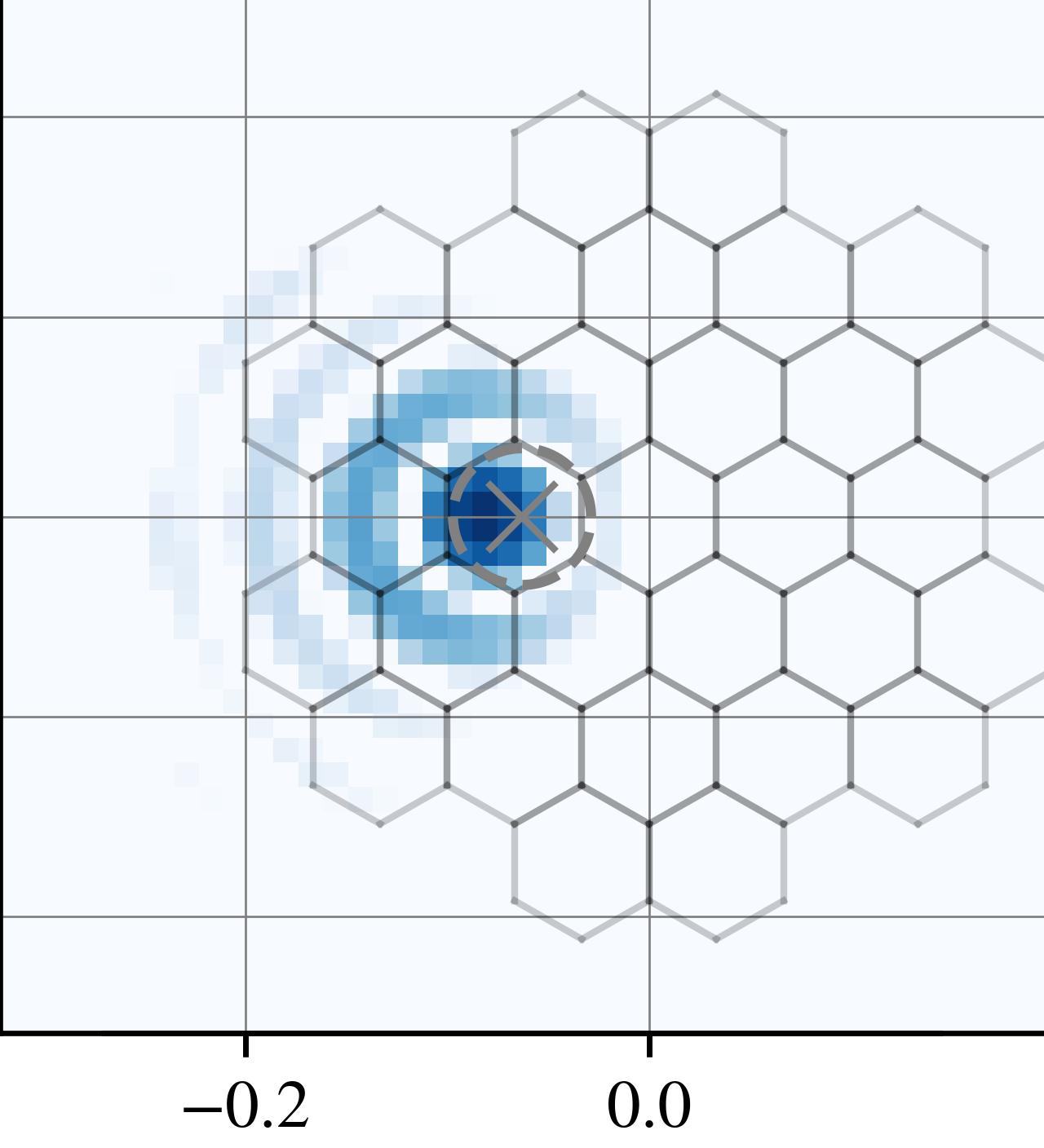}
            \includegraphics[width=0.3\textwidth]{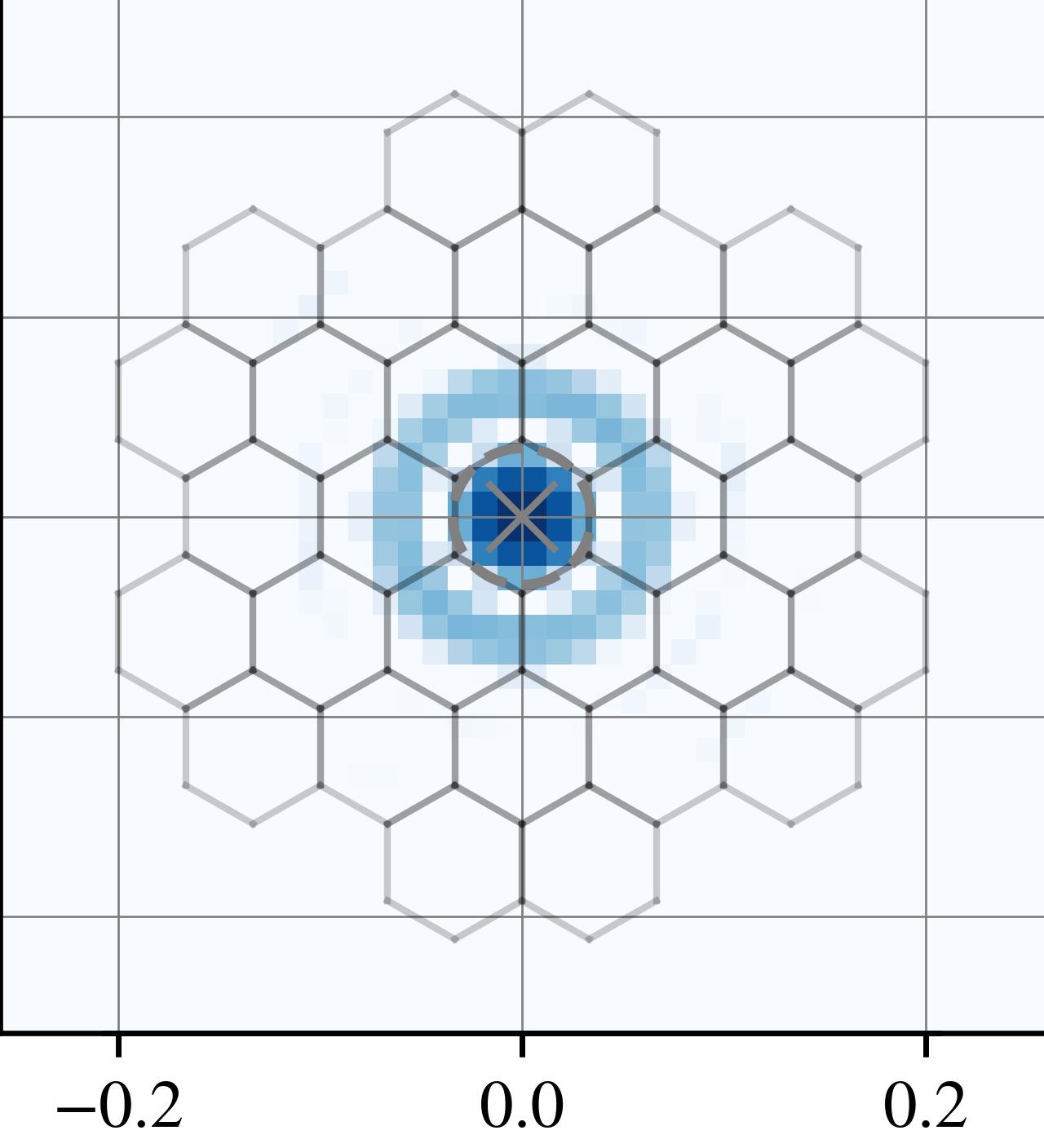}
            \label{FigCromeGuideStars}
        }
        \\
        \subfloat[$\mathbf{M}$edium telescope]{
            \centering
            \includegraphics[width=0.3\textwidth]{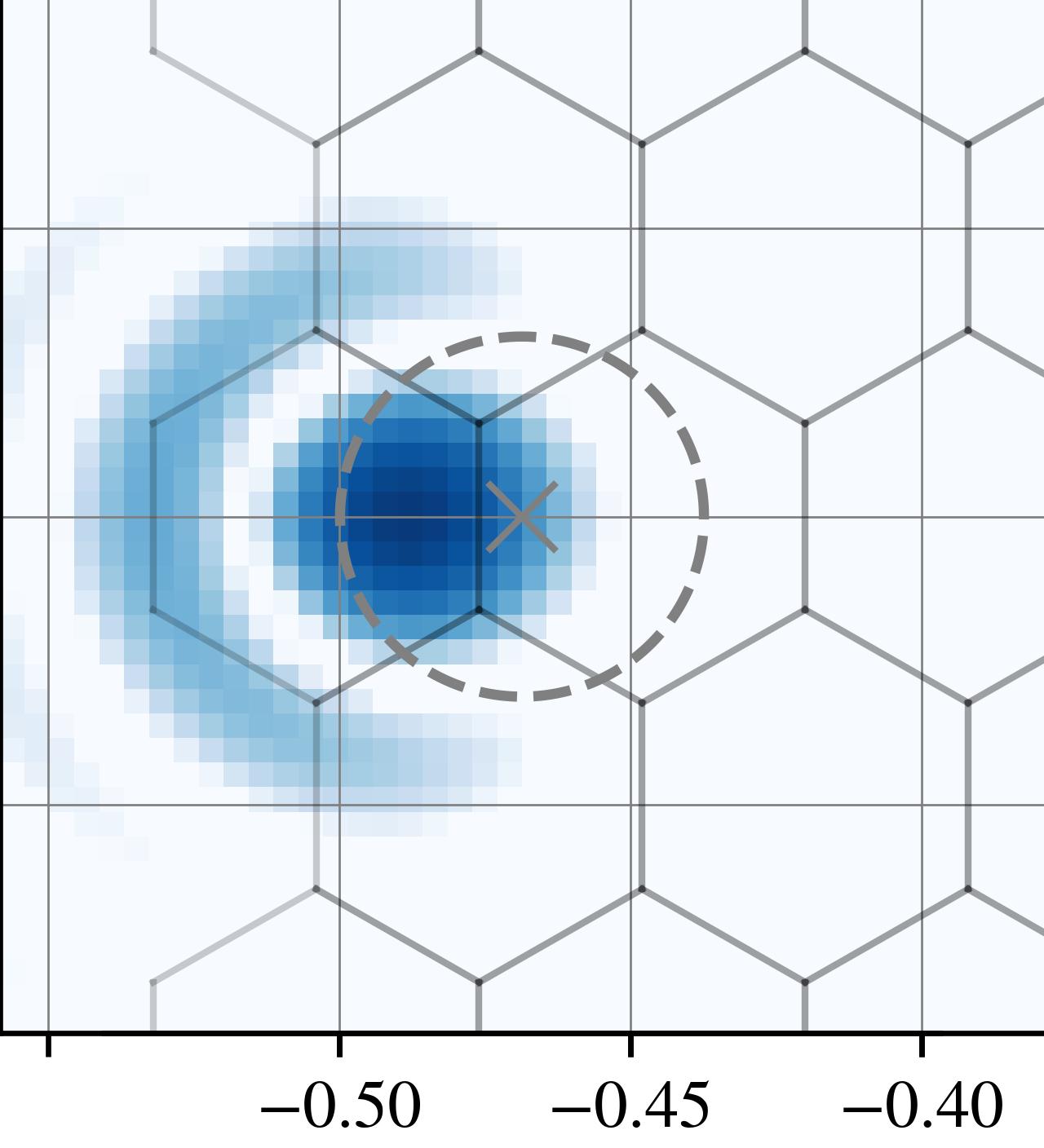}
            \includegraphics[width=0.3\textwidth]{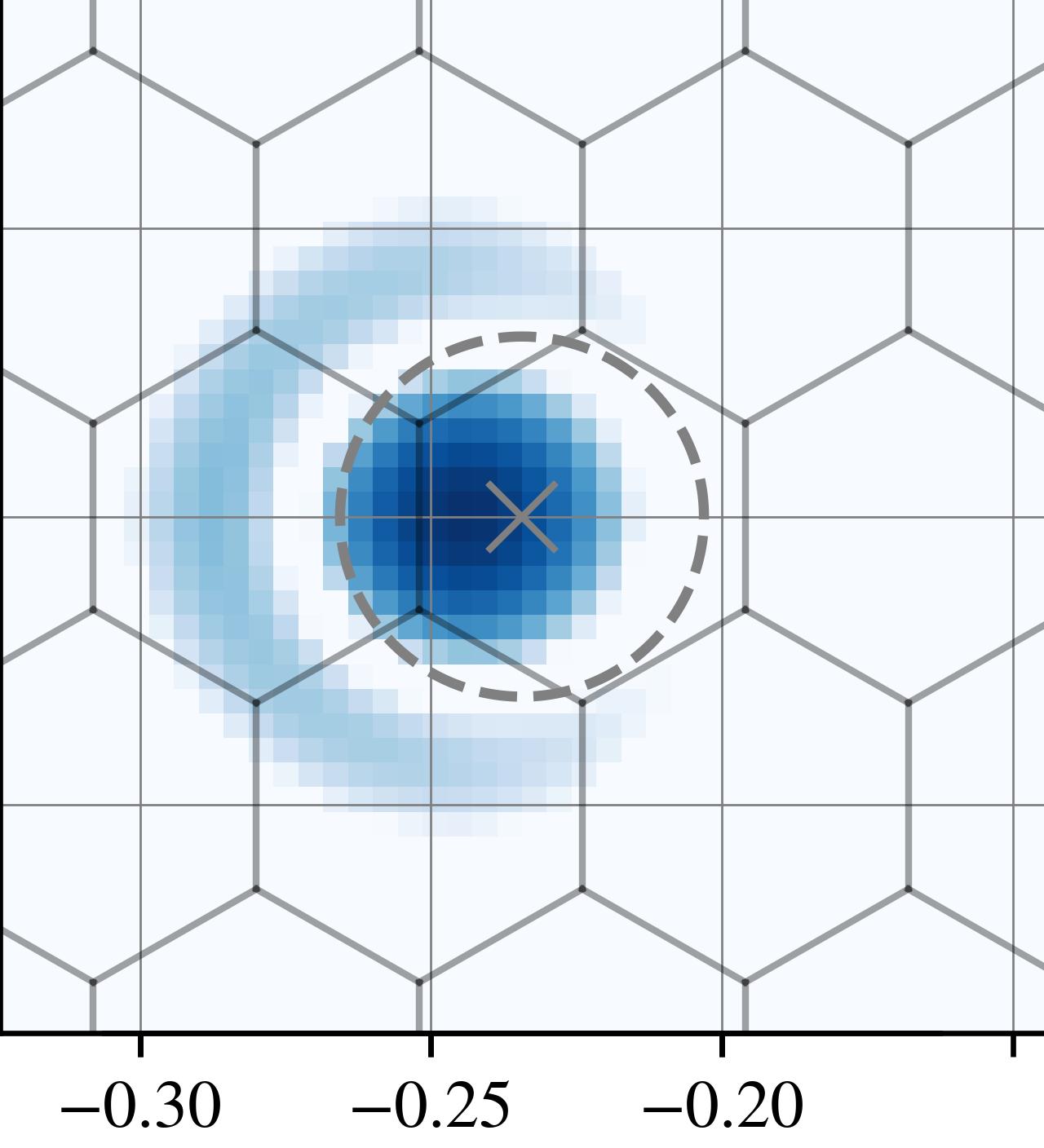}
            \includegraphics[width=0.3\textwidth]{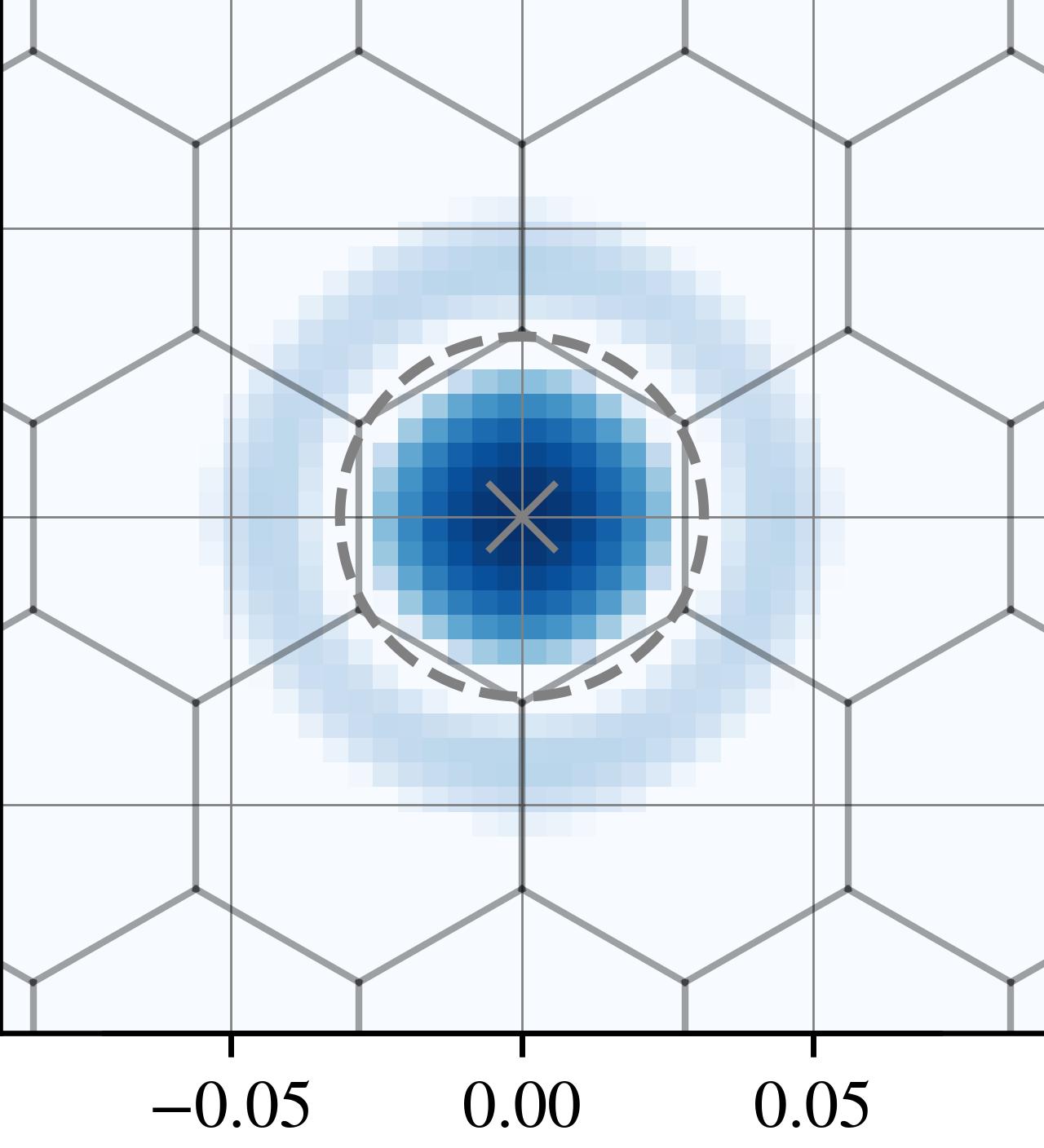}
            \label{FigMediumGuideStars}
        }
        \\
        \subfloat[$\mathbf{L}$arge telescope]{
            \centering
            \includegraphics[width=0.3\textwidth]{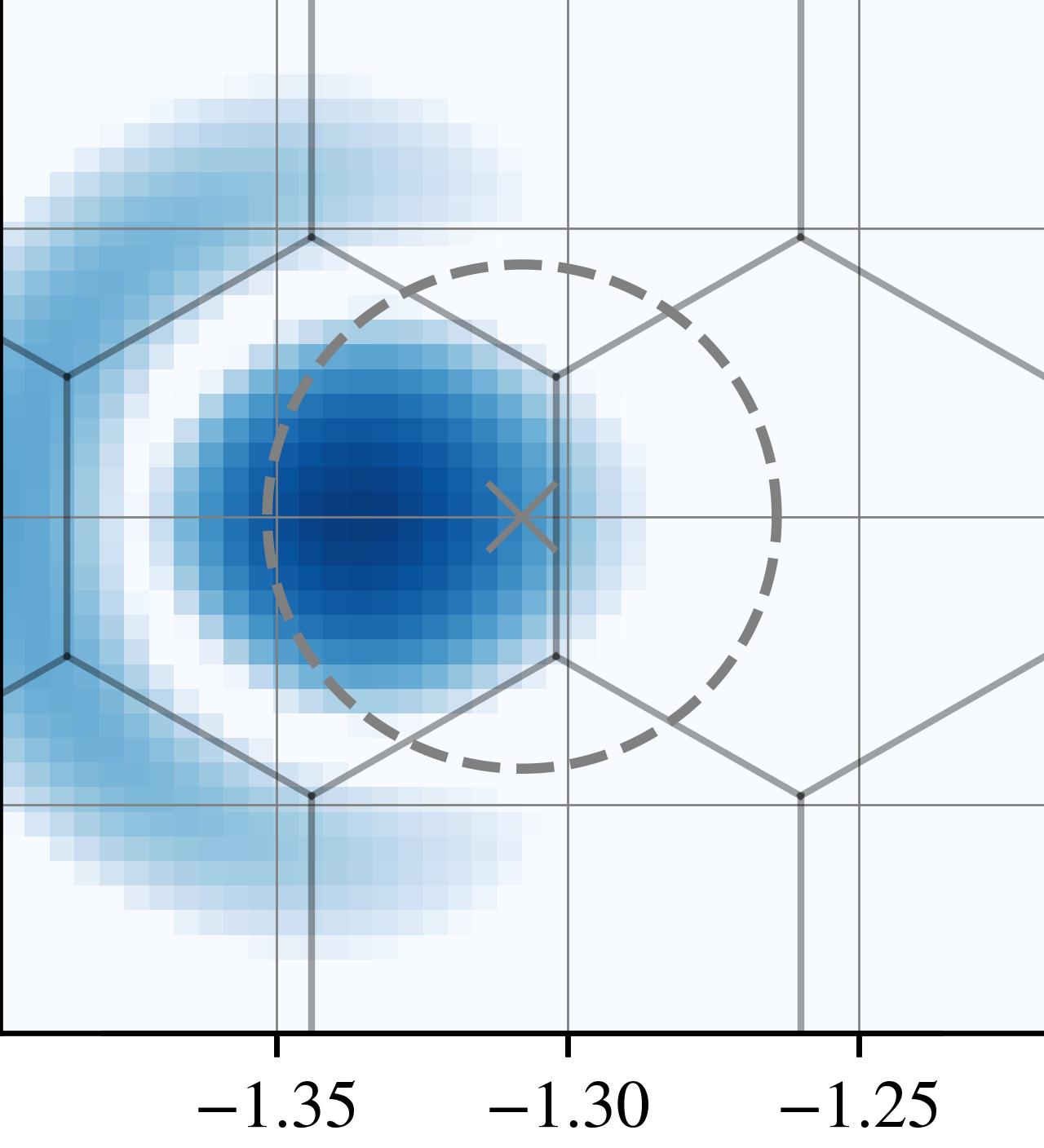}
            \includegraphics[width=0.3\textwidth]{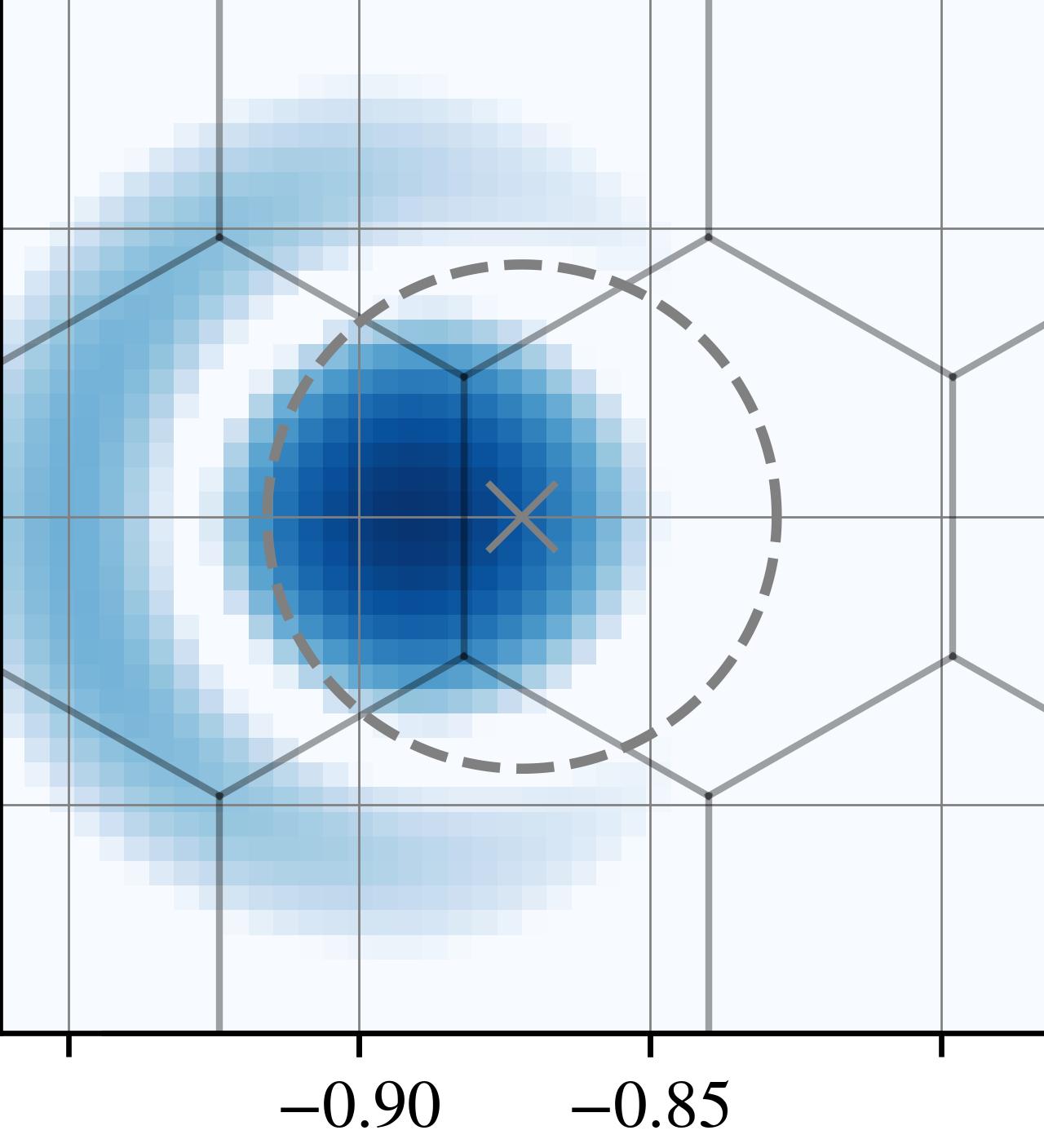}
            \includegraphics[width=0.3\textwidth]{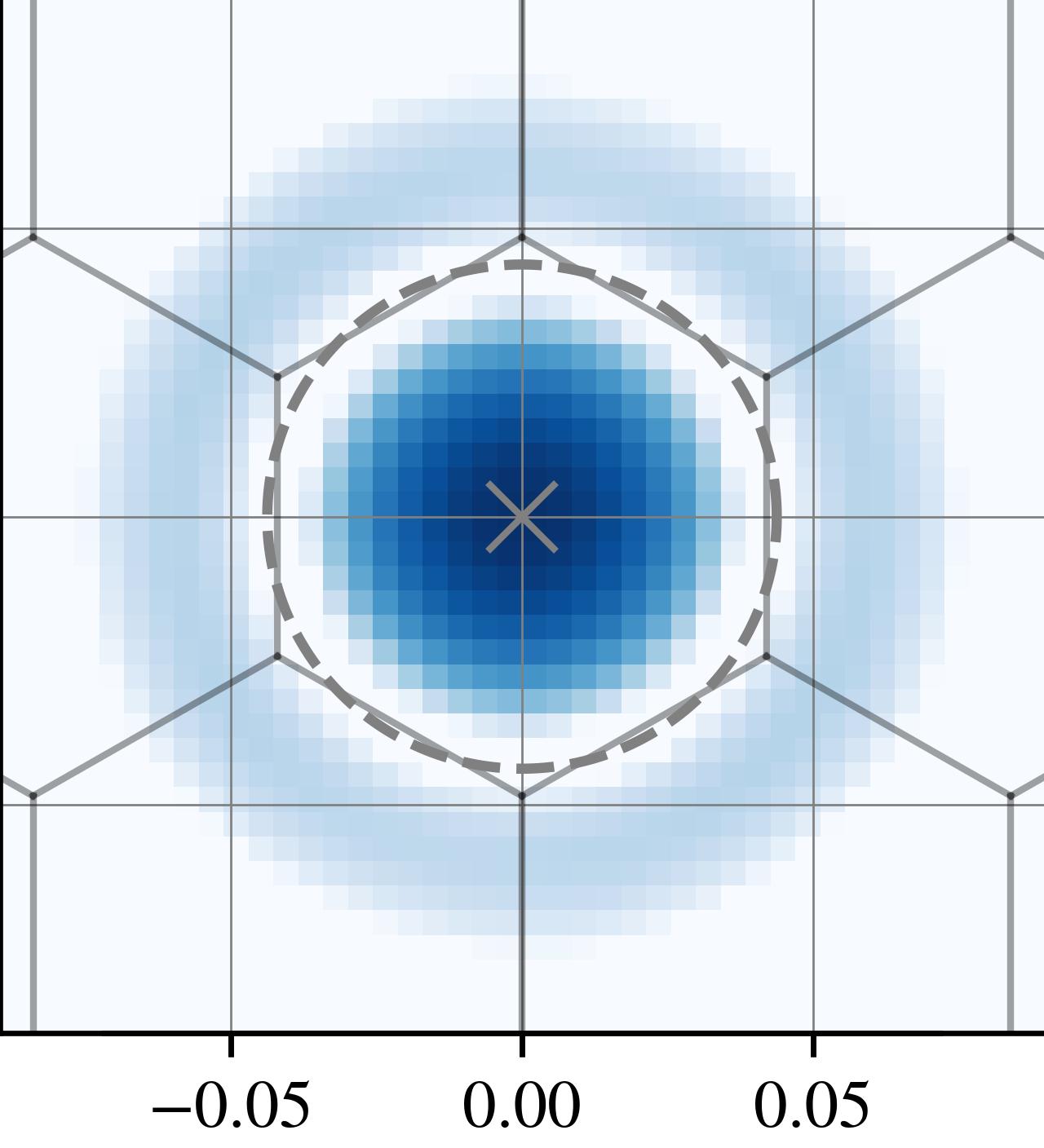}
            \label{FigLargeGuideStars}
        }
        \\
        \centering
        \includegraphics[width=0.3\textwidth]{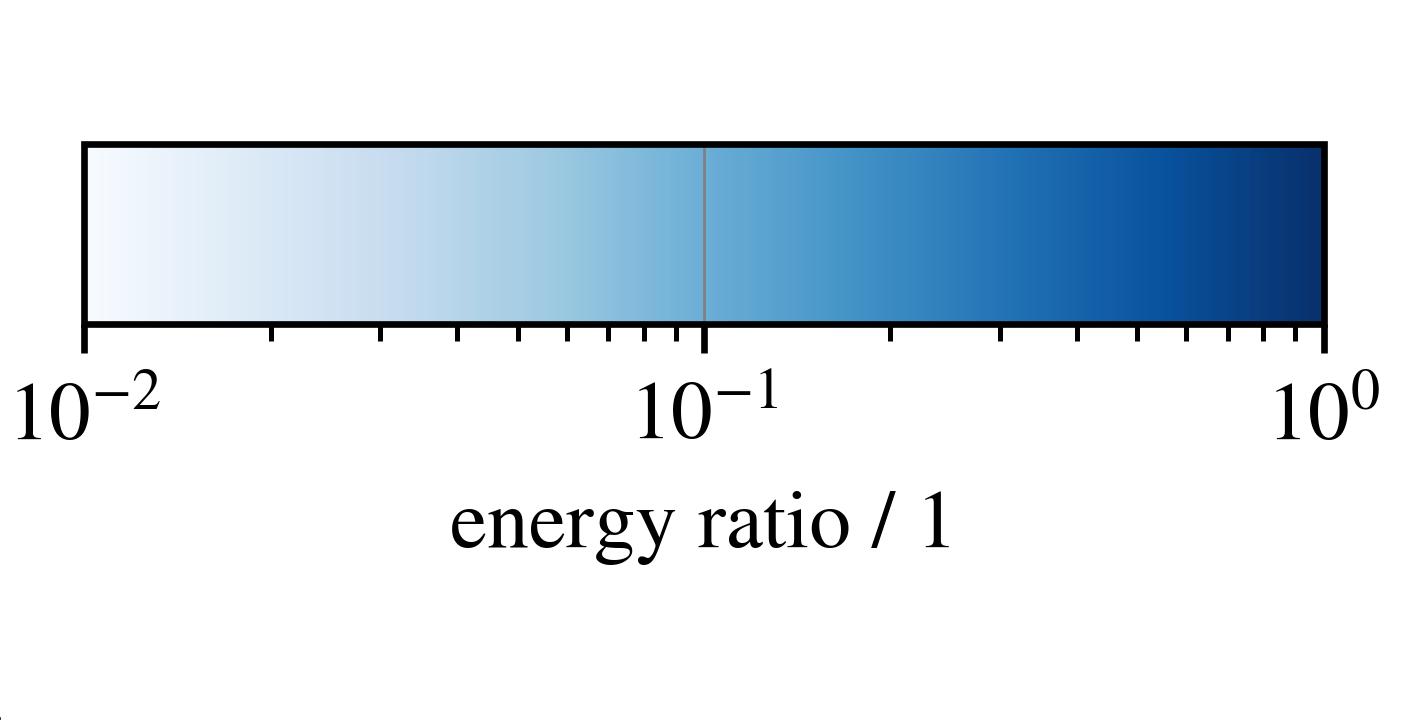}
        \caption{
            Three panels show images of plane waves.
            Each plane wave comes at a different angle off the optical axis.
            Left panels are at the outer edge of the field-of-view, right panels are on the optical axis at the center of the field-of-view.
            Axes are camera screen\,/\,m.
            Hexagons show the feed horn openings.
            Dashed circle shows Airy's disk according to \autoref{EqAiry}.
            Cross in the center of each panel marks the expected position of the
            radio light if the mirror was not causing distortions.
        }
        \label{FigGuideStars}
    \end{figure*}

        The Huygens image formation presented here is able to have the area of its point spread function always above the area of Airy's disk while getting close to Airy's disk in the central field-of-views of the Medium and Large telescopes.
        This can also be seen in the right most panels of \autoref{FigGuideStars}.
        Clearly, the Huygens image formation can reproduce the far field expectations.

        Going off the optical axis one finds that the Huygens image formation also can reproduce geometric optics as the areas of the point spread functions from Huygens and ray tracing approach each other once the off axis aberrations of the mirror take over.
        
        In Crome's case in panel \autoref{FigCromePerformanceSpread}, our ray tracing simulation has its point spread function area shrink towards the edge of the reasonable useful field-of-view as light spills over the edge of the screen.
        The spread size of the Huygens based image formation is still growing here but also suffers from a spill over effect as the `region of interest' in which the image of the point spread function is computed (compare \autoref{FigCromeGuideStars}) can no longer fully contain the ever growing point spread function.
        Because of this, the panel \autoref{FigCromePerformanceSpread} is only meaningful up to $\approx 6.0^{\circ}$ off the optical axis.

        Here we compare the area $A_\text{Spot80\%}$ what is the area of a circle surrounding the brightest spot and containing 80\% of the energy.

        \autoref{TabSpotSize} shows this area in units of the mirror's Airy disk area $A_\text{Airy}$ for radio light, compare \autoref{EqSpotSize}.
        \begin{eqnarray}
            \frac{A_\text{Spot80\%}}{(A_\text{Airy})} &=& \text{PSF}_a \frac{\theta}{(1^\circ)} + \text{PSF}_b
            \label{EqSpotSize}
        \end{eqnarray}
        \begin{table}
            \begin{tabular}{lrrr}
                 & $\mathbf{C}$ & $\mathbf{M}$ & $\mathbf{L}$\\
                \toprule
                $\text{PSF}_a$ & $1.47\pm0.06$ & $0.41\pm0.02$ & $0.15\pm0.02$ \\
                $\text{PSF}_b$ & $9.4\pm0.4$ & $1.21\pm0.04$ & $1.30\pm0.05$ \\
            \end{tabular}
            \caption{
                Size of the point spread function according to \autoref{EqSpotSize}.
                The large focal-ratios of Medium and Large allow these telescopes to almost reach the limit of Airy's disk ($\text{PSF}_b \approx 1.0$).
            }
            \label{TabSpotSize}
        \end{table}
        The spot size areas are not estimated using the image created by the telescope's feed horns but using a finer rectangular array of scatter centers in the camera's screen as it is also used to create the images seen in \autoref{FigGuideStars}.

\begin{figure}{}
            \subfloat[$\mathbf{C}$rome telescope]{
                \centering
                \includegraphics[width=1\columnwidth]{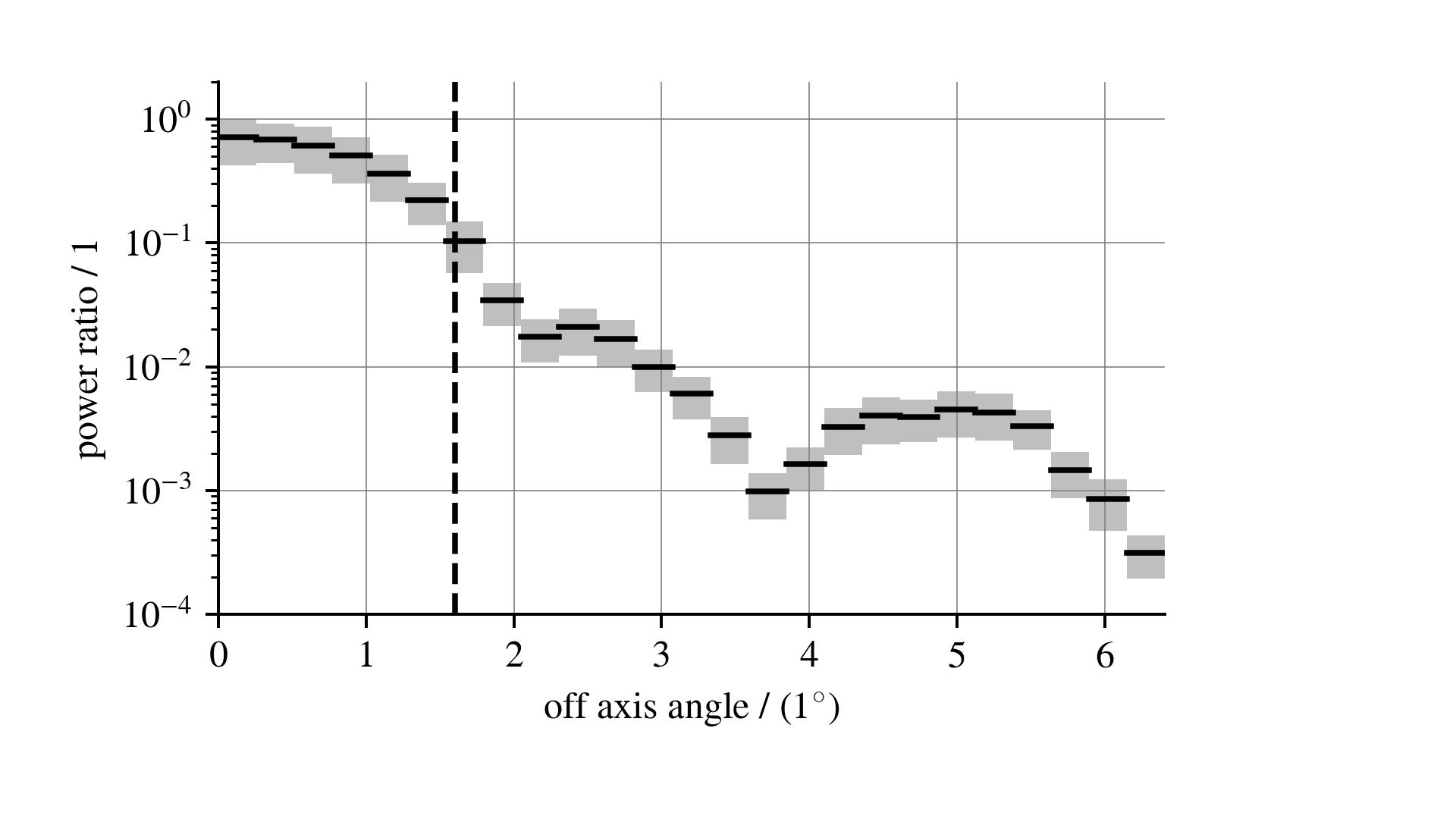}
                \label{FigCromeFeedHornScan}
            }
            \\
            \subfloat[$\mathbf{M}$edium telescope]{
                \centering
                \includegraphics[width=1\columnwidth]{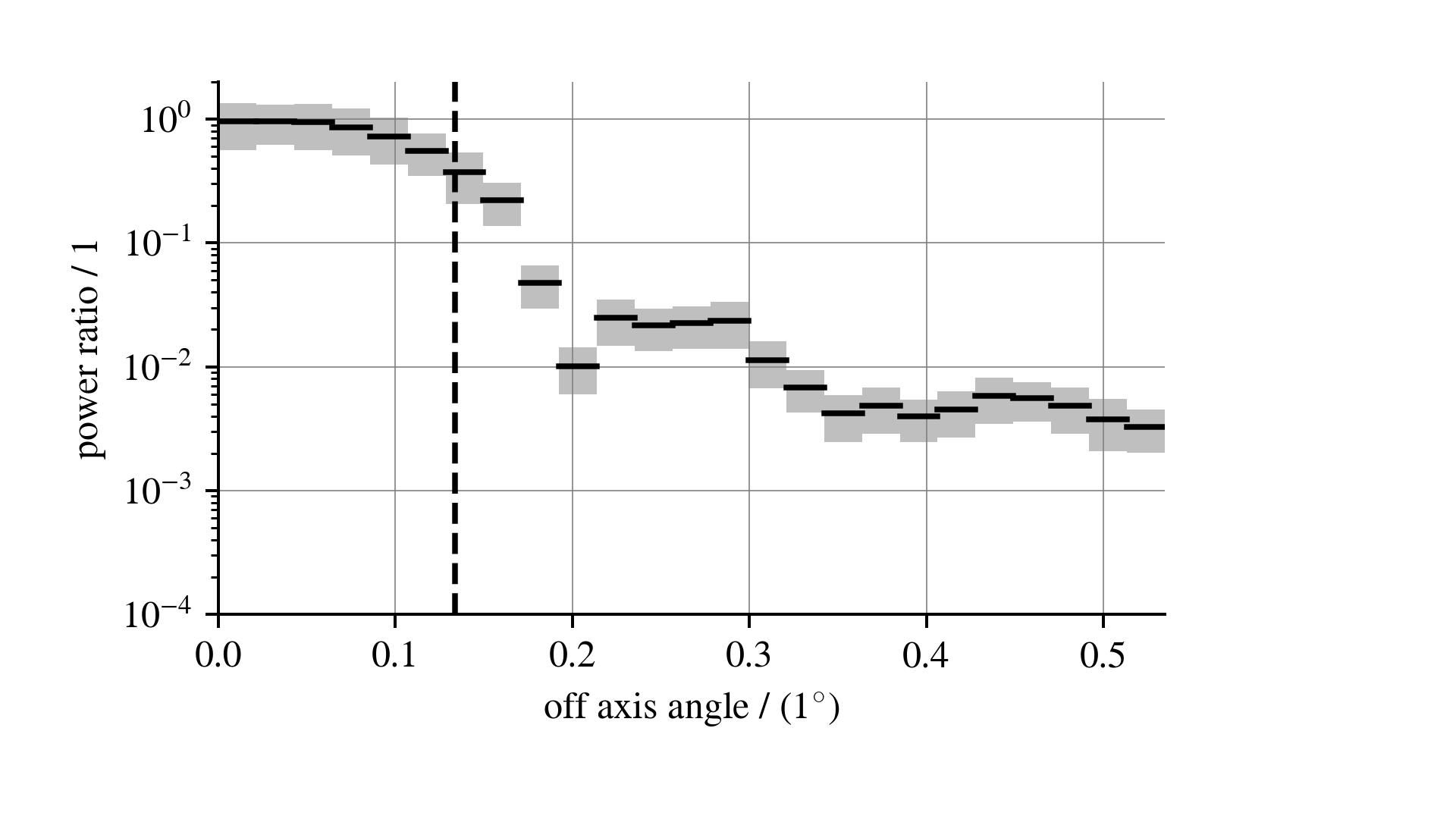}
                \label{FigMediumFeedHornScan}
            }
            \\
            \subfloat[$\mathbf{L}$arge telescope]{
                \centering
                \includegraphics[width=1\columnwidth]{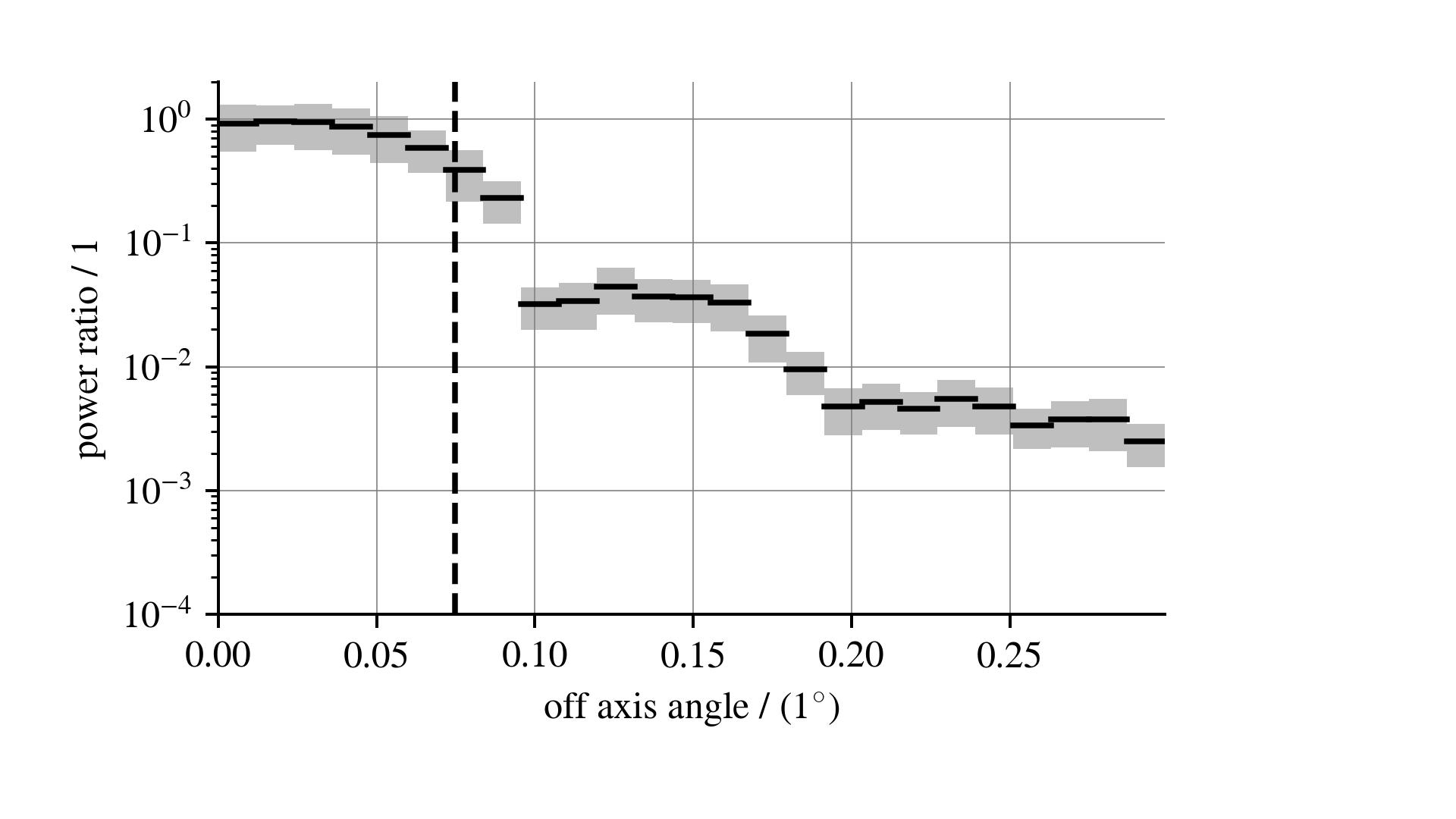}
                \label{FigLargeFeedHornScan}
            }
            \caption{
                Scanning the central feed horn in the camera up to an angle four times the feed horn's half angle.
                Black dashed vertical line is the feed horn's half angle.
                The curve shows the fraction of a plane waves energy which ends up in the central feed horn in the camera's screen.
                The plane waves have random azimuth orientations but same frequencies.
                This scan corresponds to a radial representation of the point spread functions shown in the most right panels of \autoref{FigGuideStars}.
            }
            \label{FigFeedHornScan}
        \end{figure}

        On radio telescopes, one often discusses the spread of a mirror using so called `antenna lobes' and `field patterns' which are off axis scans of the central feed horn over a bright and distant source.
        \autoref{FigFeedHornScan} shows such field patterns where one can identify individual antenna lobes.

    \subsection{Distortion}
        \label{SecDistortion}
        Both the Huygens image formation and the ray tracing agree on a `Pincushion' distortion where the actual spot of the radio light is further out on the screen as one would expect it from the thin lens \autoref{EqThinLens}. 
        The figure panels \autoref{FigCromePerformanceDistortion}, \autoref{FigMediumPerformanceDistortion}, and \autoref{FigLargePerformanceDistortion} show the distortion on the three telescopes $\mathbf{C}$, $\mathbf{M}$, and $\mathbf{L}$.
        Note the panels $y$-axes narrow range but large offset which was chosen to highlight a small but consistent disagreement between the Huygens image formation and ray tracing.
        One does not expect the Huygens image formation to consistently underestimate the ray tracing predictions by a small but significant factor of a few $1\%$.

        But on the absolute scale, both the Huygens and the ray tracing predictions are close to each other and show a similar trend along the angle off the optical axis.
        \autoref{TabDistortion} shows the fitted distortion 
        \begin{eqnarray}
            \label{EqDistortion}
            \text{DIS} &=& \frac{\theta_\text{actual}}{\theta}
        \end{eqnarray}
        across the field-of-view, where $\theta$ is the angle off the optical axis.
        \begin{table}
            \begin{tabular}{lrrr}
                 & $\mathbf{C}$ & $\mathbf{M}$ & $\mathbf{L}$\\
                \toprule
                $\text{DIS}$ & $1.281$ & $1.0394$ & $1.01896$ \\
                $\Delta\text{DIS}$ & $\pm0.002$ & $\pm0.0001$ & $\pm0.00006$ \\
            \end{tabular}
            \caption{
                Distortions caused by Huygens image formation according to \autoref{EqDistortion} and fitted on black curves in figure panels \autoref{FigCromePerformanceDistortion}, \autoref{FigMediumPerformanceDistortion}, and \autoref{FigLargePerformanceDistortion}.
            }
            \label{TabDistortion}
        \end{table}
        While $\mathbf{M}$edium and $\mathbf{L}$arge have Pincushion distortions in the $4\%$ and $2\%$ regime, $\mathbf{C}$rome's distortion is about $28\%$.
    \subsection{Observing two plane waves simultaneously}
        \label{SecObservingMultiplePlaneWaves}
\begin{figure}
            \centering
            \subfloat[image]{
                \includegraphics[width=1\columnwidth]{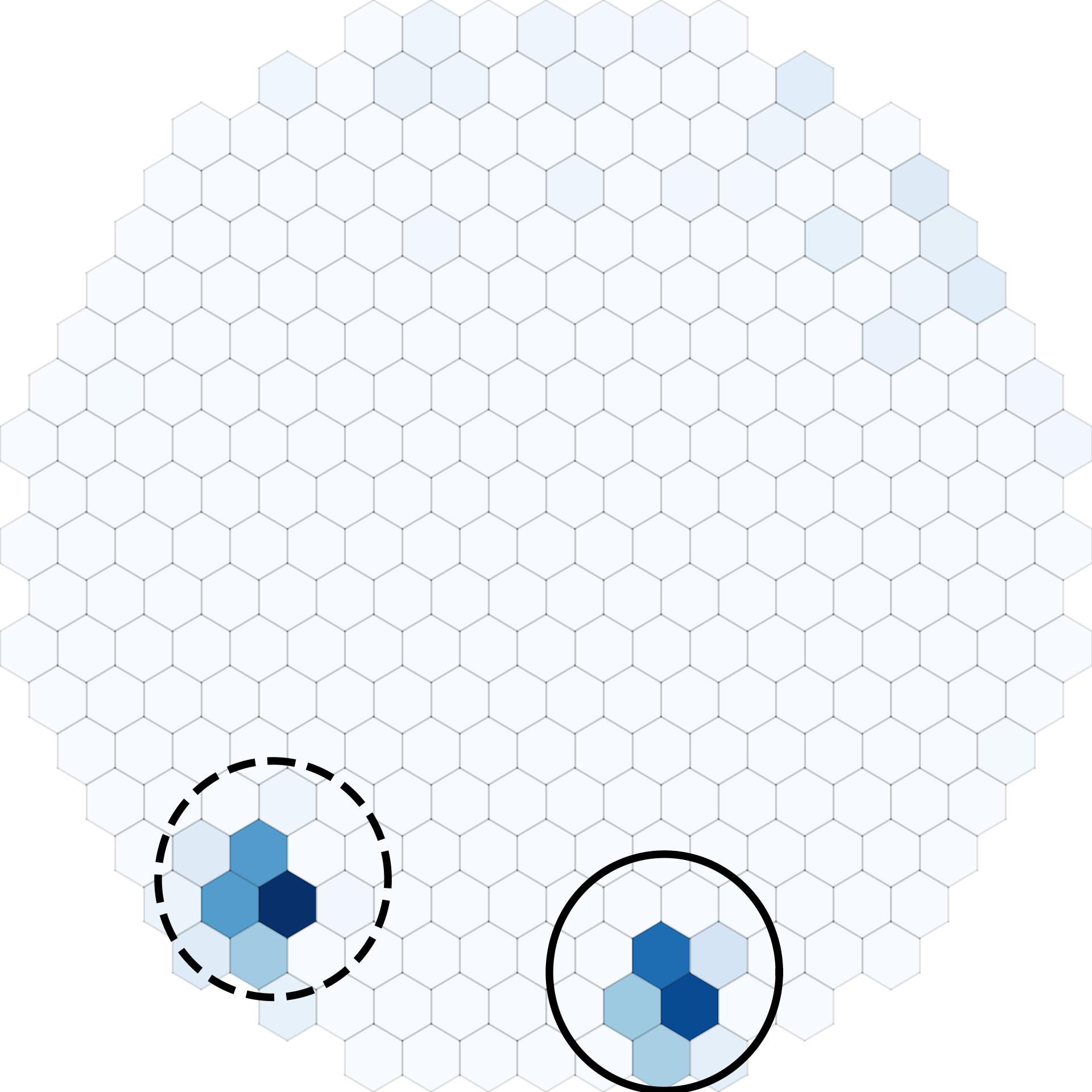}
                \label{FigMediumMulitExampleImage}
            }
            \\
            \includegraphics[width=1\columnwidth]{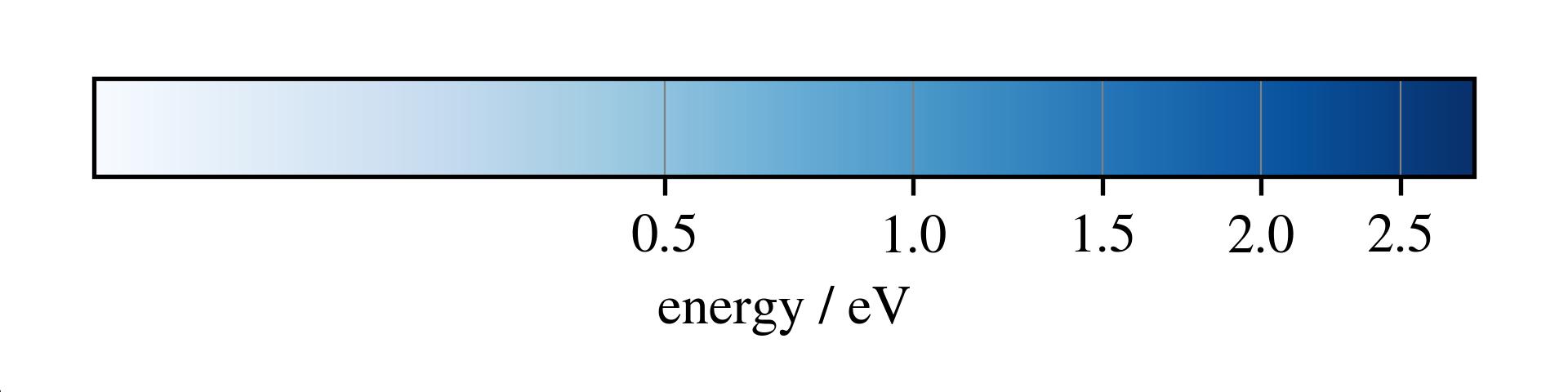}
            \\
            \includegraphics[width=1\columnwidth]{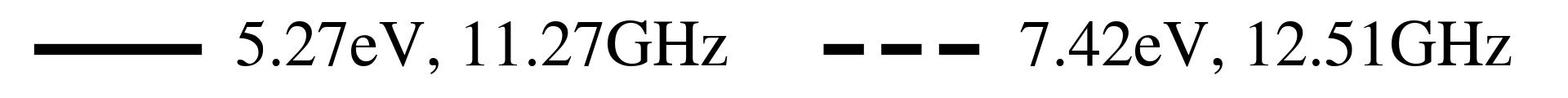}
            \\
            \subfloat[frequency]{
                \centering
                \includegraphics[width=1\columnwidth]{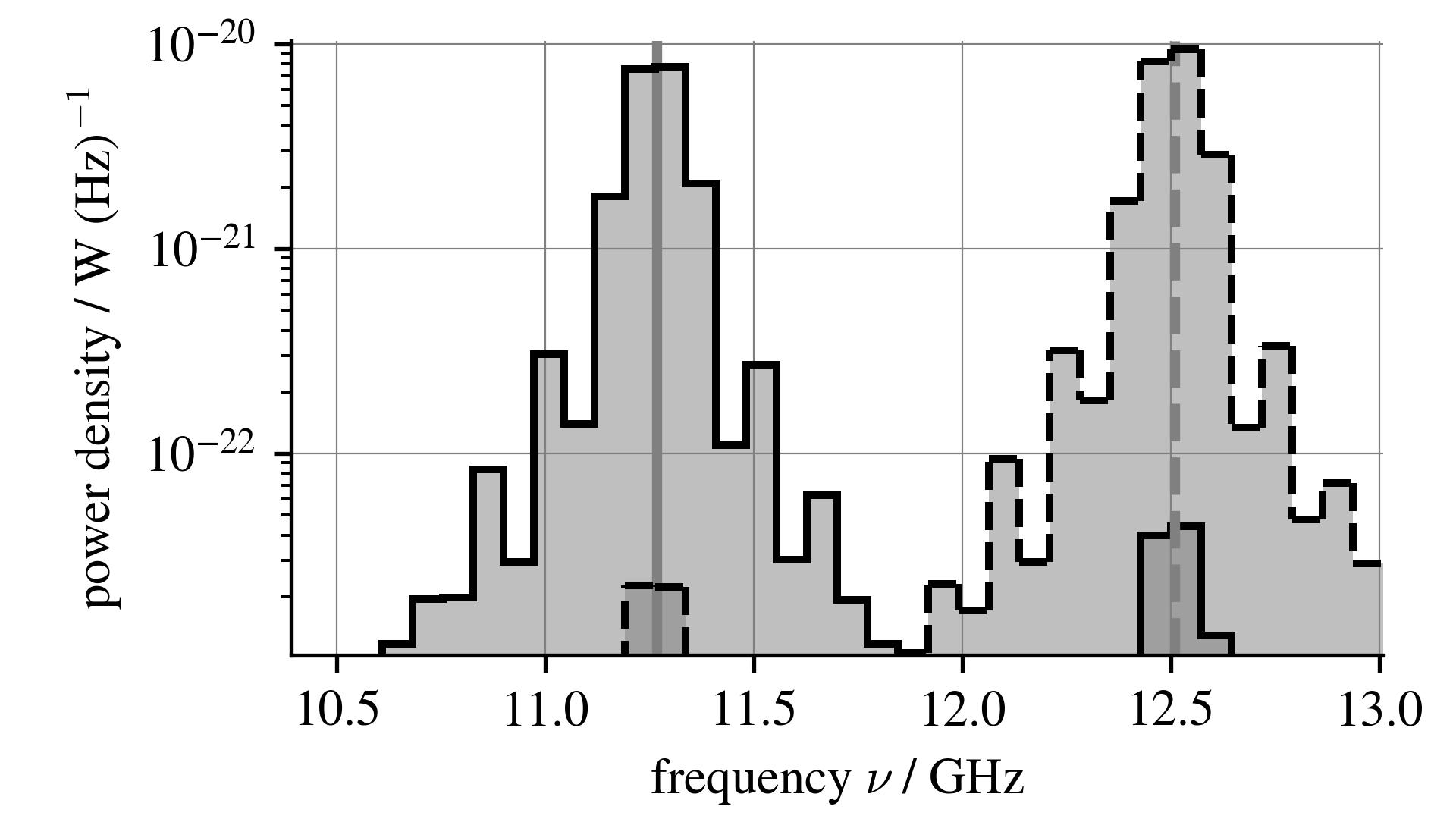}
                \label{FigMediumMulitExampleFrequency}
            }
            \caption{
                Observing two plane waves with the $\mathbf{M}$edium telescope.
                The two frequency spectra contain only the feed horns in close proximity to the corresponding spots in the image and are shown with a solid and a dashed line respectively.
                The full and dashed ring in the image show the expected position of the images for the two plane waves.
                The legend with the full and dashed line shows the true energy and frequency of the incoming plane wave package.
            }
            \label{FigMediumMulitExample}
        \end{figure}

\begin{figure}
            \centering
            \subfloat[image]{
                \includegraphics[width=1\columnwidth]{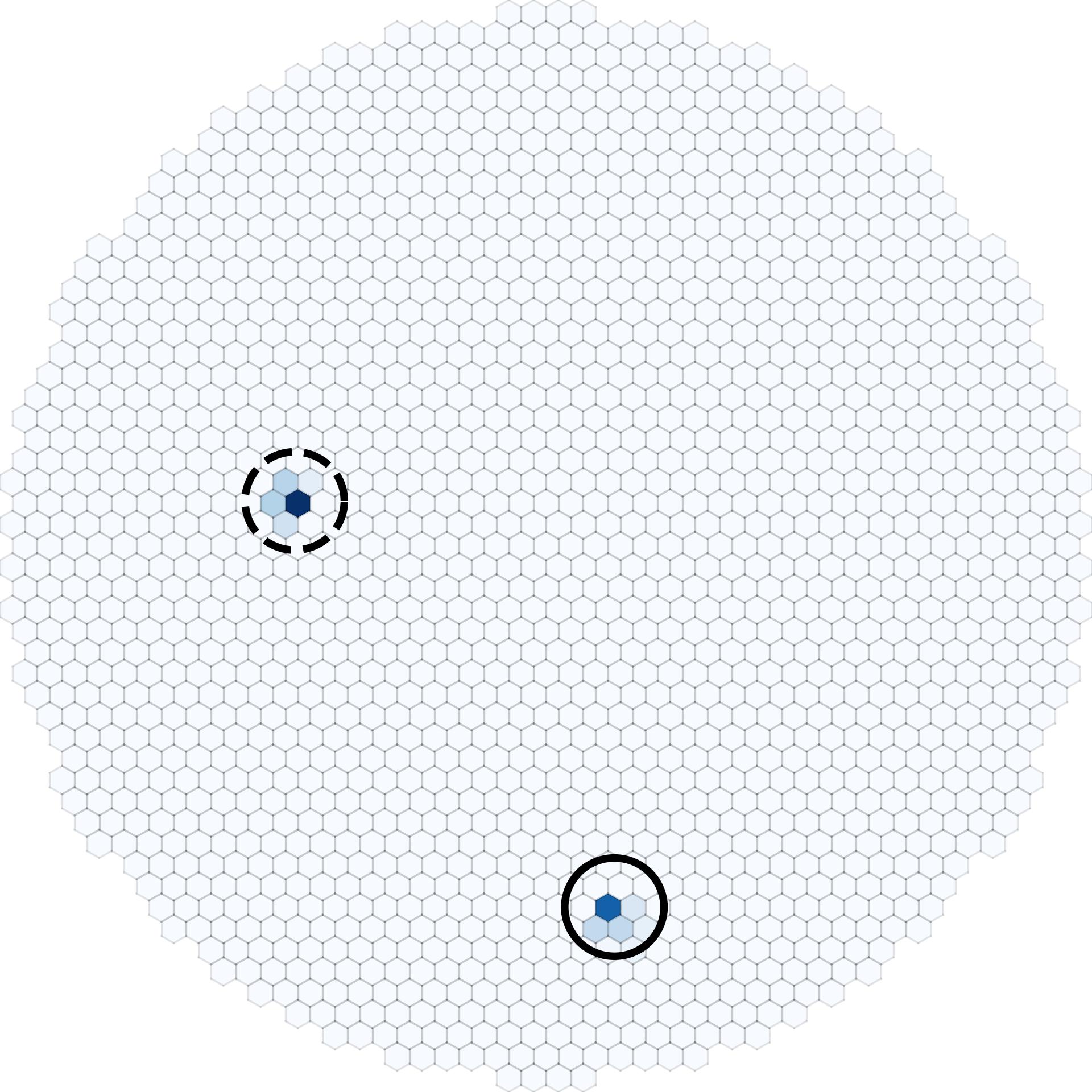}
                \label{FigLargeMulitExampleImage}
            }
            \\
            \includegraphics[width=1\columnwidth]{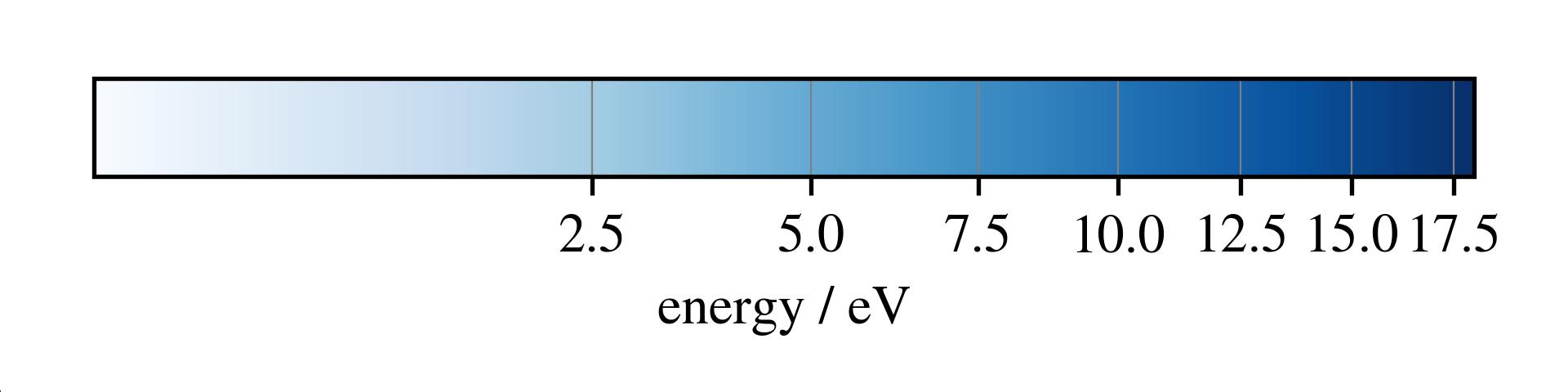}
            \\
            \includegraphics[width=1\columnwidth]{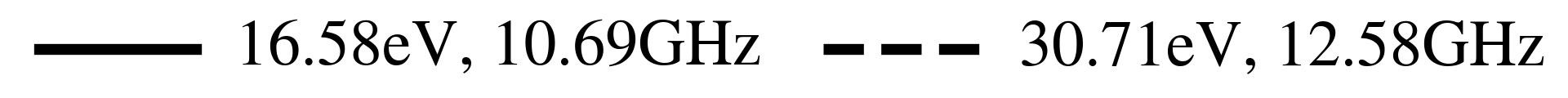}
            \\
            \subfloat[frequency]{
                \centering
                \includegraphics[width=1\columnwidth]{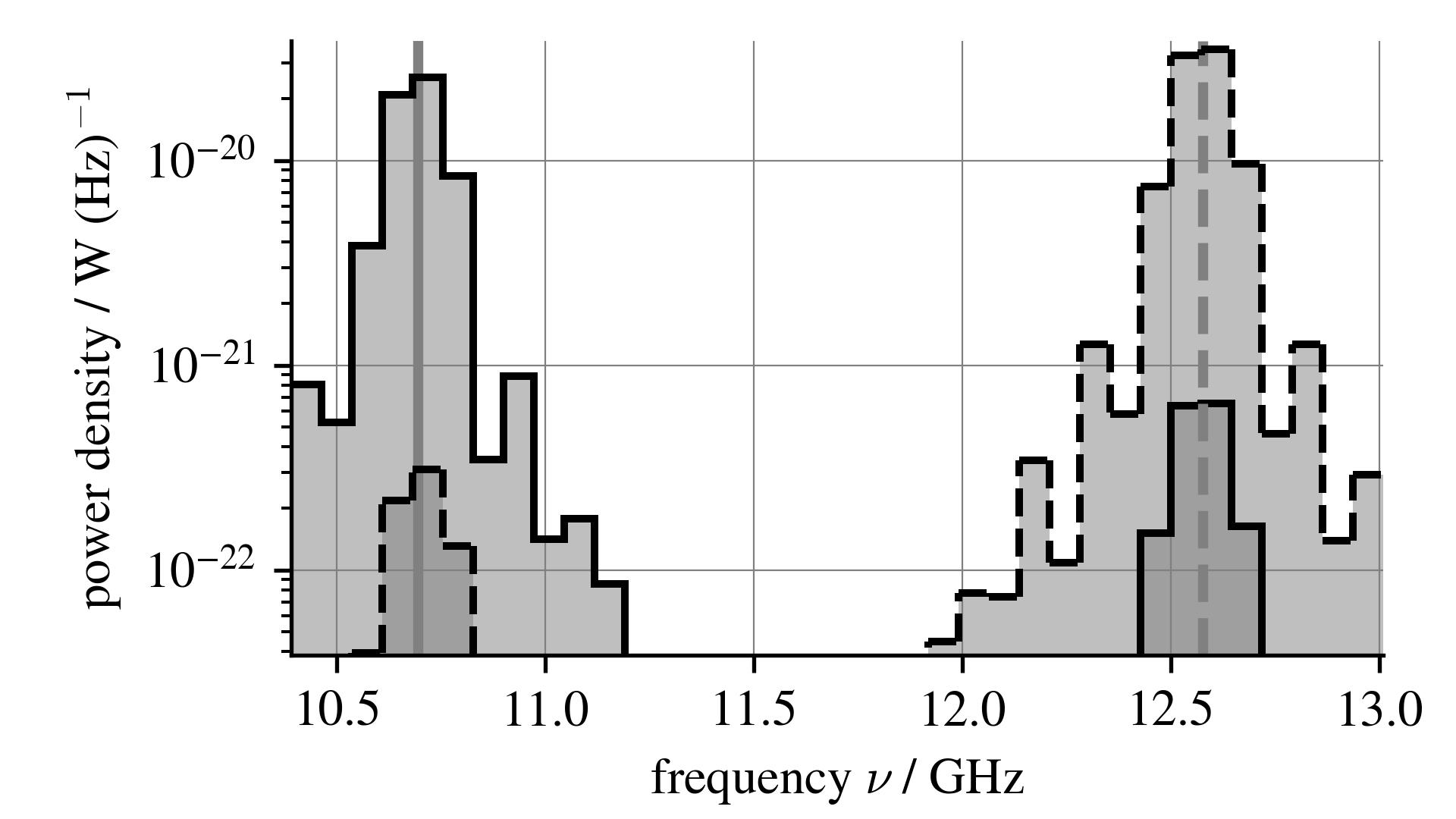}
                \label{FigLargeMulitExampleFrequency}
            }
            \caption{
                Observing two plane waves with the $\mathbf{L}$arge telescope.
                See \autoref{FigMediumMulitExample} for explanation.
            }
            \label{FigLargeMulitExample}
        \end{figure}

        An important test for the Huygens based image formation is the observation of multiple plane waves where the electric fields from the individual sources overlap in time when they reach the mirror.
        The plane waves come from different directions, carry different energies, and have different frequencies and polarization angles.
        \autoref{FigMediumMulitExample} and \autoref{FigLargeMulitExample} show both the $\mathbf{M}$edium and $\mathbf{L}$arge telescope observing each two simultaneously arriving plane wave packages.
        Figure panels \autoref{FigMediumMulitExampleImage} and \autoref{FigLargeMulitExampleImage} shows the energy carried by the electric fields present at the feed horn entrances.
        The panels \autoref{FigMediumMulitExampleFrequency} and \autoref{FigLargeMulitExampleFrequency} show the two power density spectra of the electric fields for selected feed horns in the two bright spots proximity.
        The electric fields at the feed horn entrances are calculated according to \autoref{SecSimulatingTheFeedHorn} and thus have been processed using the approximation from \autoref{EqFeedHornCombine}.
        
        The two examples on the Medium and Large telescope produce the expected results and indicate that a superposition of electric fields at the mirror level is possible.
        This is crucial as it indicates that this Huygens image formation can be used to process the electric field output of air shower simulations such as \CorsikaCoreas{} or \Aires{} where the emissions of many particle interactions are superimposed.
    \subsection{Observing out-of-focus}
        \label{SecObservingOutOfFocus}
\begin{figure}
        \centering
        \includegraphics[width=0.48\columnwidth]{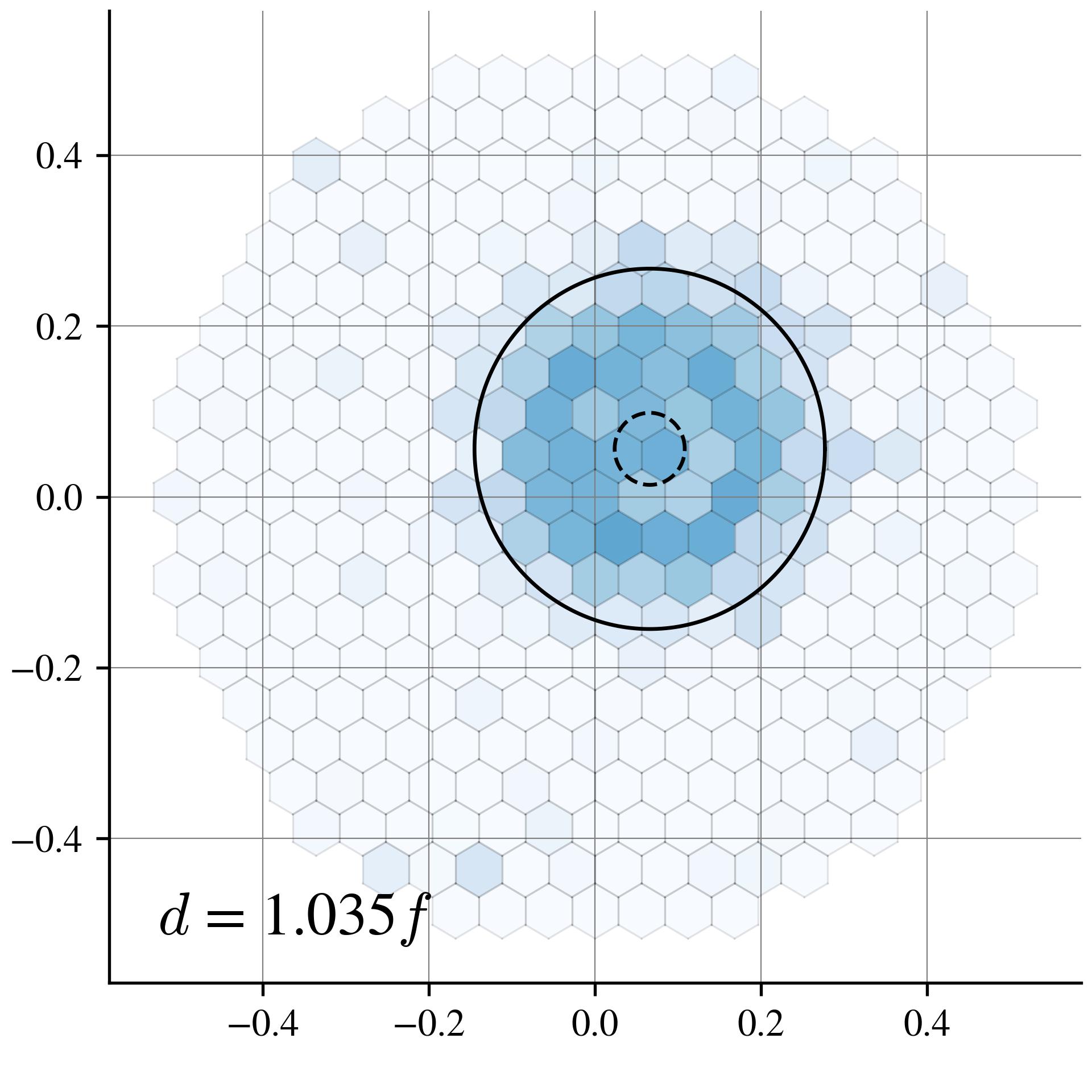}
        \includegraphics[width=0.48\columnwidth]{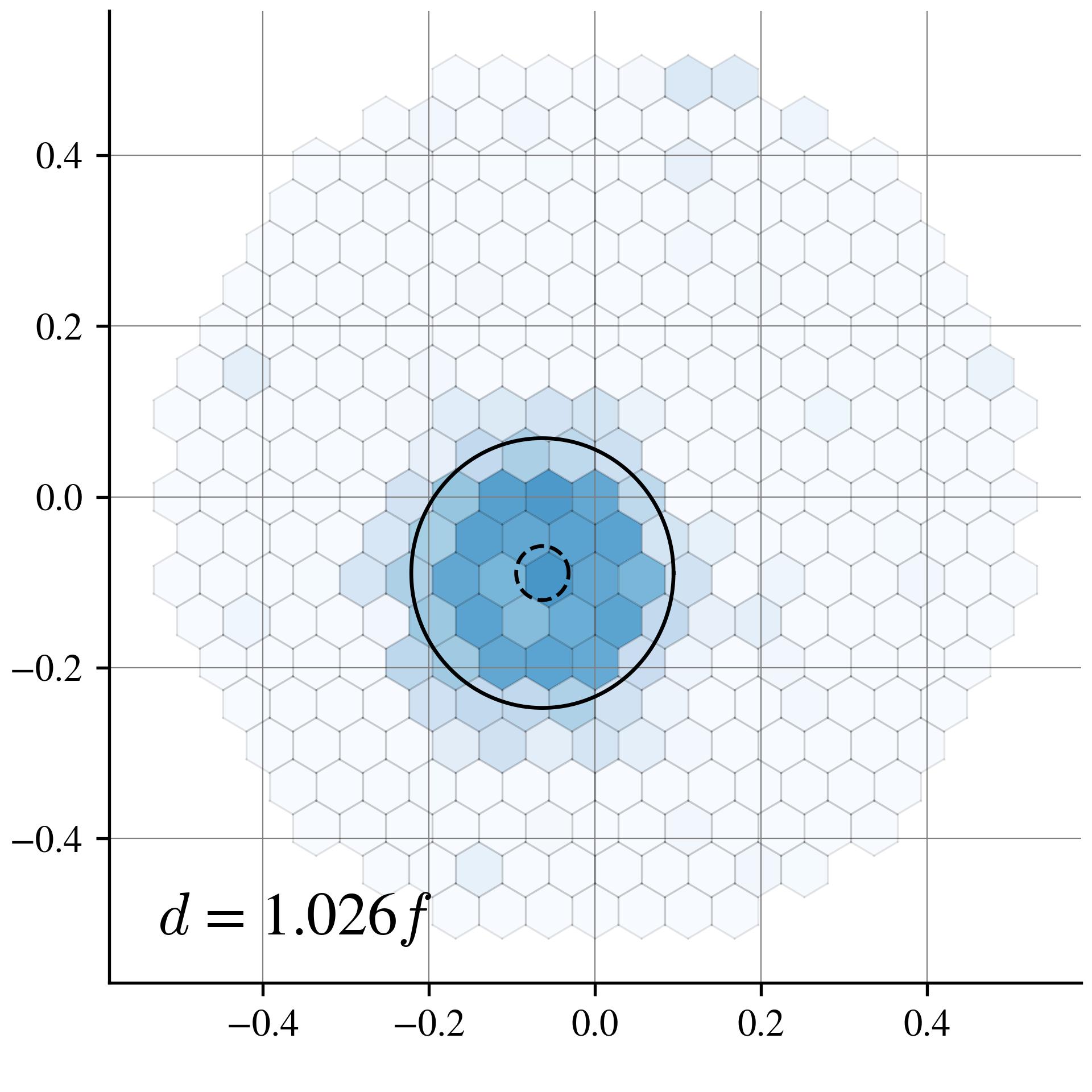}
        \includegraphics[width=0.48\columnwidth]{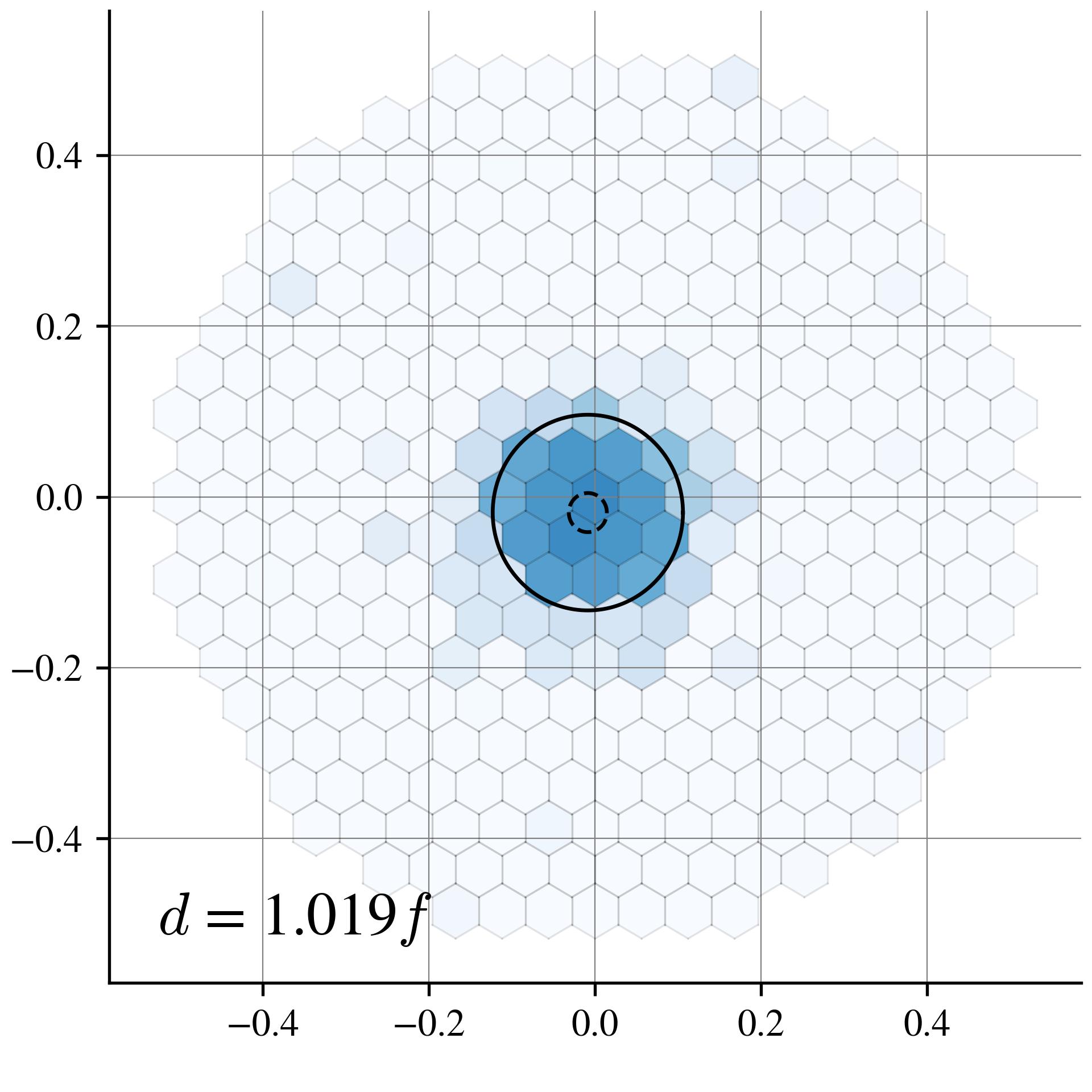}
        \includegraphics[width=0.48\columnwidth]{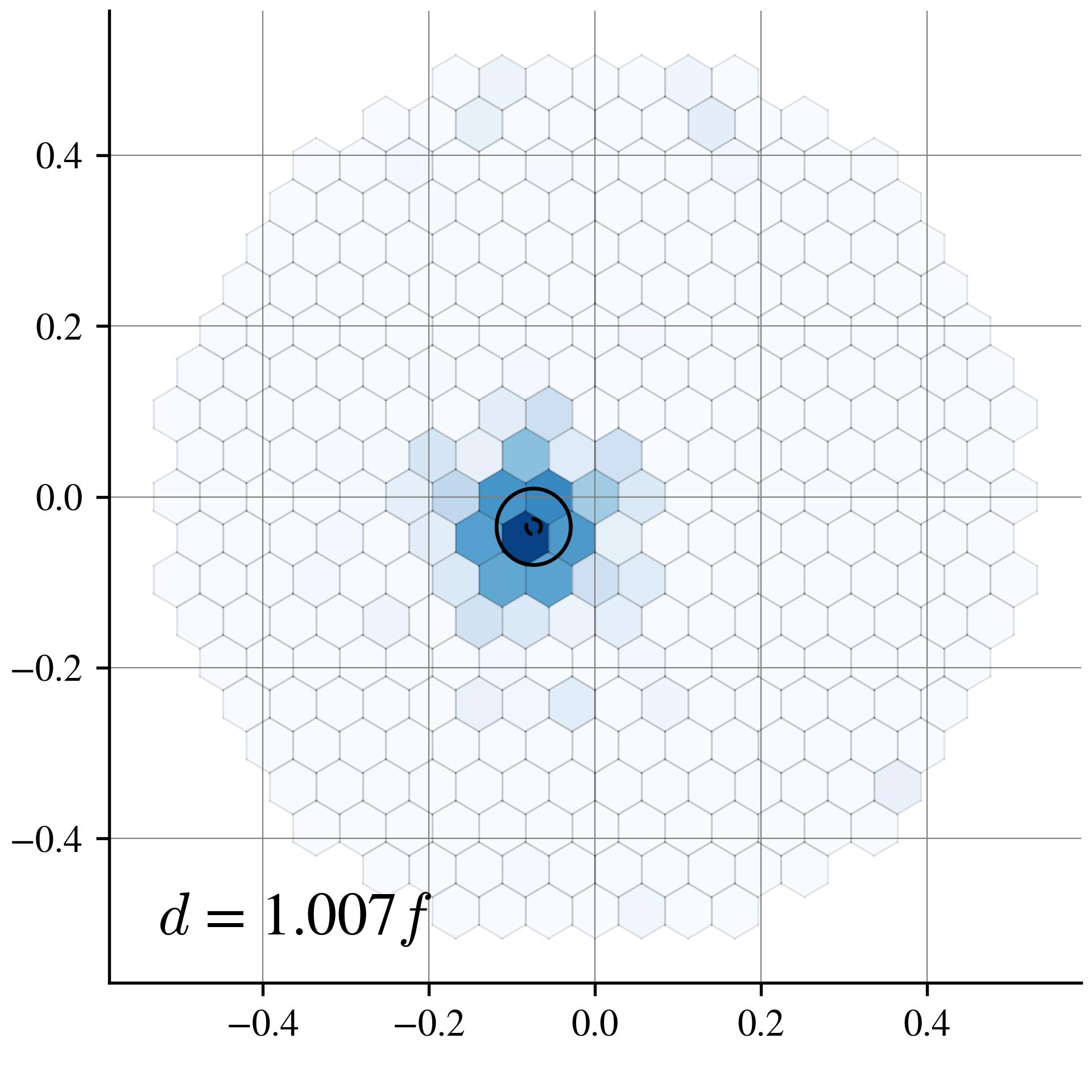}
        \includegraphics[width=0.8\columnwidth]{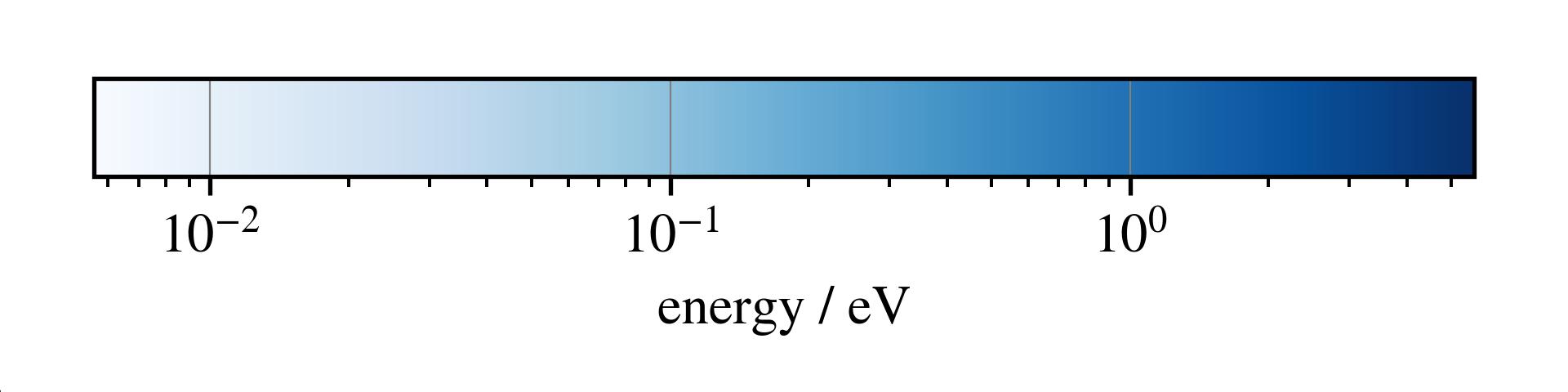}
        \caption{
            Observing a plane wave with the $\mathbf{M}$edium telescope while the camera screen is out-of-focus.
            Big black circle marks the expected image of the mirror's outer edge and small dashed black circle marks the expected inner edge.
        }
        \label{FigMediumDefocusExamples}
    \end{figure}

\begin{figure}
        \centering
        \includegraphics[width=0.8\columnwidth]{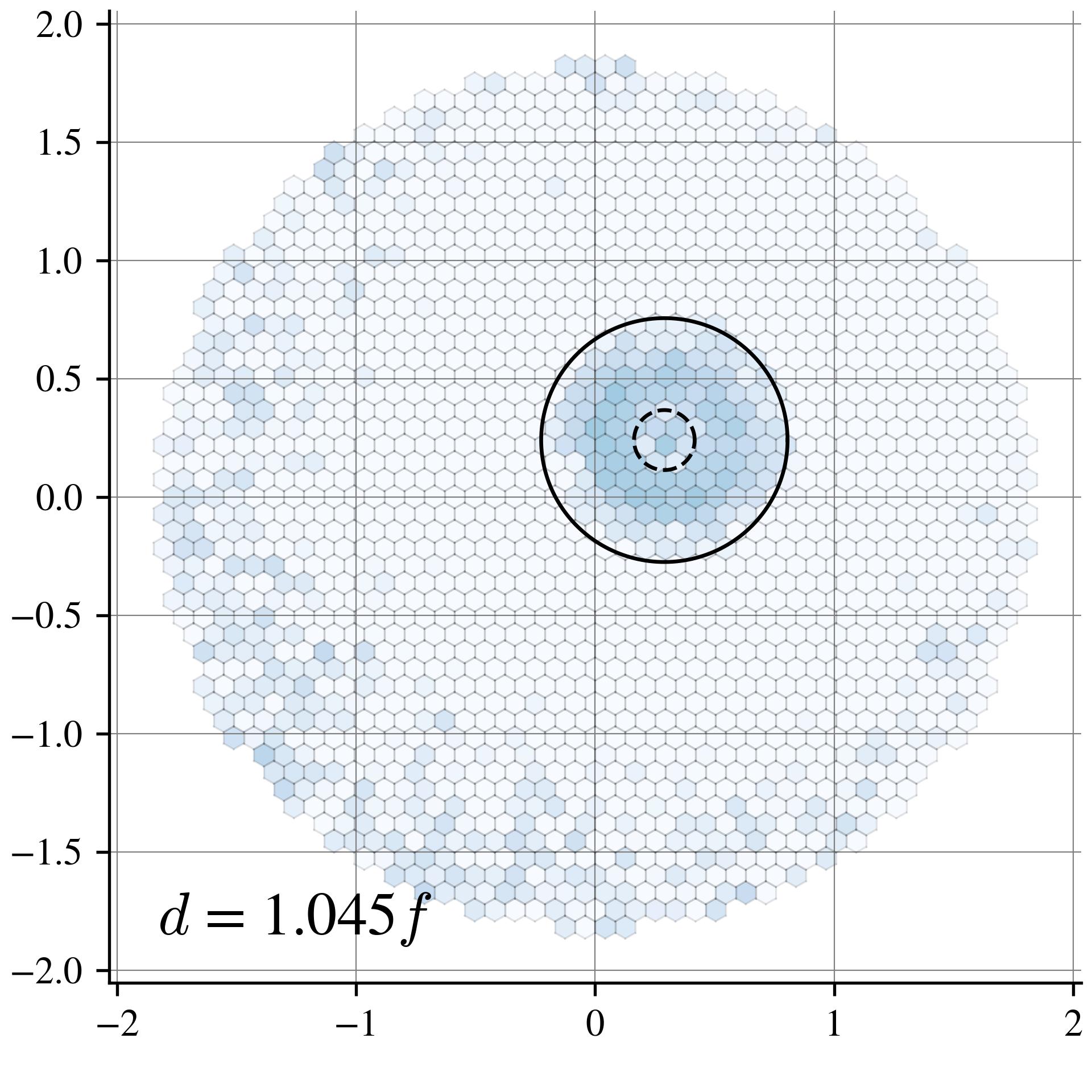}
        \includegraphics[width=0.8\columnwidth]{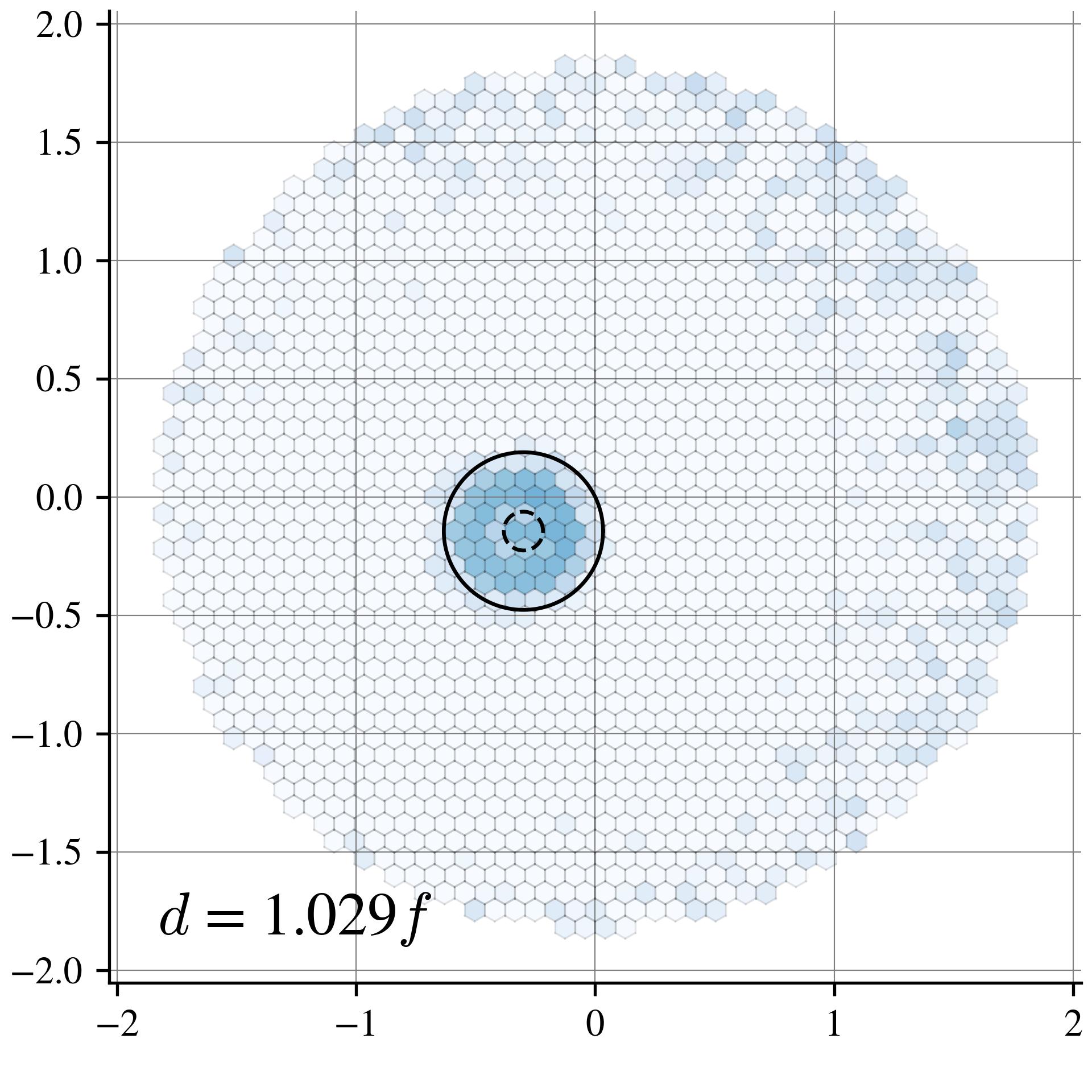}
        \includegraphics[width=0.8\columnwidth]{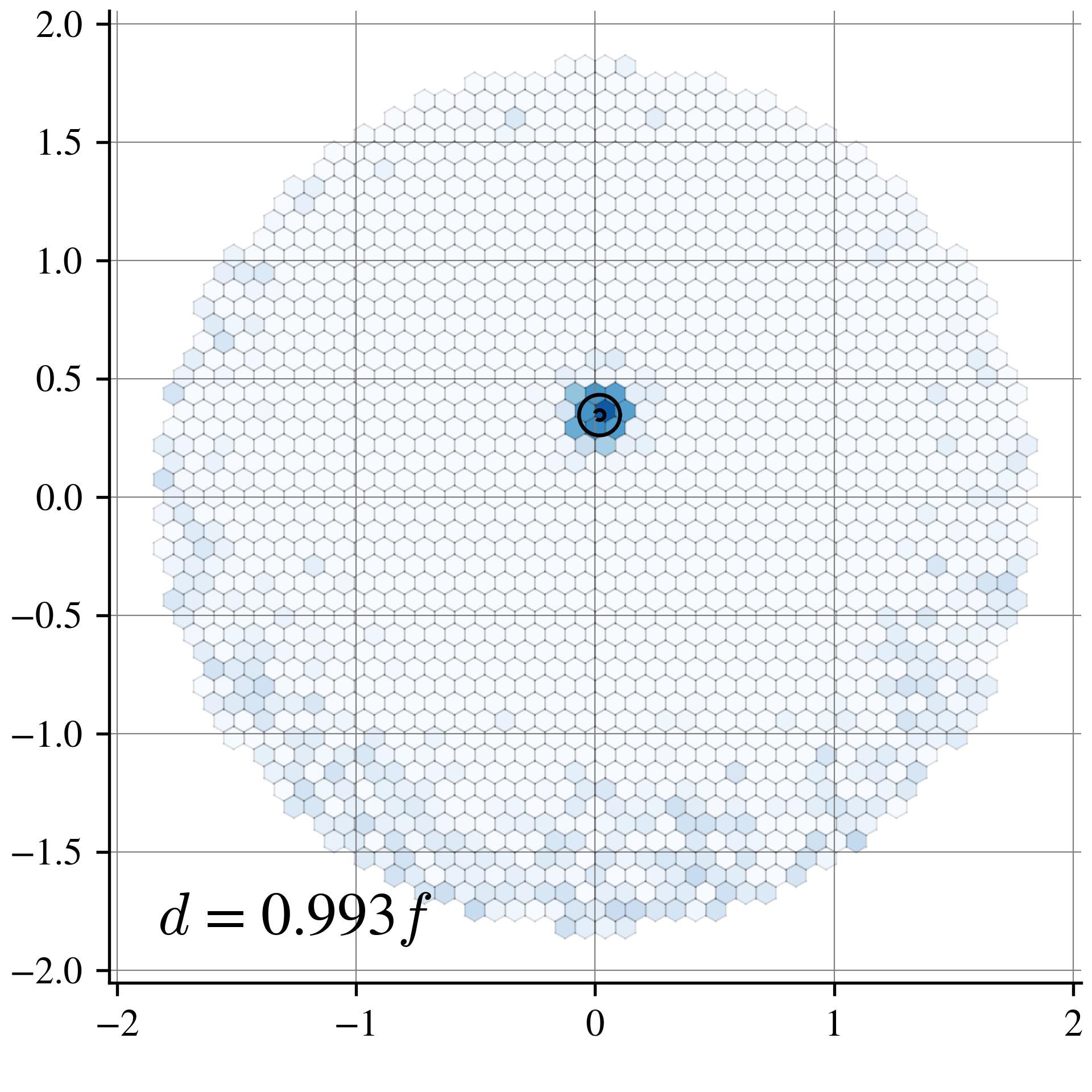}
        \includegraphics[width=0.8\columnwidth]{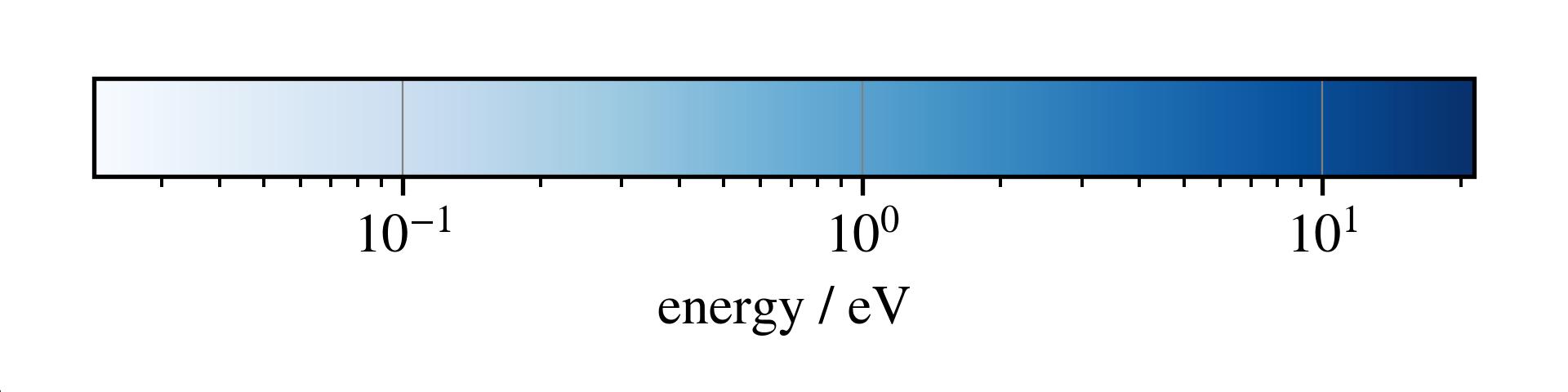}
        \caption{
            Observing a plane wave with the $\mathbf{L}$arge telescope while the camera screen is out-of-focus.
            See \autoref{FigMediumDefocusExamples} for legend.
        }
        \label{FigLargeDefocusExamples}
    \end{figure}

        For a wider test of the image formation one observes a plane wave while the camera screen is out of focus $d \neq{} f$.
        Here one expects to see the aperture function of the mirror showing up in the out-of-focus images.
        Ideally, the annulus of the mirror should show up in this case \cite{ahnen2016bokeh}.
        \autoref{FigMediumDefocusExamples} shows how this phenomena on the $\mathbf{M}$edium telescope and \autoref{FigLargeDefocusExamples} shows it for the $\mathbf{L}$arge telescope.
        On the $\mathbf{L}$arge telescope, one can clearly see the mirror's annulus in the images as one would expect it.

\begin{figure}
        \centering
        \includegraphics[width=1.0\columnwidth]{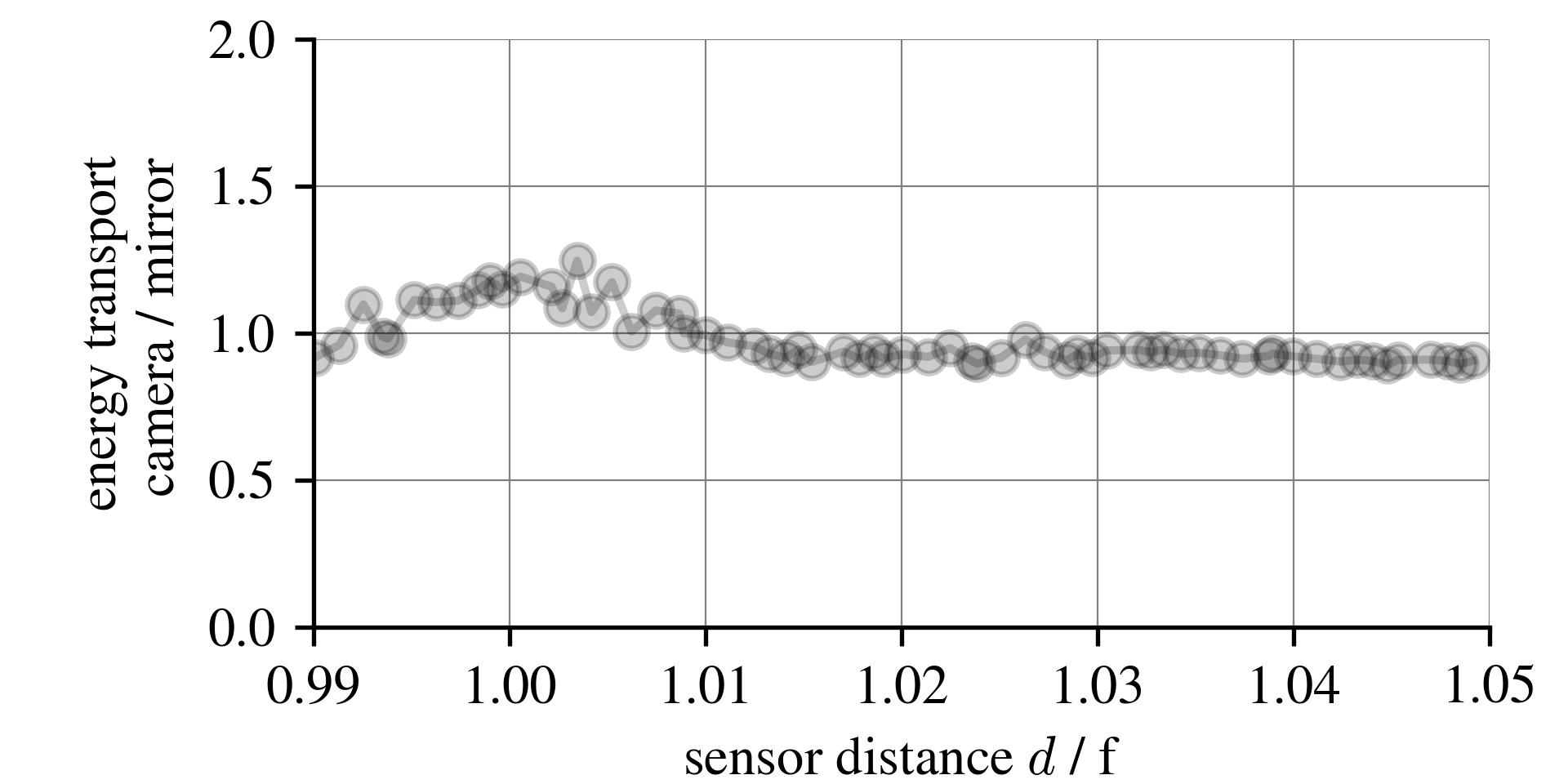}
        \caption{
            Out-of-focus energy conservation on the $\mathbf{M}$edium telescope. Dots corresponds to out-of-focus observation of which some are shown in \autoref{FigMediumDefocusExamples}.
            Here the output energy is taken from all the feed horns and not only from few specific feed horns located in the proximity of the bright spot as it is done for the energy calibration in \autoref{SecEnergyCalibration}.
            Because of this, here the energy conservation can peak above 1.0.
        }
        \label{FigMediumDefocusEnergy}
    \end{figure}

        With the logarithmic color scale in \autoref{FigLargeDefocusExamples}, an artifact pattern becomes visible what is not expected from a perfect imaging mirror.
        A diffuse, ring like structure surrounds the out-of-focus spot.
        In this simulation, the mirror's electric field is always the same and only the sensor distance $d$ is changed.
        Further, one would expect every out-of-focus image to carry the same energy with just a different distribution across the feed horns, but a sensor distance scan in \autoref{FigMediumDefocusEnergy} shows fluctuations.
    \subsection{Artifacts}
        \label{SecArtifacts}
\begin{figure}{}
            \subfloat[$\mathbf{C}$rome telescope]{
                \centering
                \includegraphics[width=0.9\columnwidth]{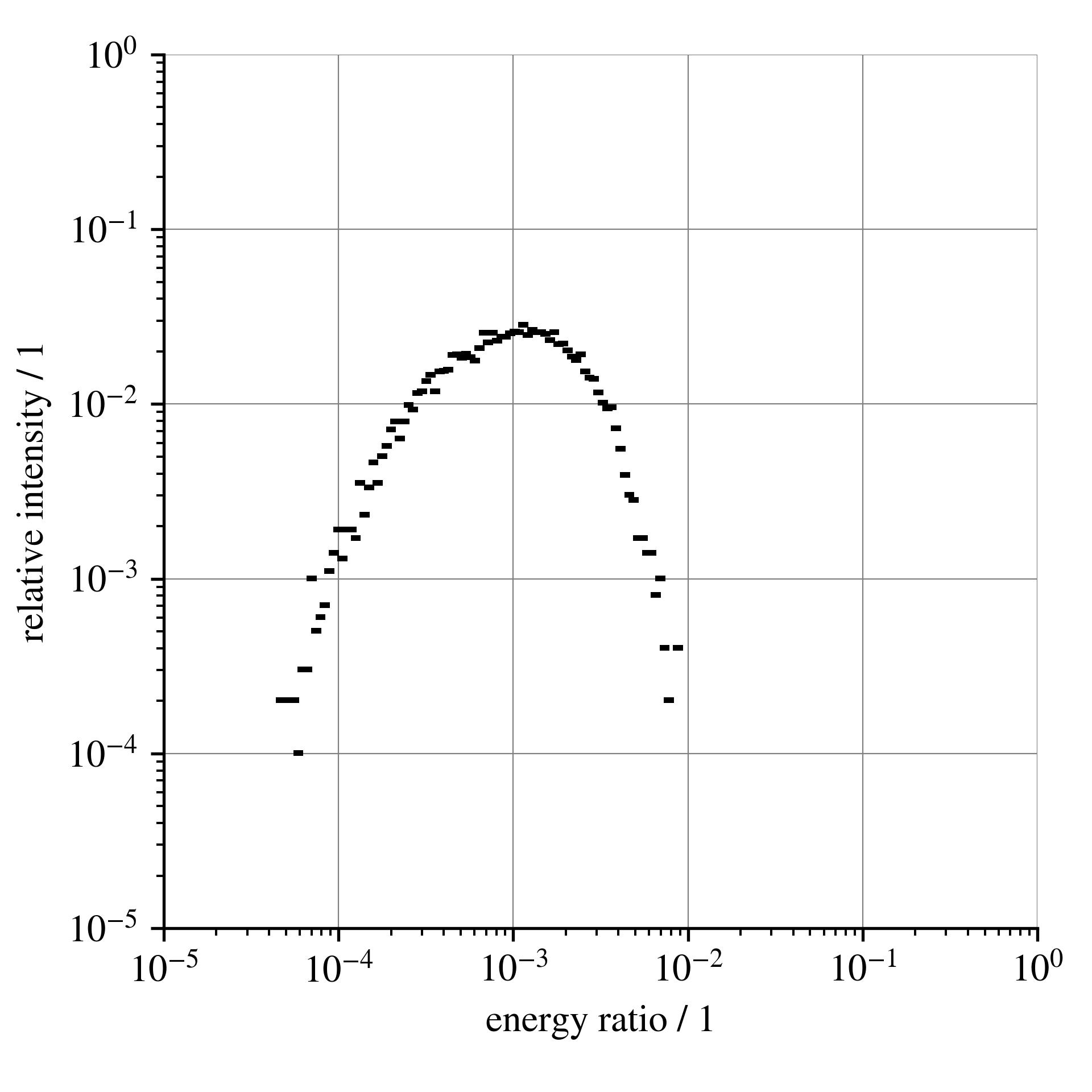}
                \label{FigCromeArtifactPowerRatio}
            }
            \\
            \subfloat[$\mathbf{M}$edium telescope]{
                \centering
                \includegraphics[width=0.9\columnwidth]{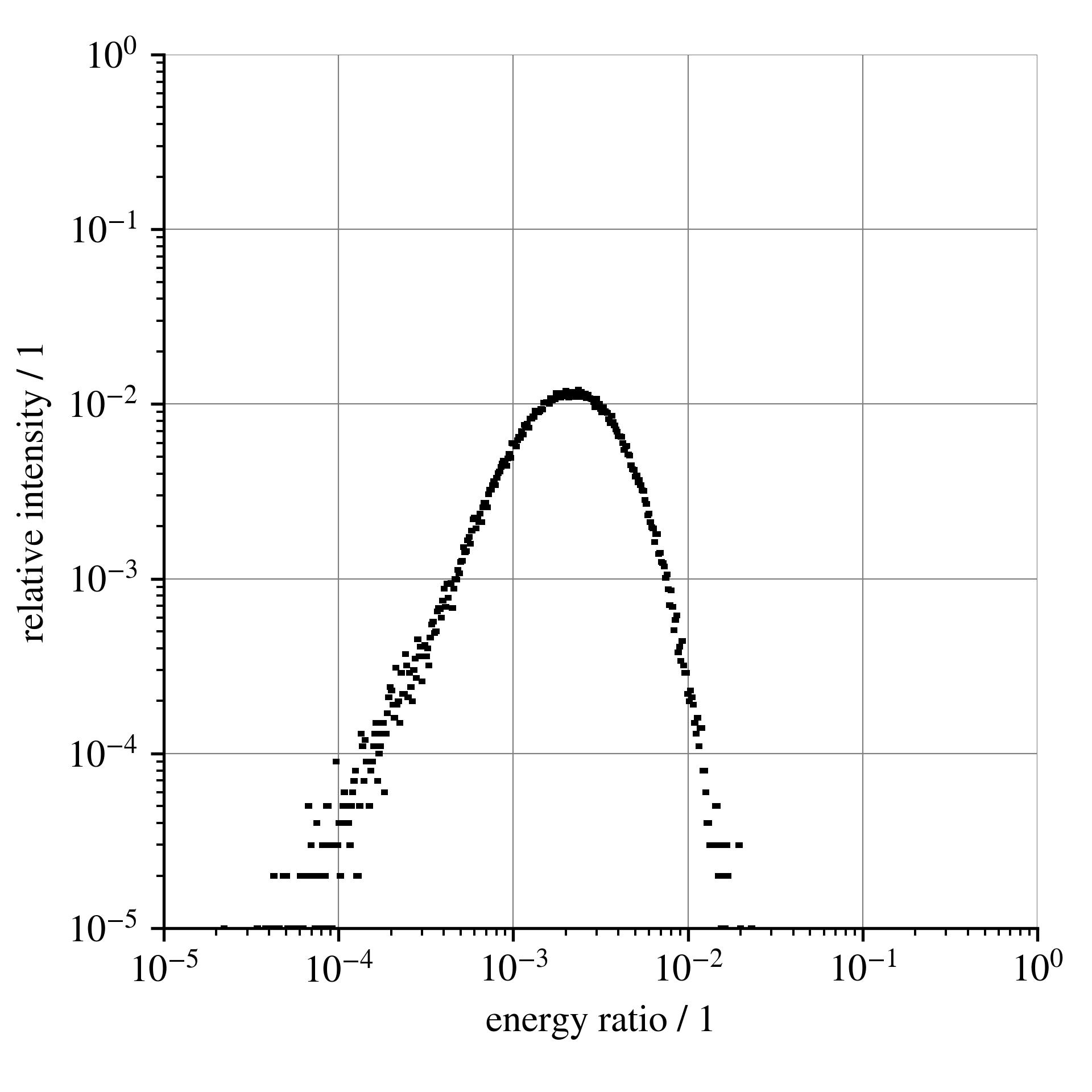}
                \label{FigMediumArtifactPowerRatio}
            }
            \\
            \subfloat[$\mathbf{L}$arge telescope]{
                \centering
                \includegraphics[width=0.9\columnwidth]{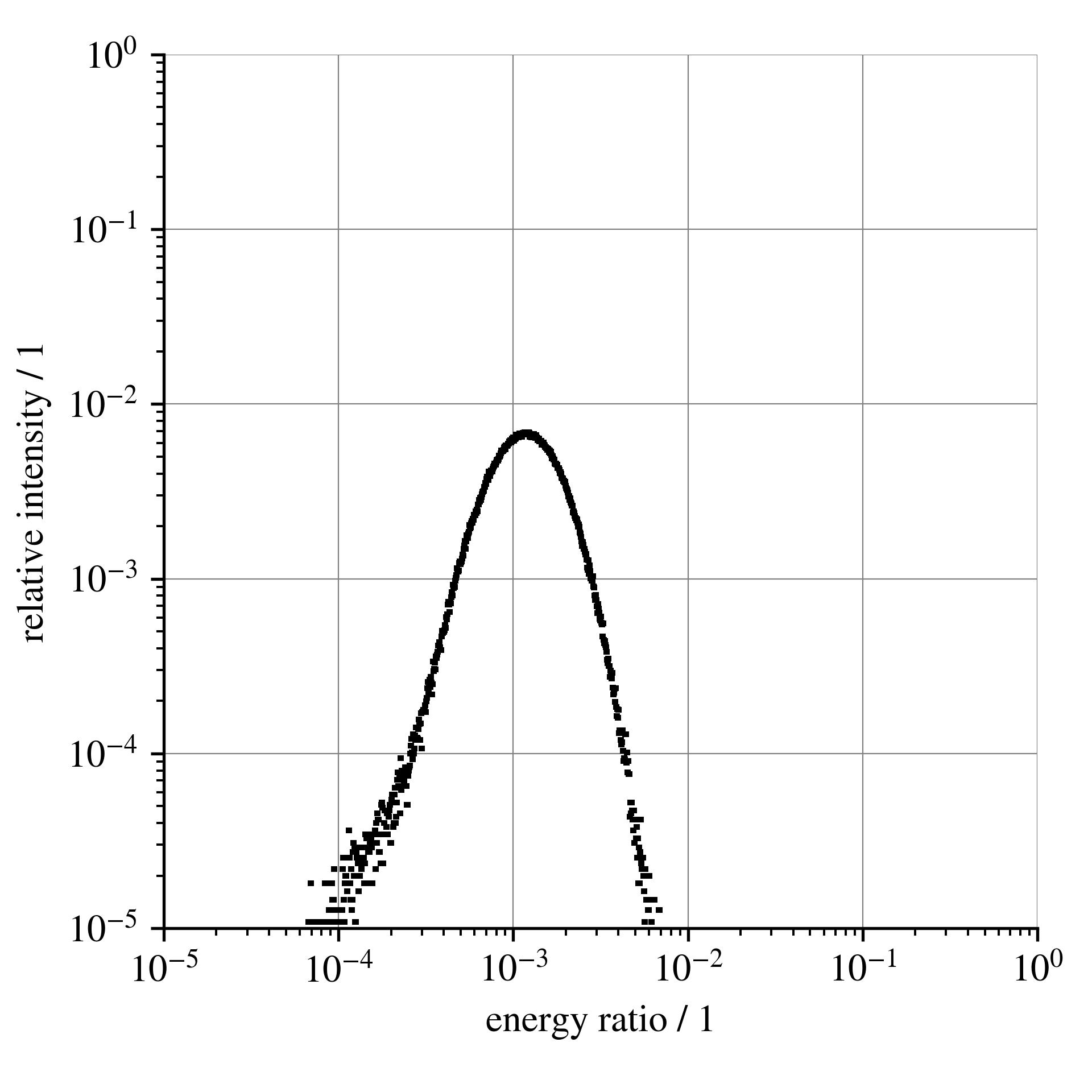}
                \label{FigLargeArtifactPowerRatio}
            }
            \caption{
                Relative energy distribution among camera screen scatter centers when the plane wave is coming from significantly outside of the telescope's field-of-view.
            }
            \label{FigArtifactEnergyRatio}
        \end{figure}

        To investigate artifacts induced by the Huygens image formation, one can set up the calibration source to shine plane wave packages which run significantly outside of the telescope's field-of-view.
        In this case, one expects the response of all the feed horns to be zero.
        \autoref{FigArtifactEnergyRatio} histograms the relative energy received by the camera screen's scatter centers.
        Here one finds that most scatter centers in the camera's screen contain an energy artifact which is roughly the energy carried by the plane wave divided by the number of mirror scatter centers $M$.
        This is expected giving the summation over $M$ scatter centers in \autoref{EqHuygensSuperposition} as it implies that the best possible contrast in the images is also in the order of $M$.

\begin{figure}
            \centering
            \includegraphics[width=0.9\columnwidth]{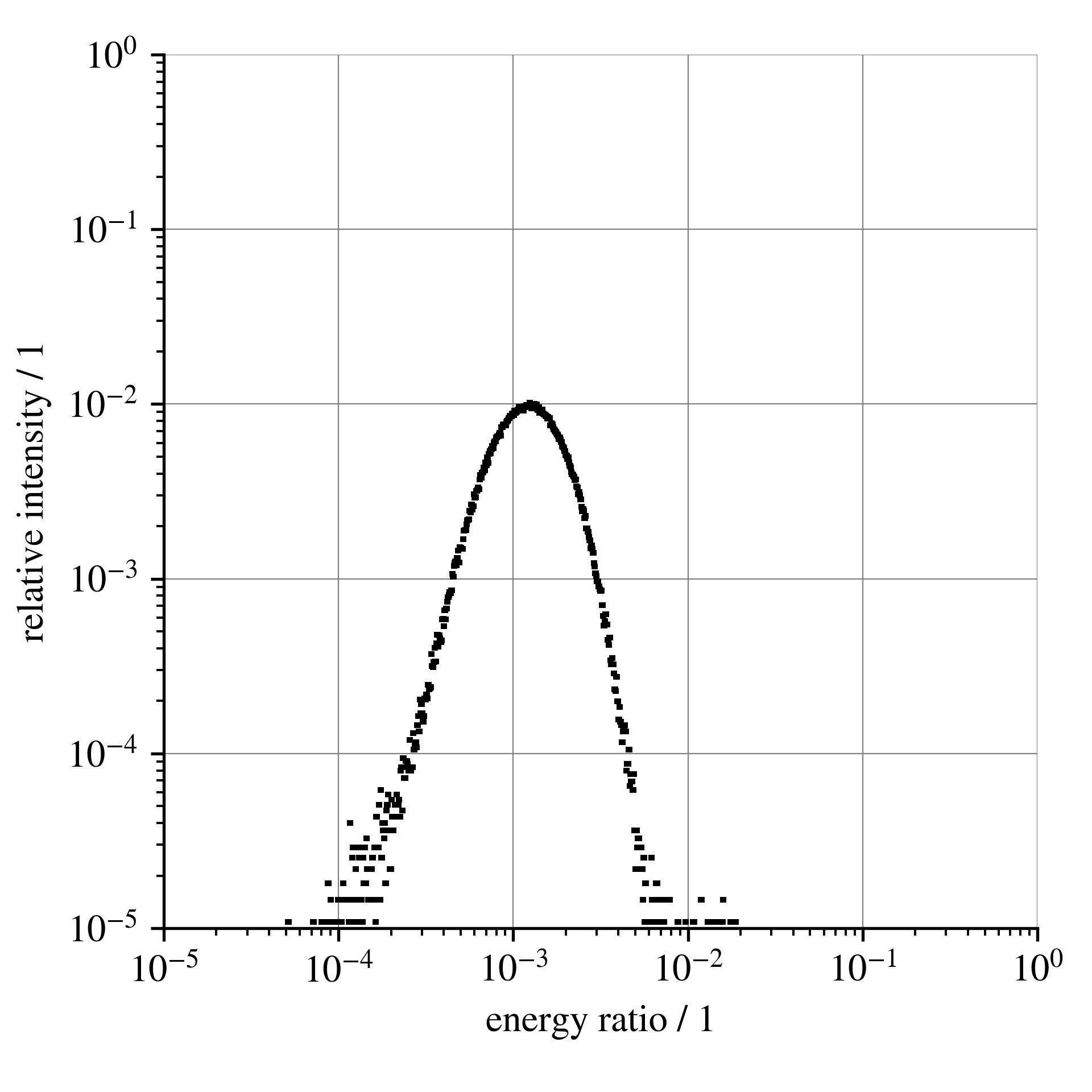}
            \caption{
                Same as panel \autoref{FigLargeArtifactPowerRatio} but the mirror's scatter centers are placed using a square grid with randomness.
                Apparently the randomness here was not strong enough and thus outliers contain over $1\%$ of the plane wave packages energy.
                Resulting image artifacts are shown in \autoref{FigLargeArtifactExampleSquareGrid}.
            }
            \label{FigLargeArtifactEnergyRatioSquareGrid}
        \end{figure}

\begin{figure}
            \centering
            \includegraphics[width=1\columnwidth]{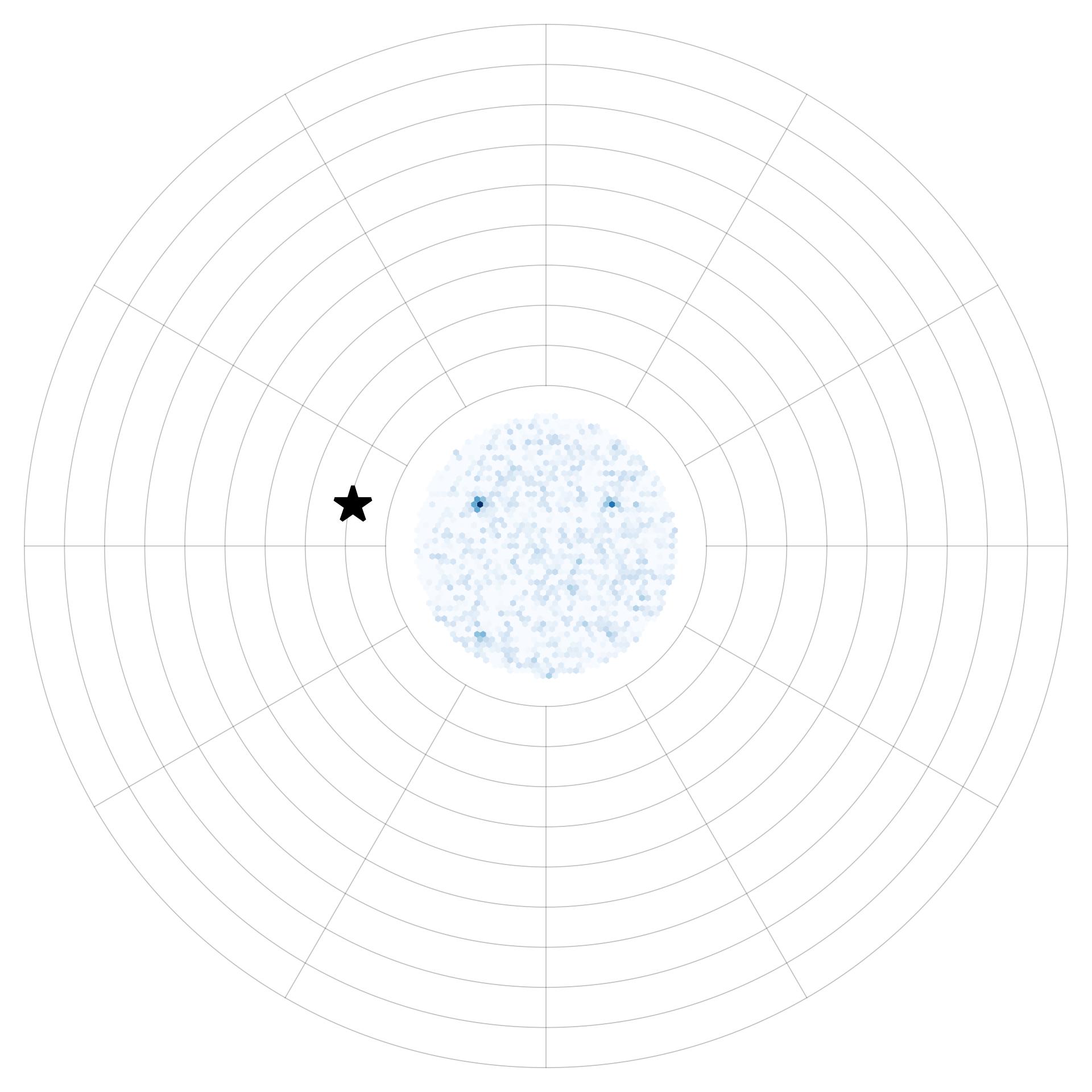}
            \includegraphics[width=1\columnwidth]{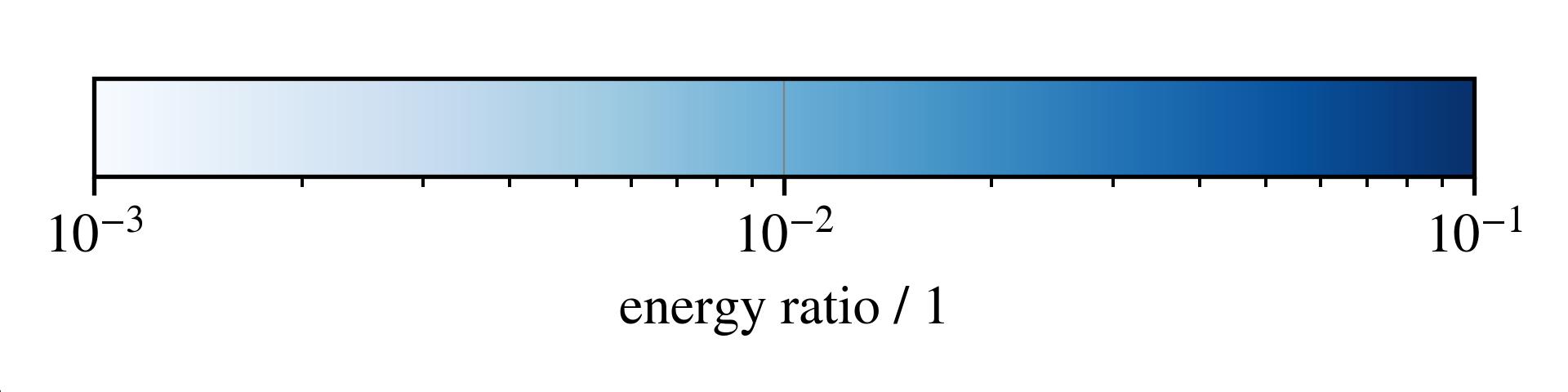}
            \caption{
                Ghosting artifact on the $\mathbf{L}$arge telescope caused by interference due to the mirror's scatter centers being placed using a square grid with randomness.
                The source emitting the plane wave is located at the black star what is over $1.5^{\circ}$ outside of the field-of-view.
                The concentric black lines show the sky in $1^{\circ}$ radial steps.
                Camera shows the energy deposited in the feed horns in blue.
                Three bright spots, noticeable being arranged in a square grid, carry each over $1\%$ of the plane wave package's energy.
                Compare Fig.\,\autoref{FigLargeArtifactEnergyRatioSquareGrid}.
            }
            \label{FigLargeArtifactExampleSquareGrid}
        \end{figure}

        The histograms in \autoref{FigArtifactEnergyRatio} do not show long tails towards higher energy containment but instead vanish quickly.
        Otherwise, scatter centers with a high relative energy containment would introduce an artifact pattern in the image.
        With the Huygens image formation and telescope representation discussed here, the occurrence of visible artifact patterns in the image noticeably depends on the arrangement of the scatter centers in the mirror.
        For example, arranging the mirror's scatter centers in a square grid, even with added randomness, causes significant artifacts as shown in \autoref{FigLargeArtifactEnergyRatioSquareGrid} and \autoref{FigLargeArtifactExampleSquareGrid}.
        \autoref{SecArrangingScatterCenters} discusses different arrangement strategies for the mirror's scatter centers.
    \section{Concluding on Huygens image formation}
    \label{SecConcludingHuygens}
    The Huygens based image formation presented in \autoref{SecHuygensPrinciple} leaves room for improvement.
    Deviations from expectations can be found, energy conservation is delicate, and finding a balance between the simulation's resolution and compute effort is challenging.

    Yet the Huygens based image formation presented here reproduces not only predictions by wave mechanics but also predictions by geometric optics.
    Features and limitations of an actual telescope such as Airy's disk, aberrations, distortions, and the narrowing depth-of-field are all reproduced naturally.
    Observations of two simultaneous plane wave packages, as shown in \autoref{FigMediumMulitExample} and \autoref{FigLargeMulitExample}, indicate that the imaging of complex air showers might be possible.
    \section{Air-shower images}
    \label{SecAirshowerImages}
    To investigate the imaging of extensive air showers, radio emission simulations were performed using \textsc{CORSIKA}--\textsc{CoREAS}. The observation level was set to 110\,m above sea level, with an Earth magnetic field of $B_x = 20.4\,\mu\mathrm{T}$ and $B_z = 43.23\,\mu\mathrm{T}$, similar to the site of CROME. As a proof of concept, the simulated showers were imaged using the Large telescope configuration.

    \subsection{Validation of the Huygens image formation}

\begin{figure*}
            \centering
            \includegraphics[width=0.8\textwidth]{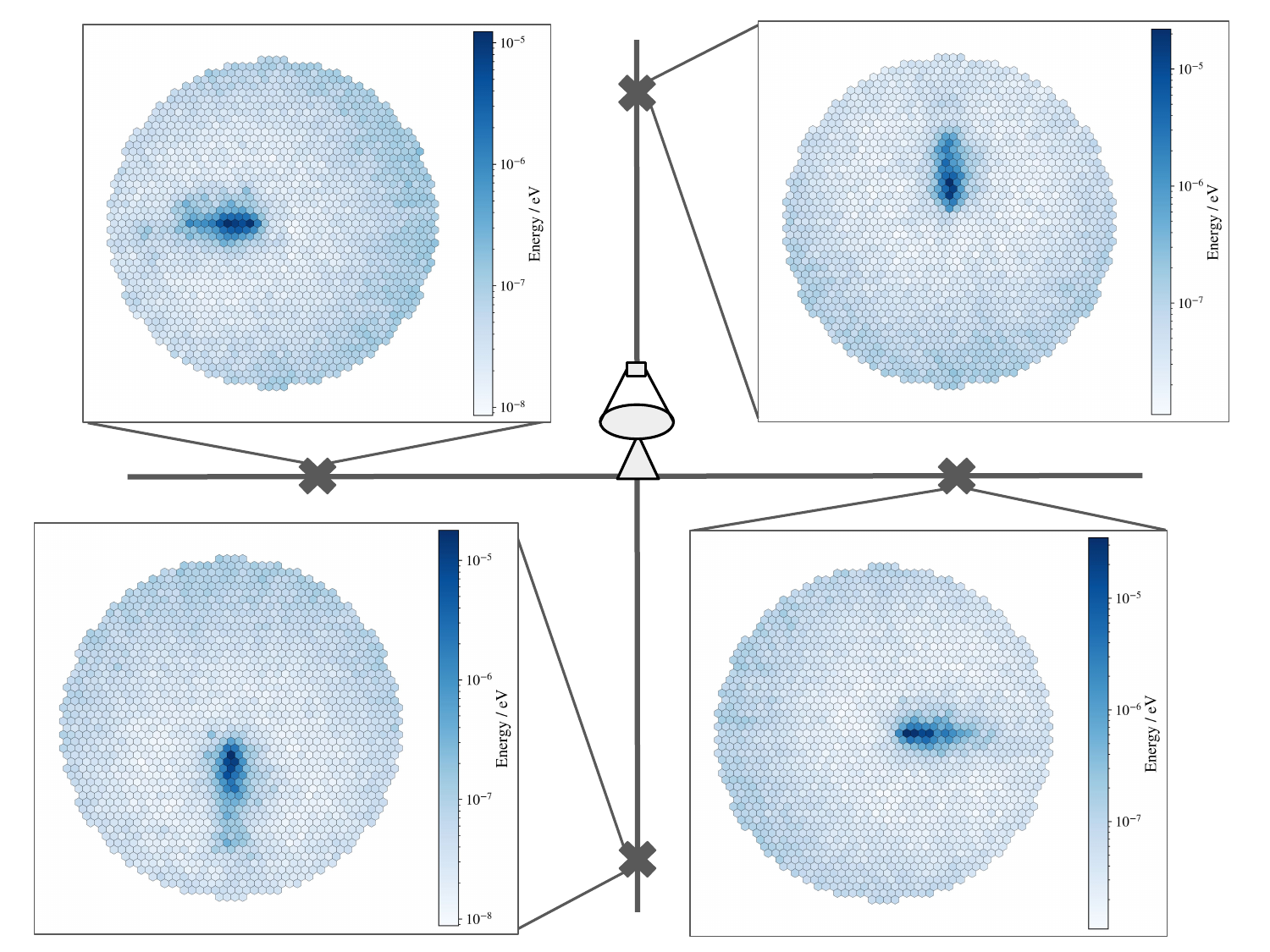}
            \caption{
                Camera images for gamma-ray initiated air showers with a primary energy of 1~TeV arriving from different directions.
                The respective orientations of the image across the camera plane reflects the change in arrival direction and validates the Huygens based image formation approach.
            }
            \label{fig:IART_directional}
        \end{figure*}

        As a first validation step, the Huygens based image formation approach is tested by imaging air showers arriving from different directions.
        ~\autoref{fig:IART_directional} shows camera images of gamma-ray–induced air showers with a primary energy of 1 TeV. The showers were simulated to arrive from positions displaced by $\pm100$ m in the north–south and east–west directions relative to the telescope. In all cases, the zenith angle was $0^\circ$, and the telescope was pointed toward zenith.
        For the camera images, the electric field is restricted to frequencies between 7 and 10 GHz using a Butterworth bandpass filter. The filtered electric field traces are then used to calculate the energy deposited in the antenna.

        The systematic displacement of the image across the camera plane with changing arrival direction demonstrates that the Huygens image formation correctly preserves directional information.
        This confirms that the imaging approach produces physically meaningful images and provides a consistent mapping between shower geometry and camera coordinates.

    \subsection{Particle-dependent images at the Cherenkov cone}

        The coherent radio emission from extensive air showers is strongly enhanced near the Cherenkov angle due to relativistic time-compression effects caused by the refractive index of the atmosphere.
        This leads to a characteristic ring-like intensity pattern on the ground, where the radio signal is strongest and the pulse duration is shortest.
        For vertical air showers, this Cherenkov ring is expected at lateral distances of order 100--120~m from the shower axis, depending on the height of the shower maximum. 
        Such Cherenkov rings have been observed in radio measurements above 100~MHz and are well reproduced by simulations that include atmospheric refractive-index effects~\cite{Nelles_2015}. 

\begin{figure*}
            \centering
            \includegraphics[width=0.9\textwidth]{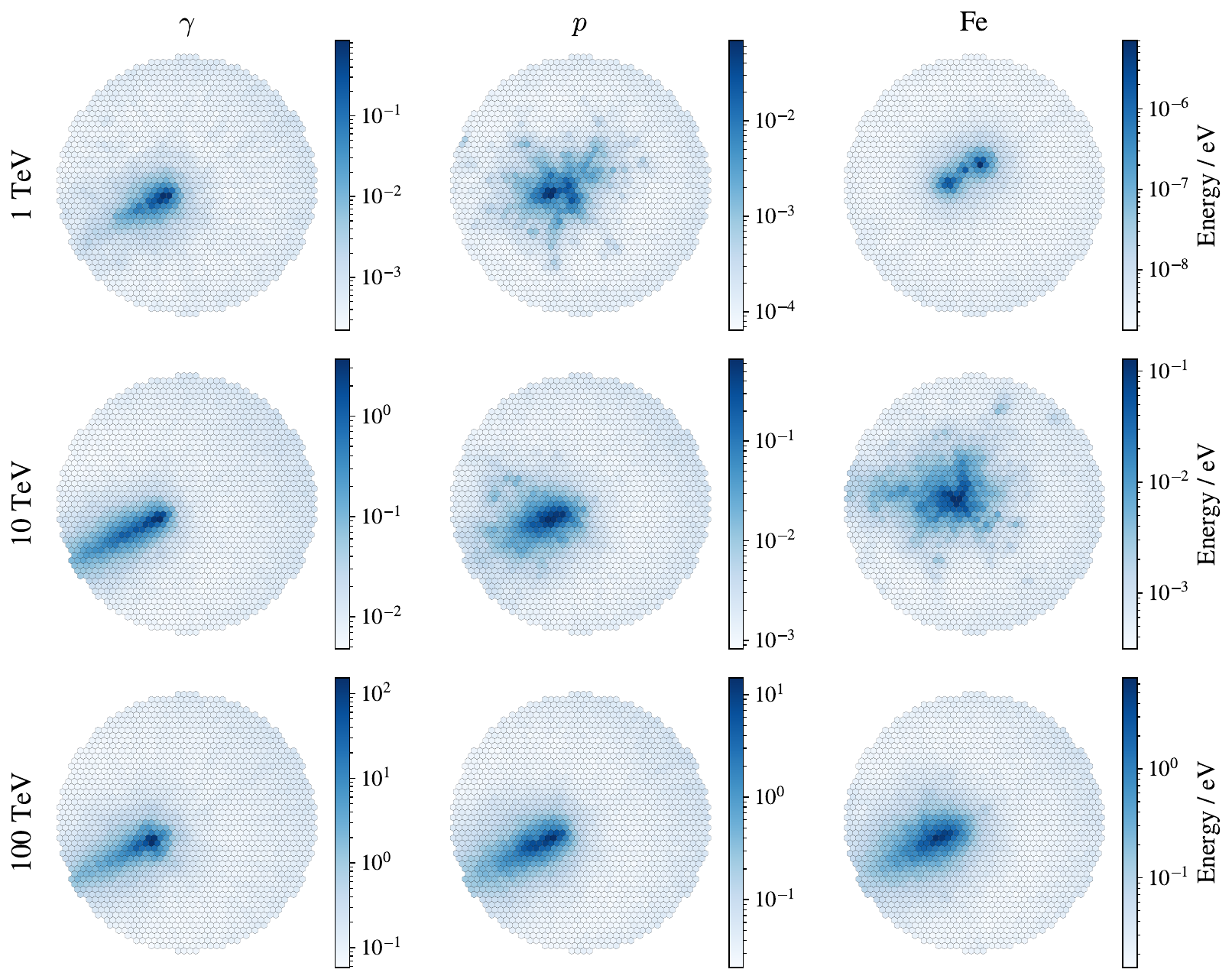}
            \caption{
                Deposited energy in each feedhorn for gamma-ray-, proton-, and iron-initiated air showers with primary energies of 1~TeV, 10~TeV, and 100~TeV, simulated at a lateral distance of approximately 112~m from the shower axis.
            }
            \label{fig:deposited_power_comparison_particles}
        \end{figure*}

        The telescope was therefore positioned at a lateral distance of approximately 112~m from the shower impact point on the ground, for all following simulations, placing it close to the expected Cherenkov cone.
        ~\autoref{fig:deposited_power_comparison_particles} shows the deposited power in each feedhorn for gamma-ray-, proton-, and iron-initiated air showers with primary energies of 1~TeV, 10~TeV, and 100~TeV.
        The resulting camera images show characteristic differences between electromagnetic and hadronic showers. 
        For a given primary particle type, increasing the primary energy leads to a proportional increase in the radio signal amplitude. 
        At fixed primary energy, the signal strength decreases with increasing nuclear charge of the primary particle, as the electromagnetic component is reduced and fragmented by multiple hadronic subshowers, leading to broader and more irregular radio morphologies. 
        At higher energies, the differences in image morphology between primary particle types become less pronounced in the simulated images.

        As a result, higher-energy proton or iron showers can generate radio images with amplitudes and morphologies comparable to those of lower-energy gamma-ray showers, creating a partial degeneracy between primary energy and composition. 
        These effects are evident in the simulated camera images and emphasize the importance of both morphological and frequency-dependent information for disentangling shower energy from the primary particle type.

    \subsection{Frequency-dependent imaging with idealized bandpass filters}

\begin{figure*}
            \centering
            \includegraphics[width=1\textwidth]{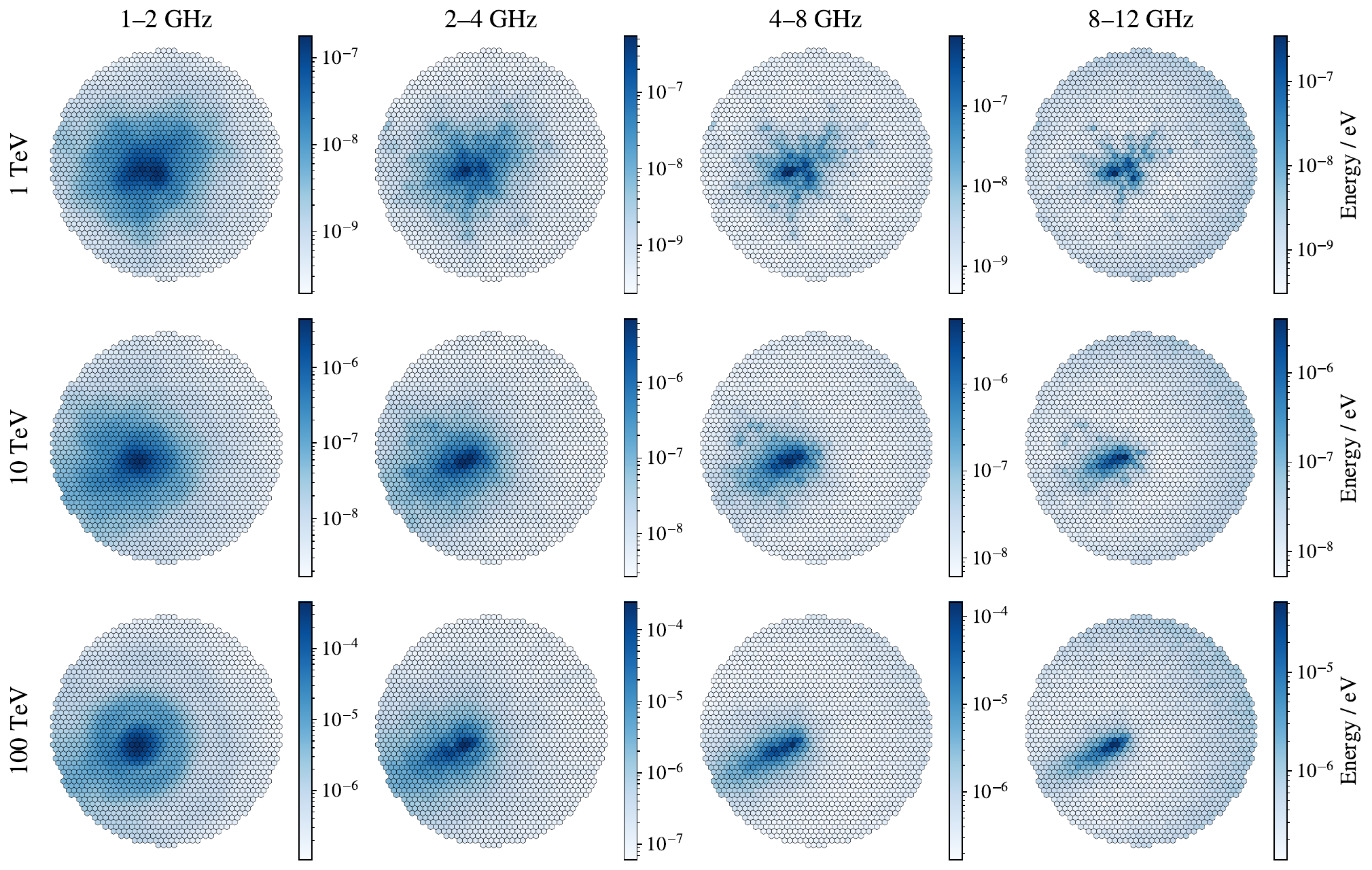}
            \caption{
                Frequency-resolved camera images for proton-initiated air showers obtained using idealized bandpass filters.
                From left to right, the frequency bands are 1--2~GHz, 2--4~GHz, 4--8~GHz, and 8--12~GHz and primary energies of 1~TeV, 10~TeV, and 100~TeV, simulated at a lateral distance of approximately 112~m from the shower axis.
            }
            \label{fig:proton_field_amplitude_freq_grid}
        \end{figure*}

        To isolate the intrinsic frequency dependence of the radio emission, the raw electric-fields are analysed using idealized bandpass filters. 
        ~\autoref{fig:proton_field_amplitude_freq_grid} shows frequency-resolved camera images for proton-initiated air showers in four frequency bands: 1-2~GHz, 2-4~GHz, 4-8~GHz, and 8--12~GHz and primary energies of 1~TeV, 10~TeV, and 100~TeV. The octave-spaced frequency bands are chosen like this as they provide approximately self-similar feed scaling and match existing microwave receiver technologies.

        At the lowest frequencies, 1-2~GHz, the emission is coherent across a large portions of the shower. 
        This produces broad, smooth illumination patterns that primarily reflect the total shower energy rather than detailed geometry, as contributions from most of the shower add constructively.

        In the 2-4~GHz band, large-scale coherence is reduced, and the emission becomes more sensitive to the shower geometry. 
        This partial coherence leads to more localized images, with asymmetries that begin to reveal the shower orientation and variations in the emission.

        At 4-8~GHz, coherence is maintained only over small regions. 
        The emission is concentrated along the projected shower axis, highlighting fine-scale structure and subshower contributions, while incoherent emission from distant parts of the shower is suppressed.

        At the highest frequencies, 8-12~GHz, coherence is largely lost except over very limited regions. 
        The emission sharply resolves the shower core and axis orientation, but the overall amplitude drops, indicating a trade-off between spatial resolution and detectability.
    \section{Concluding on air-shower imaging}
    \label{SecConcludingAirShowerImages}
    
    The simulated radio images of extensive air showers demonstrate that the Huygens-based image-formation approach preserves key features of extensive air showers throughout the full simulation chain.
    Directional information and particle-dependent morphologies are correctly reproduced, providing a clear proof of principle for imaging atmospheric radio telescopes under idealized, noise-free conditions.

    Gamma-ray- and hadron-induced shower images resemble those known from imaging atmospheric Cherenkov telescopes, suggesting that established IACT reconstruction techniques may be transferable to the radio imaging domain.
    Frequency-dependent imaging further enables disentangling shower geometry, highlighting the potential of multi-band radio observations for detailed air-shower studies.
    These results demonstrate the promise of imaging extensive air showers with an atmospheric radio telescope and motivate future studies including noise, detector performance, and quantitative reconstruction accuracy.

    \section{Outlook}
    \label{SecOutlook}
    A natural next step of this study is the inclusion of a realistic receiver chain and its noise contributions. 
    While this present study focuses on noise-free signals to establish a proof of principle, electronic noise from the receiver chain and diffuse Galactic emission will influence both the detectability and the morphology of the shower images. 
    Incorporating these noise sources will allow for more realistic estimates of signal-to-noise ratios, image cleaning procedures, and trigger thresholds, and will provide a more accurate assessment of the achievable energy threshold.
    Extending the simulations to air showers arriving at different zenith angles will further clarify the imaging performance of an atmospheric radio telescope. 
    Inclined showers are expected to produce different image morphologies due to projection effects, changes in the distance to the shower maximum, and the increased extent of the radio footprint. 
    Studying these dependencies is essential for understanding the stability of image parameters under varying observational geometries.
    In addition, simulations covering a broader range of primary energies and particle types will enable a systematic investigation of gamma-hadron separation using radio images. 
    Differences in image shape, intensity distribution, and substructure may provide discrimination power analogous to techniques employed in imaging atmospheric Cherenkov telescopes.
    Also, the impact of different telescope designs should be explored. 
    Varying the telescope size and hence the angular resolution and collection power will directly affect image quality and sensitivity. 
    Additionally, the telescope's position relative to the expected Cherenkov ring should be systematically studied. 
    Optimal placement can maximize signal strength and preserve fine-scale image features, while positions far from the ring may result in weaker, more diffuse images. 
    Understanding this dependence is crucial for both telescope array layout and event reconstruction.
    Such studies are required to optimize the telescope design and to assess the feasibility and scientific reach of future imaging atmospheric radio telescope concepts.
    Finally, a new use case in neutrino detection could be explored.
    The energy threshold in the PeV range makes the detection of Earth-skimming tau neutrinos in this energy regime a compelling case to investigate.
    Adapting the telescope structure to a design similar to MACHETE or Trinity could enable the imaging of showers induced by such Earth-skimming tau neutrinos.
    Such an extension would broaden the applicability of this technique and open a new avenue of research at the intersection of radio and neutrino astronomy.
    \section*{Acknowledgments}
    This work would not exist without the backing provided by Jim Hinton (MPIK-Heidelberg).
    We acknowledge the discussions with Christian Monstein (ETH-Zurich) on radio~antennas and analog electronics as well as the discussions with Kai Bruegge (TU-Dortmund).
    S.A.M. fondly recalls when Wolfgang Rhode (TU-Dortmund) introduced him to radio astronomy at the Astropeiler Stockert, Germany.
    \bibliographystyle{elsarticle-harv}
    \bibliography{sebastians_references,annes_references}
    \section*{Appendix}
    \appendix
    \section{Motivating the components of the global weight $w$}
        \label{SecSuggestingEnergyWeight}
        One demands that the power received by the central feed horn must be the total power which arrives on the mirror except for the containment fraction $Q$
        \begin{eqnarray}
            P_\text{central-feed-horn}{} &=& Q \sum_{m}^{M} {P_\text{SM}}_\MirrorIndex{}.
            \label{EqPointSpreadFunctionQuantileContainedInFeedHorn}
        \end{eqnarray}
        As each feed horn is represented by $\NumScatterCentersPerFeedHorn{}$ scatter centers one can express the power received by a feed horn as the sum of the powers received by its scatter centers
        \begin{eqnarray}
            \sum_{n}^{N} {P_\text{SF}}_n
             &=& Q \sum_{m}^{M} {P_\text{SM}}_m.
            \label{Eq14}
        \end{eqnarray}
        Next one uses that in the far field approximation the power received by a scatter center is
        \begin{eqnarray}
            {P_\text{SM}} &=& A_\text{SM} \frac{(\ElectricField{}_\text{SM})^2}{Z}
            \label{Eq15}
        \end{eqnarray}
        and
        \begin{eqnarray}
            {P_\text{SF}} &=& A_\text{SF} \frac{(\ElectricField{}_\text{SF})^2}{Z}
            \label{Eq16}
        \end{eqnarray}
        respectively, what leads to
        \begin{eqnarray}
            A_\text{SF} \sum_{n}^{N} {{\epsilon{}_\text{SF}}_n}^2
             &=& Q A_\text{SM} \sum_{m}^{M} {{\epsilon{}_\text{SM}}_m}^2.
            \label{Eq18}
        \end{eqnarray}
        To work around the squaring of electric fields, one uses that the variance among the electric field amplitudes on the mirror and the screen
        \begin{eqnarray}
            \text{VAR}[\epsilon_\text{SM}] &=& 0\\
            \text{VAR}[\epsilon_\text{SF}] &=& 0
            \label{EqMirrorElectricFieldVariance}
        \end{eqnarray}
        can be neglected.
        This implies that the squared mean of the mirror's and camera's electric filed amplitudes
        \begin{eqnarray}
            \label{EqMirrorElectricFieldAverageSquared}
            {\text{E}[\epsilon_\text{SM}]}^2
            &=&
            \frac{1}{M}
            \sum_{m}^{M}
            {\epsilon_\text{SM}}_m^2\\
            \label{EqCameraElectricFieldAverageSquared}
            {\text{E}[\epsilon_\text{SF}]}^2
            &=&
            \frac{1}{N}
            \sum_{n}^{N}
            {\epsilon_\text{SF}}_n^2
        \end{eqnarray}
        is equal to the sum of individually squared electric field amplitudes.
        Putting \autoref{EqMirrorElectricFieldAverageSquared} and \autoref{EqCameraElectricFieldAverageSquared} into \autoref{Eq18} yields
        \begin{eqnarray}
            N \text{E}[{\epsilon{}_\text{SF}}]^2
             &=& Q \frac{A_\text{SM}}{A_\text{SF}} M \text{E}[{\epsilon{}_\text{SM}}]^2.
            \label{Eq20}
        \end{eqnarray}
        \begin{eqnarray}
            \text{E}[{\epsilon{}_\text{SF}}]^2
             &=& Q \frac{A_\text{SM}}{A_\text{SF}} \frac{M}{N} \text{E}[{\epsilon{}_\text{SM}}]^2
            \label{Eq21}
        \end{eqnarray}
        \begin{eqnarray}
            \text{E}[\epsilon{}_\text{SF}]
             &=& \sqrt{Q \frac{A_\text{SM}}{A_\text{SF}} \frac{M}{N}} \text{E}[{\epsilon{}_\text{SM}}]
            \label{Eq21p1}
        \end{eqnarray}
        using the definition of the mean yields
        \begin{eqnarray}
            \text{E}[\epsilon{}_\text{SF}]
             &=& \sqrt{Q \frac{A_\text{SM}}{A_\text{SF}} \frac{M}{N}} \frac{1}{M} \sum_{m}^{M} {\epsilon{}_\text{SM}}_m
            \label{Eq21p2}
        \end{eqnarray}
        and using VAR$[\epsilon{}_\text{SF}] = 0$ again implies that E[$\epsilon{}_\text{SF}$] can be replaced by any of the feed horns scatter centers $n$.
        \begin{eqnarray}
            {\epsilon{}_\text{SF}}_n
             &=& \sqrt{\frac{A_\text{SM}}{A_\text{SF}} \frac{Q}{NM}} \sum_{m}^{M} {\epsilon{}_\text{SM}}_m
            \label{Eq22}
        \end{eqnarray}
        Comparing \autoref{Eq22} with \autoref{EqHuygensSuperposition} implies that the weight $w$ must be
        \begin{eqnarray}
            w &=& \sqrt{\frac{A_\text{SM}}{A_\text{SF}} \frac{Q}{NM}}
            \label{Eq23}
        \end{eqnarray}
        in order to conserve the energy.
    \section{Arranging scatter centers in the mirror}
        \label{SecArrangingScatterCenters}
        We experimented with different strategies to arrange the scatter centers in the mirror.
        This is challenging because there are two mechanisms working against each other.
        On the one hand side one wants to minimize the compute effort by using a small number of $M$ scatter centers which sample the mirror as uniformly as possible.
        Uniformity is important for the Huygens image formation in \autoref{EqHuygensSuperposition} what assumes that all scatter centers in the mirror represent areas of the mirror with equal sizes.
        But on the other hand side one wants to minimize the risk of introducing interference patterns which show up as ghosting artifacts in the images as shown in \autoref{SecArtifacts}.
        Here is a qualitative comparison of the arranging methods we have used.

        The best compromise so far is what we call `uniform randomness while rejecting high density'.
        It is shown in Fig.\,\autoref{FigCromeGeometryCloseUpMirror} and it is used for the studies shown here.
        One fills the mirror's annulus with random points drawn one by one from a uniform distribution in the $x$-$y$-plane.
        But one rejects points which already have too many neighbors in their proximity.
        The resulting images show the least artifacts so far but they are not the most flat and uniform ones.

        Second best is a Fibonacci spiral with additional uniform randomness.
        This arrangement induces ring like artifacts in the images but beside from this it gives the best results.
        Especially the out-of-focus images discussed in \autoref{SecObservingOutOfFocus} are noticeable more uniform and flat.

        The worst arrangement so far was a square grid with added uniform randomness.
        The resulting artifacts are shown in \autoref{FigLargeArtifactEnergyRatioSquareGrid} and \autoref{FigLargeArtifactExampleSquareGrid}.
        Especially with the plane wave calibration source there are noticeable artifact spots in the image which look like the expected spot of the source but are arranged around it in a square grid pattern.
    \section*{Author Contributions}
        %
        \textbf{Sebastian Achim Mueller:} Project administration, Writing, Conceptualization, Methodology,
        Investigation, Formal analysis, Visualization, Validation, Software.

        \textbf{Anne Timmermans:} Writing, Editing, Conceptualization, Methodology, Investigation,
        Formal analysis, Visualization, Validation, Software.

        \textbf{Juan Ammerman-Yebra:} Conceptualization, Investigation, Validation and Writing–review, Editing.

        \textbf{Harm Schoorlemmer:} Conceptualization, Investigation, Validation and Writing–review, Editing.
    \section*{Additional Information}
        Computer~simulations are public on \url{https://github.com/relleums/imaging-atmospheric-askaryan-telescope}.
        Correspondence and requests for materials should be addressed to S.A.M.
    \section*{Competing Financial Interests}
        The authors declare no competing financial interests.
\end{document}